\documentclass[aps,prx,twocolumn,preprintnumbers,superscriptaddress,10pt,showkeys]{revtex4-2}
\usepackage{graphicx}
\usepackage{dcolumn}
\usepackage{bm}
\usepackage{colortbl}
\usepackage{hyperref}
\hypersetup{
  colorlinks=true,        
  linkcolor=blue,         
  citecolor=blue,         
  urlcolor=blue           
}
\usepackage{amsmath}
\usepackage{times}
\usepackage{tensor}

\begin{document}

\title{Prompt Black Hole Formation in Binary Neutron Star Mergers}
\author{Christian Ecker}
\email{ecker@itp.uni-frankfurt.de}
\affiliation{Institut f\"ur Theoretische Physik, Goethe Universit\"at, Max-von-Laue-Str. 1, 60438 Frankfurt am Main, Germany}
\author{Konrad Topolski}
\email{topolski@itp.uni-frankfurt.de}
\affiliation{Institut f\"ur Theoretische Physik, Goethe Universit\"at,  Max-von-Laue-Str. 1, 60438 Frankfurt am Main, Germany}
\author{Matti J\"arvinen}
\email{matti.jarvinen@apctp.org}
\affiliation{Asia Pacific Center for Theoretical Physics, Pohang, 37673, Korea}
\affiliation{Department of Physics, Pohang University of Science and Technology, Pohang, 37673, Korea}
\author{Alina Stehr}
\email{stehr@itp.uni-frankfurt.de}
\affiliation{Institut f\"ur Theoretische Physik, Goethe Universit\"at,  Max-von-Laue-Str. 1, 60438 Frankfurt am Main, Germany}

\begin{abstract}
\noindent
We carry out an in-depth analysis of the prompt-collapse behaviour of binary neutron star (BNS) mergers. 
To this end, we perform more than $80$ general relativistic BNS merger simulations using a family of realistic Equations of State (EOS) with different stiffness, 
which feature a first order deconfinement phase transition between hadronic and quark matter. 
From these simulations we infer the critical binary mass $M_{\rm crit}$ that separates the prompt from the non-prompt collapse regime.
We show that the critical mass increases with the stiffness of the EOS
and obeys 
a tight quasi-universal relation, $M_{\rm crit}/M_{\rm TOV}\approx 1.41\pm 0.06$, which links it to the maximum mass $M_{\rm TOV}$ of static neutron stars, and therefore provides a straightforward estimate for the total binary mass beyond which prompt collapse becomes inevitable.
In addition, we introduce a novel gauge independent definition for a one-parameter family of threshold masses in terms of curvature invariants of the Riemann tensor which characterizes the development toward a more rapid collapse with increasing binary mass. 
Using these diagnostics, we find that the amount of matter remaining outside the black hole sharply drops in supercritical mass mergers compared to subcritical ones and is further reduced in mergers where the black hole collapse is induced by the formation of a quark matter core. 
This implies that $M_{\rm crit}$, particularly for merger remnants featuring quark matter cores, imposes a strict upper limit on the emission of any detectable electromagnetic counterpart in BNS mergers.

\end{abstract}
\preprint{APCTP Pre2024 - 006}
\maketitle

\section{Introduction}
Neutron stars have become leading laboratories to study dense matter in the course of the past decade. 
As a key component in this development, multi-messenger signals from binary neutron star (BNS) mergers include information about the properties of strongly interacting matter at densities not accessible to terrestrial experiments.
The detection of gravitational waves (GW) from the GW170817 event~\cite{LIGOScientific:2017vwq} already provided constraints on the neutron star Equation of State (EOS)~\cite{LIGOScientific:2018cki} that complement those from direct mass~\cite{Antoniadis:2013pzd,NANOGrav:2019jur,Fonseca:2021wxt,Romani:2022jhd} and radius measurements~\cite{Miller:2019cac,Riley:2019yda,Riley:2021pdl,Miller:2021qha,Barr:2024wwl} as well as from microphysics calculations in nuclear theory~\cite{Hebeler:2013nza,Gandolfi:2019zpj,Keller:2020qhx,Drischler:2020yad} and perturbative Quantum Chromodynamics (QCD)~\cite{Freedman:1976ub,Vuorinen:2003fs,Gorda:2021kme,Gorda:2021znl}.
Neutron stars with masses close to the maximal limiting value set by collapse into a black hole are particularly interesting for EOS inference, since their cores reach densities multiple times larger than the nuclear saturation density $n_s=0.16/{\rm fm}^{3}$.
Our theoretical understanding of QCD, the fundamental theory for matter, at these densities is very limited, which makes observations of highly massive neutron stars and their mergers currently the only reliable source of information about their material properties.

There exists a simple and well defined notion for the limiting mass of isolated static ($M_{\rm TOV}$) and rotating stars ($M_{\rm max}$), beyond which they become unstable against gravitational attraction and collapse into a black hole, in terms of the so-called turning-point criterion~\cite{1988ApJ...325..722F}.
While the values of $M_{\rm TOV}$ and $M_{\rm max}$ individually depend strongly on the EOS, in case of uniformly rotating stars they were shown to satisfy the quasi-universal (EOS independent) relation $M_{\rm max}\approx 1.2~M_{\rm TOV}$~\cite{Breu:2016ufb,Demircik:2020jkc,Musolino:2023edi}.
In contrast to isolated neutron stars, which are well approximated by a stationary 
and axially symmetric spacetime, the notion of a limiting mass is more intricate for dynamic systems like BNS mergers, because they evolve non-linearly in time and possess essentially no symmetries. Crucially, the 
time-dependent differential angular velocity profile of the merger remnant influences the stability and changes the simplified picture observed for isolated, uniformly rotating stars, so that simple recipes like the turning-point criterion do not apply.
Depending on the total mass and its division into its binary constituents, a BNS merger can either result in the prompt formation of a black hole or a (meta-)stable neutron star merger remnant, which may collapse into a black hole at later times.
These two scenarios feature different patterns of GW emission and electromagnetic (EM) radiation~\cite{Gottlieb:2023b}.

Significant effort has been put in the field towards  investigating the collapse of isolated neutron stars~\cite{Novak:2001ck,Noble:2007vf,Radice:2010rw} and black hole formation in neutron star head-on collisions~\cite{Jin:2006gm,Kellermann:2010rt}.
More recently the mass-threshold to prompt black hole formation has also been investigated in numerical simulations of BNS mergers starting from more realistic inspiral initial conditions~\cite{Bauswein:2013jpa,Koppel:2019pys,Agathos:2019sah,Bauswein:2020xlt,Tootle:2021umi,Kashyap:2021wzs} rather than head-on conditions. 
Such threshold mass analysis, in combination with observational data from the GW170817 event, lead to estimated lower bounds of neutron star radii~\cite{Bauswein:2017vtn,Koppel:2019pys} that agree surprisingly well with those from entirely different theoretical approaches~\cite{Altiparmak:2022bke}.
The aforementioned studies provided important insights about the collapse behaviour of BNS mergers, but also rely on various simplifying assumptions to estimate the threshold mass.
These assumptions include for example an approximate description for the gravitational field~\cite{Bauswein:2017vtn}, simple gauge dependent criteria based on a Newtonian notion of free-fall~\cite{Koppel:2019pys} or the use of EOS models that are in tension with some of the recent neutron star measurements that constrain the EOS at the densities and temperatures reached right before and during the collapse.

In this article we address these limitations by introducing the notion of the \textit{critical mass} $M_{\rm crit}$ that separates the \textit{prompt} from the \textit{non-prompt collapse} regime, as well as by suggesting a gauge independent definition for a one parameter family of threshold masses $M^{(p)}_{\rm th}$ in terms of curvature invariants of the Riemann tensor which does not rely on a Newtonian free-fall criterion.
We deduce the critical mass from a delay in the black hole formation time that appears generically in a small mass range close to the prompt collapse regime, while the threshold masses of promptness $p$ are inferred by demanding positivity of the $p$th time derivative of the spatial maximum of the Kretschmann scalar in the prompt collapse regime, therefore characterizing the competition between the gravitational pull and internal pressure of the hypermassive neutron star formed in the merger. 
We demonstrate the robustness and analyse the consequences of these definitions with a set of more than $80$ individual large-scale general relativistic hydrodynamic simulations of BNS mergers with different total masses populating the prompt and non-prompt collapse regime.
We find that especially the critical mass $M_{\rm crit}$ plays an important role, since it sets a strict upper bound beyond which essentially all matter ends up inside the black hole during the merger and where the possibility of any subsequent EM emission is strongly suppressed.

To account for uncertainties in the matter description we employ three different variants (soft, intermediate, stiff) of the holographic V-QCD hybrid EOS model~\cite{Demircik:2021zll}.
All models in this article are by construction consistent with nuclear theory and perturbative QCD at low and high densities, respectively, and in-between faithfully cover the central region of the EOS-uncertainty band of Ref.~\citep{Ecker:2022dlg} which has been obtained by combining recent constraints from neutron star mass-radius measurements and GW data.
A distinguished characteristic of the V-QCD framework is that it is non-perturbative and provides a unified description of both the nuclear and quark matter phases, and therefore also the (first order) phase transition between these phases. 
This feature allows us to track the amount of quark matter formed and study its impact on the lifetime of the post-merger remnant and the various limiting masses introduced in this article.  

To determine the impact of the phase transition on our results we construct and simulate for comparison also a model without quark matter phase that is otherwise identical to the intermediate variant of the V-QCD EOSs.
As we will show, near critical-mass mergers produce remnants that have the highest fraction of quark matter inside their cores.
For mergers with binary mass slightly above the critical value, this in turn leads to a quicker collapse with approximately one order of magnitude less residual matter outside the black hole and therefore a stronger suppression of EM emission in the model with quark matter than in the model without quark matter.
We argue that this suppression-mechanism may be useful to infer the existence of quark matter in neutron stars from future multi-messenger observations of BNS mergers.
Furthermore, we find that short-lived hypermassive neutron stars formed in binaries with total mass 
slightly below the critical value 
can experience periodic episodes of quark matter formation in their cores for several milliseconds before they collapse into a black hole.

\begin{figure}[bht]
   \centering
    \includegraphics[width=0.5\textwidth]{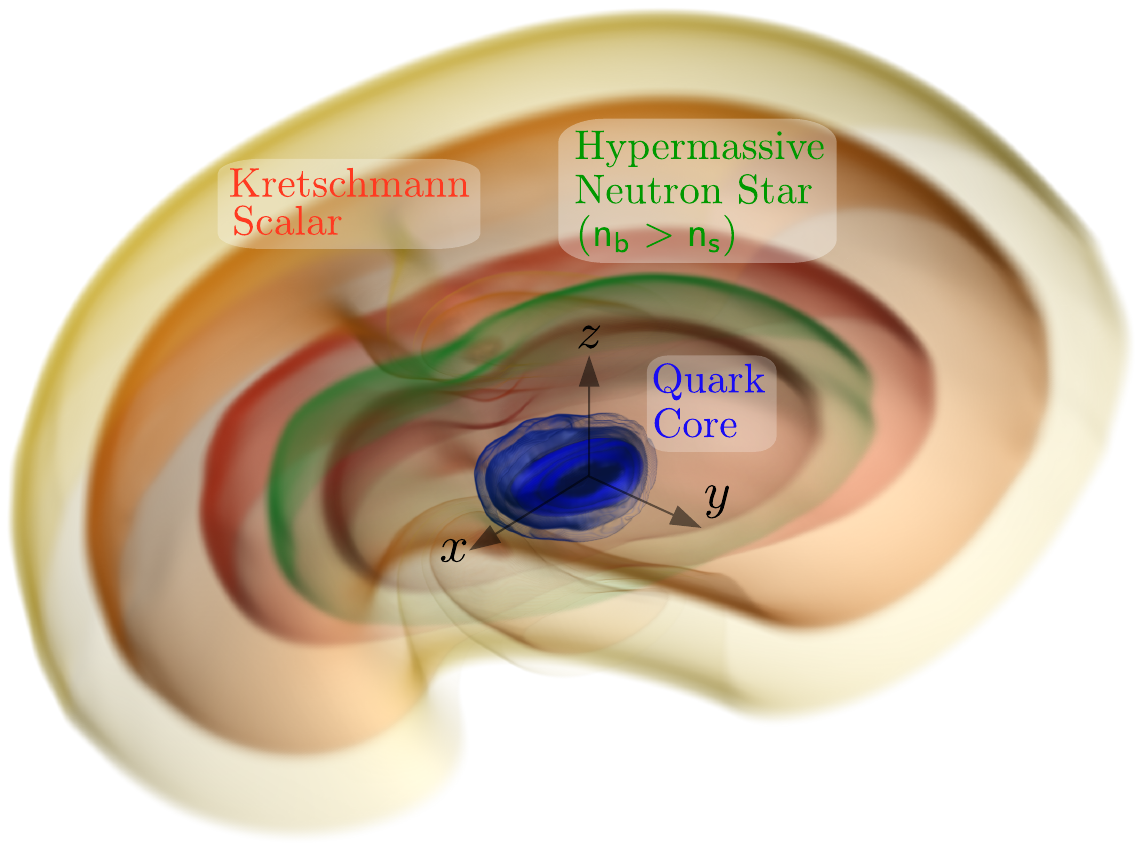}
    \caption{Iso-contours of the Kretschmann scalar (orange), the baryon number density enclosing values larger than the nuclear saturation density (green) and the fraction of quark matter (blue), for a remnant formed in a binary neutron star merger moments before collapse into a black hole.
    Light and dark colours indicate small and large values of these quantities, respectively (only results for negative values of $y$ are shown for clarity).}
\label{fig:setup}
\end{figure}

The 3D-rendering in Fig.~\ref{fig:setup} summarises the main ingredients of our analysis. 
Shown in orange are the iso-contours of the Kretschmann scalar, in green those of baryon number densities close to $n_s$ and in blue those of the volume fraction of deconfined quark matter in the core region, as they manifest in a post-merger remnant close to the critical mass only moments before it collapses into a black hole. 

The rest of this article is organized as follows.
Section~\ref{sec:EOS} summarises the properties of the V-QCD hybrid EOS models used.
In Sec.~\ref{sec:Simulation} we review the relevant definitions and properties of the curvature invariants used in our analysis, give definitions for $M_{\rm crit}$ and $M_{\rm th}^{(p)}$ and explain how to compute them in BNS merger simulations.
In Sec.~\ref{sec:Results} we present and interpret our results and end with a summary and conclusions in Sec.~\ref{sec:Conclusion}. 
For completeness, we provide four appendices with additional details on the EOS construction (App.~\ref{app:NMext}), a number of simulation results that are not shown in the main text (App.~\ref{app:MoreResults}), numeric checks with the Reissner-Nordstr{\"o}m geometry of the curvature invariants implemented in our numeric code (App.~\ref{app:RN}) as well as an comparison of our simulation results with the free-fall criterion (App.~\ref{app:freeFall}).  

Throughout this article we use natural units where $c=\hbar=k_B=G_N=1$.

\section{Equation of State}\label{sec:EOS}
Currently, calculations from-first-principles of even the most basic thermodynamic quantities as the phase diagram and the EOS at finite temperature and baryon density directly in QCD are illusive because of the non-perturbatively large coupling strength as well as the infamous fermionic sign-problem in lattice QCD.
Traditional approaches to tackle this problem include effective field theory models (see, e.g.,~\cite{Tews:2012fj,Keller:2020qhx,Oertel:2016bki}), which however need to be extrapolated to densities way beyond their range of applicability~\cite{Tews:2018iwm}, or generic (model independent) parametrizations of the EOS that are tuned to satisfy theoretical and observational constraints~\cite{Altiparmak:2022bke,Ecker:2022xxj,Ecker:2022dlg,Jiang:2022tps}, but do not contain any information about the particle composition.
A promising alternative are non-perturbative models for QCD that are based on the holographic gauge/gravity duality and where it is feasible to rigorously compute the phase structure and the EOS of theories that closely mimic the expected behaviour of QCD~\cite{BitaghsirFadafan:2020otb,Demircik:2021zll,Ghoroku:2021fos,Jarvinen:2021jbd,Kovensky:2021kzl,Hoyos:2021uff,Bartolini:2022rkl}.
\begin{figure*}
\center
 \includegraphics[width=0.95\textwidth]{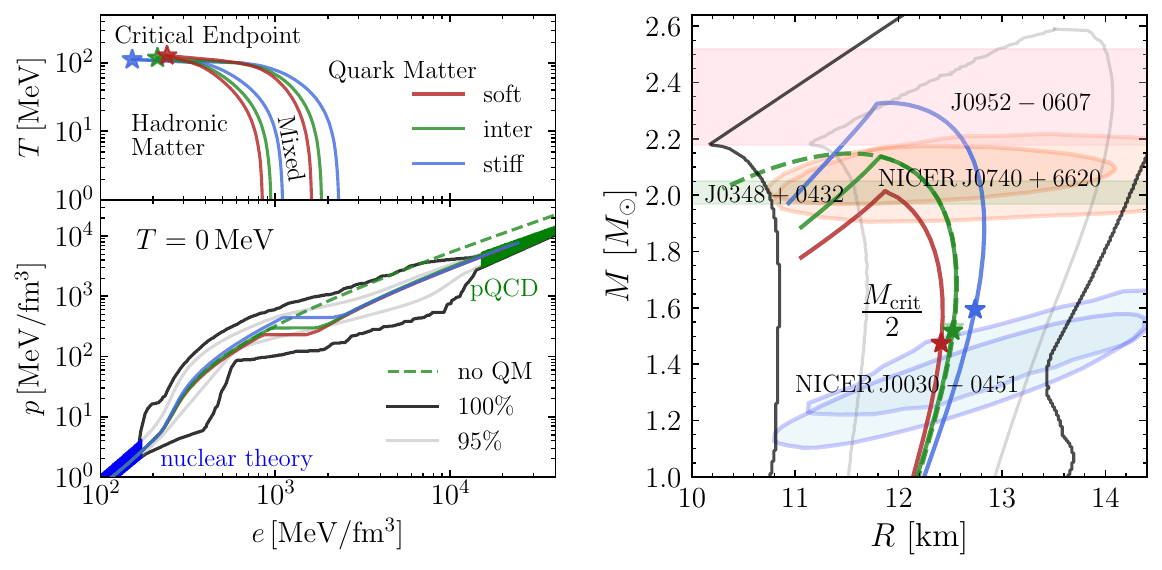}
 \caption{
  Top-Left: Phase boundaries between hadronic, mixed and quark matter together with the critical end points of the soft (red), intermediate (green) and stiff (blue) VQCD variants in the energy density and temperature plane.
  Bottom-Left: Cold, beta-equilibrium slices of VQCD together with a variant (dashed, labeled ``no QM'') of the intermediate model with suppressed first order phase transition.
  Blue and green areas denote the uncertainty bands from nuclear theory~\citep{Hebeler:2013nza} and perturbative QCD~\citep{Fraga:2013qra}, respectively.
  The solid black and grey lines enclose the $100\%$ and $95\%$ confidence intervals constructed from a large ensemble of generic EOS models interpolating between the colored bands~\citep{Ecker:2022dlg}, respectively.
  Right: Corresponding Mass-Radius relations where coloured stars mark (half) the critical mass and ellipses indicate the radius measurements  of J0030+0451~\cite{Miller:2019cac,Riley:2019yda} (blue) and J0740+6620~ \cite{Riley:2021pdl,Miller:2021qha} (orange) by the NICER experiment, while colored bands are direct mass measurements of J0348+0432~\cite{Antoniadis:2013pzd} (green) and J0952-0607~\cite{Romani:2022jhd} (pink).
 }
 \label{fig:EOS}
\end{figure*}

Formation of quark matter during and after the merger of two neutron stars is expected to affect significantly the properties of the resulting merger remnant, because quark matter may have drastically different compressibility properties compared to nuclear matter.
In particular, the presence of quark matter is expected to impact the limiting mass at which a BNS merger collapses promptly into a black hole as well as modify the observable GW longevity and spectrum~\cite{Radice:2016rys, Most:2018eaw,Bauswein:2018bma,Weih:2019xvw,Bauswein:2020ggy,Tootle:2022pvd,Huang:2022mqp,Blacker:2023afl,Haque:2022dsc} and influence the conditions dictating the emergence of an EM signal~\cite{Prakash:2021wpz}.
In order to study this effect reliably, we opt to use a set of three EOS variants (soft, intermediate, and stiff) constructed in~\cite{Demircik:2021zll}, which contain both nuclear and quark matter phases. In this setup, based on the gauge/gravity duality, both the EOS of the nuclear and the quark matter phase arises from the same model (V-QCD,~\cite{Jarvinen:2011qe}), making it possible to obtain sound predictions for the location and the strength of the nuclear to quark matter phase transition. This is demonstrated in the top left panel of Fig.~\ref{fig:EOS}. At low temperatures the phase transition is relatively strong with latent heat $\sim 10^3$~MeV/fm$^3$~\cite{Jokela:2018ers,Jokela:2020piw,Demircik:2021zll}.
With increasing temperature, the phase transition becomes weaker, ending in a critical point above $T=100$~MeV.
The model for the quark matter phase follows the construction in~\cite{Alho:2012mh,Alho:2015zua}, with the details of the five-dimensional gravity action pinned down through fit to lattice data in~\cite{Jokela:2018ers}. The model for the nuclear matter phase, within the same V-QCD setup, was established in~\cite{Ishii:2019gta} by using the approach of~\cite{Rozali:2007rx}. 

The three EOS variants at zero temperature
are shown in Fig.~\ref{fig:EOS} (bottom left panel) as the colored solid curves. We point out that they compare well with the probability distribution of EOSs obtained by sampling a large set of interpolations between the low and high density regimes~\cite{Ecker:2022dlg}, where the EOS is known to a good precision. 
In addition, the intermediate variant without a quark matter phase (see App.~\ref{app:NMext}) is shown as the dashed green curve.
A characteristic feature of the EOS in the regime of the phase transitions is that the nuclear matter EOS is rather stiff, with the speed of sound squared clearly above the value $1/3$ of conformal theories, whereas the quark matter EOS is soft, with the speed of sound below the conformal value. Therefore there is a strong first order phase transition from nuclear to quark matter inside a neutron star which tends to destabilize the star. This holds true both for static and rotating stars~\cite{Ecker:2019xrw,Demircik:2020jkc} as well as the hypermassive neutron stars formed in neutron star mergers~\cite{Tootle:2022pvd}.

The agreement of the EOSs  at zero temperature is illustrated through comparison with the measurements by the NICER collaboration~\cite{Miller:2019cac,Riley:2019yda,Riley:2021pdl,Miller:2021qha} and direct mass measurements~\cite{Antoniadis:2013pzd,Romani:2022jhd} 
on the right panel of Fig.~\ref{fig:EOS}. The agreement is generically good~\cite{Ecker:2019xrw,Jokela:2020piw,Jokela:2021vwy}. The $M_\mathrm{TOV}$ for the soft EOS, set by the kink of the mass-radius curve where a quark core develops, is nevertheless somewhat lower when compared to the recent observations of neutron star masses~\cite{Romani:2022jhd,Barr:2024wwl}.
Therefore this variant is slightly disfavoured.

\section{Characterizing Prompt Collapse}\label{sec:Simulation}
There have been several attempts to determine the mass-threshold to prompt black hole formation in BNS mergers~\cite{Bauswein:2013jpa,Koppel:2019pys,Agathos:2019sah,Bauswein:2020xlt,Tootle:2021umi,Kashyap:2021wzs}.
Here we introduce a novel approach, which uses curvature scalars built from the Riemann tensor $R^{a}_{\;\;bcd}$ to locate the borderline between prompt and non-prompt black hole formation and to formulate a gauge independent definition for a one-parameter family of threshold masses.
There exists an infinite number of curvature scalars that can be formed by contracting tensor products of an arbitrary number of Riemann tensors.
For simplicity, our focus in this manuscript will be on the three quadratic polynomial invariants that are built by contracting Riemann and epsilon tensors.
These three quantities are the principal invariants of the Riemann tensor on four dimensional Lorenzian manifolds, known as the Kretschmann scalar $K_1$, the Chern-Pontryagin scalar $K_2$, and the Euler scalar $K_3$ and are defined as follows~\cite{Cherubini:2002gen} 
\begin{eqnarray}
	K_1&:=&R_{abcd}\,R^{abcd}\,,\\
	K_2&:=&{{}^\star R}_{abcd}\,R^{abcd}\,,\\
	K_3&:=&{{}^\star R}^\star_{abcd}\,R^{abcd}\,,
\end{eqnarray}
where the left and right duals of a generic four index tensor $A_{abcd}$ are defined by the following contractions with the epsilon tensor
\footnote{Note, that it is possible to define $K_2$ using the right dual instead of the left dual, but this difference is irrelevant for our purposes.} 
\begin{eqnarray}
	{{}^\star A}_{abcd}&:=&\epsilon_{abef}{A^{ef}}_{cd}\,,\\
	A^\star_{abcd}     &:=&{A_{ab}}^{ef}\epsilon_{efcd}\,.
\end{eqnarray}
It is useful to express these contractions in terms of the Weyl tensor, i.e., the trace free part of the Riemann tensor
\begin{align} 
 C_{abcd}=R_{abcd} -\left(g_{a[c}R_{d]b} - g_{b[c}R_{d]a}\right) + \frac{1}{3}R~g_{a[c}g_{d]b}\,,
\end{align}
which has only two linear independent principle invariants $I_1=C_{abcd}\,C^{abcd}$ and $I_2={{}^\star C}_{abcd}\,C^{abcd}$, because ${{}^\star C^\star}_{abcd}\,C^{abcd}=-I_1$. 
The principle invariants of the Riemann tensor can then be written as
\begin{eqnarray}
	K_1&=&I_1+2R_{ab}R^{ab}-\frac{1}{3}R^2\,,\\
	K_2&=&I_2\,,\\
	K_3&=&-I_1+2R_{ab}R^{ab}-\frac{2}{3}R^2\,.
\end{eqnarray}

As it turns out, the numerical values of $K_1, K_3$ in our simulations are dominated by their trace-free contributions $\propto I_1$ 
resulting, except for an overall sign, in very similar structures and numeric values of these quantities.
Both $K_{1}$ and $K_{3}$ attain their finite extremal values outside the black hole horizon in the orbital plane $z=0$, which makes them viable probes to analyse the collapse behaviour of BNS mergers.
The scalar $K_2$ has negative parity $K_2(z)=-K_2(-z)$, which makes $K_2=0$ by definition in the orbital plane $z=0$ and therefore not suitable as a probe for black hole formation in the centre of the merger (see App.~\ref{app:MoreResults}).
In order to simplify the discussion we report henceforth only results for $K_1$ and note that it can be replaced by $-K_3$ without loss of generality in the following.

For the numeric evaluation it turns out to be advantageous to use Einstein equations to replace the explicit Ricci tensor and Ricci scalar contributions in $K_1$ 
with the corresponding expressions of the energy momentum tensor $T^{ab}$, which leads to
\begin{eqnarray}
	K_1=I_1+128\pi^2\left(T_{ab}T^{ab}-\frac{1}{6}T^2\right)\,, 
\end{eqnarray}
where $T=g_{ab}T^{ab}$ denotes the trace of the energy momentum tensor.

We proceed now to outline the procedure of threshold mass determination.
As will be shown shortly, the spatial maximum of $K_1$ 
has an interesting and generic structure when plotted as a function of the total binary mass $M_{\rm total}$ and the post-merger time $t-t_{\rm merge}$, where the merger time $t_{\rm merge}$ is set by the global maximum of the GW strain amplitude.
For small values of $M_{\rm total}$ the function $K_1$ is non-monotonic in $t-t_{\rm merge}$, but rises monotonically for sufficiently large values of $M_{\rm total}$.
These two regions are identified as the  \textit{non-prompt collapse regime} and the \textit{prompt collapse regime}, respectively.
However, as it turns out, in order to precisely define the critical mass $M_\mathrm{crit}$ separating the two regions it is simpler to use the black hole formation time $t_{\rm crit}(M_{\rm total})$  instead of the properties of $K_1$. That is, 
the critical mass turns out to be clearly distinguished by a sharp local maximum 
in this critical time, 
where a trapped surface (apparent horizon) is found in the simulation: 
\begin{equation}
	\left\{M_{\rm crit}={\rm min}(M_{\rm total}) : \frac{dt_{\rm crit}}{dM_{\rm total}}<0\,\, \forall\,\, t>t_{\rm merge} \right\}\,.
 \label{eq:crit_mass_def}
\end{equation}
While this definition does not depend on $K_1$ explicitly, the local maximum of  $t_{\rm crit}$ also implies a sharp feature in the behavior of $K_1$, which grows rapidly when the black hole is formed, as we will demonstrate in the following section. 
The specific value of $M_{\rm crit}$ 
depends on the EOS~\footnote{We expect $M_{\rm crit}$ to also depend in a non-trivial way on the mass ratio, but leave an investigation of this dependence for future work.}, but the overall behaviour of $K_1(t-t_{\rm merge})$ and the sharp delay in $t_{\rm crit}(M_{\rm total})$ are generic and robust features that are EOS-independent.
Fine-tuning the total binary mass $M_{\rm total}$ to $M_{\rm crit}$ establishes a subtle dynamic balance between the gravitational attraction and the material pressure in the post-merger remnant which leads to a significant delay of black hole formation.

In order to determine when gravitational pull dominates over residual repulsive forces in neutron star matter we introduce, in addition, a one-parameter family of threshold masses $M_{\rm th}^{(p)}$ in terms of $K_1$ via the following min-max criterion~\footnote{Strictly speaking, the criterion is not entirely covariant because of the coordinate time derivatives, but can probably be phrased in an entirely covariant way by replacing the time coordinate $t$ by the parameter of any non-spacelike geodesic that passes through the point $p$ where $K_1(p)={\rm max}(K_1(x))$}
\begin{equation}\label{eq:Mth}
	\left\{M_{\rm th}^{(p)}={\rm min}(M_{\rm total}) : \frac{d^p}{dt^p}{\rm max}(K_1)\geq 0\,\, \forall\,\, t>t_{\rm merge} \right\}\,.
\end{equation}
The p-th derivative characterises the \textit{promptness} of the collapse, or in other words, the competition between the gravitational and repulsive 
forces in neutron star matter as well as the overall collision dynamics after the merger.
In principle, this definition incorporates arbitrarily high promptness, i.e., all values $p\geq 1$. 
For practical purposes, however, we restrict our analysis to $p=1,2$, as the determination of higher derivatives of $K_1$ becomes increasingly difficult given the finite temporal resolution of the simulations. The critical and threshold masses satisfy $M_{\rm crit}<M_{\rm th}^{(1)}<M_{\rm th}^{(2)}<\ldots$  as we shall demonstrate explicitly below.

We have implemented $K_1,K_2$ and $K_3$ by extending the Einstein Toolkit thorn \texttt{WeylScalar4}~\cite{Loffler:2011ay} using the Mathematica-based~\cite{Mathematica} computer-algebra package \texttt{KRANK}~\cite{Husa:2004ip} for code generation.
To construct the binary neutron star initial data we use the Frankfurt University Kadath code (FUKA)~\cite{Papenfort:2021hod} which is based on the \texttt{KADATH} spectral solver library~\cite{Grandclement:2009ju}.
For simplicity we consider here only non-spinning, equal mass binaries with an initial separation of $42\rm km$ and leave a more challenging exploration including unequal masses and/or spins of the binary components to future work.
We have checked the independence of our simulation results on the choice of $42\rm km$ initial separation by comparing to selected cases with $40\rm km$ and $45\rm km$ and found no noticeable difference. 
Furthermore, we neglect magnetic fields in our simulations, because 
the merger and collapse dynamics is predominantly determined by binary parameters and the properties of the neutron star fluid and less by the magnetic field.

To perform the binary evolution we use the Einstein-Toolkit~\cite{Loffler:2011ay} infrastructure and its fixed-mesh box-in-box refinement framework Carpet~\cite{Schnetter:2003rb}.
All simulations in this article are performed with six refinement levels, finest grid-spacing of $295\rm m$ and total domain size of $(3025{\rm km})^3$ while imposing reflection symmetry orthogonal to the orbital plane.
Opting for such intermediately fine resolution allows us to perform $\approx 20$ individual simulations per EOS which we found necessary to resolve sufficiently well the mass dependence of $K_1$ and other quantities for each of the four different EOS models.
As shown in~\cite{Tootle:2022pvd}, simulations with higher resolution result in minor quantitative changes, but generic features are expected to remain robust.
For the spacetime evolution we solve the constraint damping formulation of the Z4 system~\cite{Bernuzzi:2009ex,Alic:2011gg} with the Antelope~\cite{Most:2019kfe} code.
We use the Frankfurt-Illinois (FIL)~\cite{Most:2019kfe} code, which is based on the IllinoisGRMHD code~\cite{Etienne:2015cea}, to evolve the hydrodynamic part. 
Finally, we locate the black hole horizon by checking the geometry every $6.3\rm ns$ for trapped surfaces with the Einstein Toolkit thorn \texttt{AHFinderDirect}, which we find more than sufficient to resolve the delay in black hole formation close to $M_{\rm crit}$.

\section{Results}\label{sec:Results}

This section provides a comprehensive analysis of our simulations based on a representative selection of simulation results while postponing a discussion of the rest of our results for the EOS variants not shown here to the Appendix.
We start with analysing the generic structure of the curvature invariant $K_1$ that emerges in BNS mergers. 

\begin{figure}
    \centering
    \includegraphics[width=0.45\textwidth]{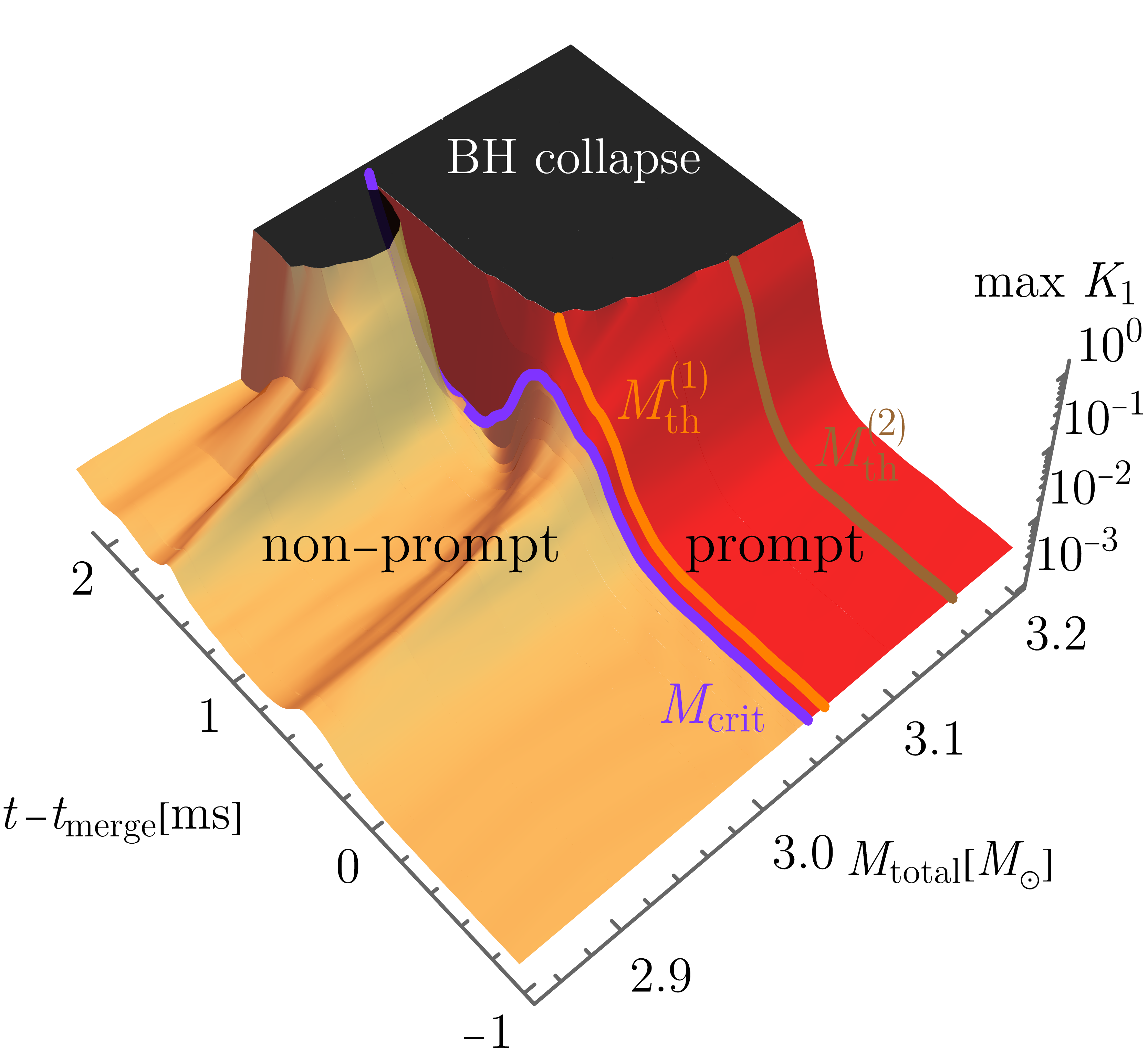}
    \caption{Maximum of $K_1$ as a function of $t-t_{\rm merge}$ and total binary mass $M_{\rm total}$ of the intermediate V-QCD model. Prompt and non-prompt collapse regions are colored in red and orange, respectively, and are separated by the critical mass $M_{\rm crit}$ marked in violet. 
    The orange and brown lines mark the threshold masses of promptness $p=1$ and $p=2$, respectively, and black regions (labeled ``BH collapse'') indicate where a black hole horizon is formed.}
    \label{fig:3D}
\end{figure}

Fig.~\ref{fig:3D} shows one of our main results, namely the dependence of ${\rm{max}}\; K_{1}$ on the total binary mass $M_{\rm total}$ when plotted as function of $t-t_{\rm merge}$.
The 2D surface of ${\rm{max}}\; K_{1}$ has an interesting structure which allows to rigorously separate merger remnants experiencing at least one bounce and therefore a prolonged lifetime from those collapsing immediately after the merger without any additional features.
This clear qualitative difference allows us to uniquely identify the critical boundary value $M_{\rm crit}$ (violet line) between the non-prompt (orange) and the prompt collapse regime (red), while regions where a black hole is formed are marked in black.
Also shown are two curves in the prompt collapse region that correspond to the threshold mass of promptness $p=1$ (orange line) and $p=2$ (brown line) as defined by  Eq.~\eqref{eq:Mth}.
The plot in Fig.~\ref{fig:3D} is based on $\approx 20$ individual BNS merger simulations with different total binary masses $M_{\rm total}$ performed using the intermediate V-QCD EOS variant.
Similar results for the other models considered in this article are given in the Appendix, which show that this structure is generic (EOS independent) and that only details, 
such as the values of $M_{\rm crit}$, depend on the EOS.

\begin{figure*}[htb]
\center
  \includegraphics[width=0.98\textwidth]{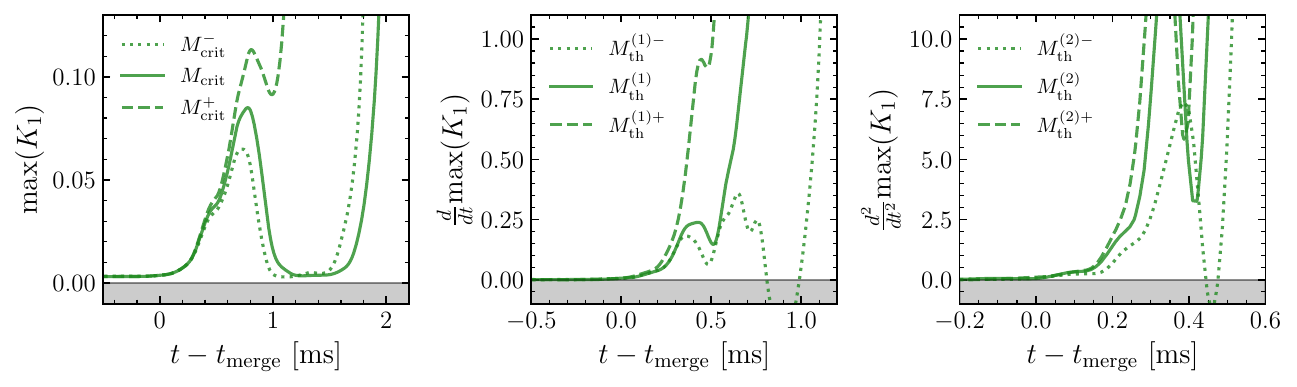}
 \caption{Spatial maxima of $K_1$ (left) and their first (middle) and second derivatives (right) as function of the post merger time $t-t_{\rm merge}$. The central values of the various limiting masses, as determined by Eq.~\eqref{eq:crit_mass_def} for $M_{\rm crit}$ in the leftmost panel and Eq.~\eqref{eq:Mth} for $M^{(p)}_{\rm th}$ in the other two, are plotted as solid lines, whereas dotted and dashed lines indicate their nearest neighbours that bracket the central values in our mass scan. Finally, the region of negative values is indicated by grey areas.
 }
 \label{fig:K1dt}
\end{figure*}

Fig.~\ref{fig:K1dt} illustrates how we identify $M_{\rm crit}$ and $M_{\rm th}^{(p)}$ from our mass scan of $K_1$. 
The left, middle and right panels here show  $K_1$, its first time derivative, and its second time derivative, respectively, at fixed values of the total mass as indicated in the plots.
The solid lines mark the central values for $M_{\rm crit}$ and $M_{\rm th}^{(p)}$, while dotted and dashed lines represent their corresponding neighbours $M_{\rm crit}^\pm$ and $M_{\rm th}^{(p)\pm}$ with slightly lower (-) and higher (+) total masses that bracket the central values.
The values of the critical and threshold masses for all EOSs as computed via Eq.~\eqref{eq:crit_mass_def} and Eq.~\eqref{eq:Mth} are collected in Table~\ref{tab:table1}, where the $\pm$ uncertainties follow from the corresponding values for $M^{\pm}_{\rm crit}$ and $M^{(p)\pm}_{\rm th}$ obtained from our mass scan.

As shown in the left panel, the solid curve for the central value of $M_{\rm crit}$ 
diverges later, indicating a delay in black hole formation, than those for subcritical (dotted) and supercritical (dashed) values $M^{\pm}_{\rm crit}$ which provide estimates for the upper and lower bounds of the central value. 
Analogously, the dotted curves in the middle and right panel for $M^{(p)-}_{\rm th}$ reach negative values at some $t-t_{\rm merge}>0$, which violates the criterion given in Eq.~\eqref{eq:Mth}, and therefore set lower bounds for $M^{(p)}_{\rm th}$, while the dashed curves for $M^{(p)+}_{\rm th}$ remain strictly positive and provide upper bounds.
\renewcommand{\arraystretch}{1.4}
\setlength\tabcolsep{0.5 pt}
\begin{table}[htb]
\begin{tabular}{ccccccc}
\centering
EOS & $M_{\rm TOV}/M_\odot$ & $C_{\rm TOV}$ & $\tilde{\Lambda}_{\rm GW}$ & $M_{\rm crit}/M_\odot$ & $M_{\rm th}^{(1)}/M_\odot$ & $M_{\rm th}^{(2)}/M_\odot$ \\
 \hline
 soft          & $2.01$ & $0.249$ & $537$ & $2.95_{-0.013}^{+0.002}$ & $2.963_{-0.006}^{+0.013}$ & $3.05_{-0.038}^{+0.075}$ \\
 interm.  & $2.14$ & $0.267$ & $565$ & $3.038_{-0.003}^{+0.002}$ & $3.05_{-0.01}^{+0.05}$ & $3.15_{-0.05}^{+0.05}$ \\
 no QM         & $2.15$ & $0.275$ & $565$ & $3.065_{-0.005}^{+0.005}$ & $3.075_{-0.005}^{+0.025}$ & $3.15_{-0.05}^{+0.05}$  \\
 stiff         & $2.34$ & $0.293$ & $617$ & $3.19_{-0.005}^{+0.005}$ & $3.195_{-0.005}^{+0.005}$ & $3.25_{-0.025}^{+0.025}$  \\
 \hline
\end{tabular}
\caption{
Static and binary neutron star properties of the four simulated EOS models. 
The $\pm$ uncertainty values are estimated from the nearest (super- and sub-critical) neighbours to the central values in our mass scan.
}\label{tab:table1}
\end{table}

We find a maximum difference of $\approx 0.25~M_\odot$ in $M_{\rm crit}$ between the soft and the stiff EOS variants. 
From this we arrive at the approximate inequality in equal-mass BNS mergers
\begin{equation}
2.95M_\odot\lesssim M_{\rm crit}\lesssim 3.19M_\odot\,,
\end{equation}
while the corresponding limits for $M_{\rm th}^{(p)}$ are slightly higher   
\begin{eqnarray}
2.963M_\odot\lesssim& M_{\rm th}^{(1)}&\lesssim 3.195M_\odot\,,\\
3.05M_\odot\lesssim& M_{\rm th}^{(2)}&\lesssim 3.25M_\odot\,,
\end{eqnarray}
with all of them consistent with the short compact binary gamma-ray burst regime reported in~\cite{Gottlieb:2023b}.
Furthermore, because $M_{\rm crit}<M_{\rm th}^{(1)}<M_{\rm th}^{(2)}<\ldots$  
and since our lower bound for $M_{\rm crit}=2.95M_\odot$ is significantly larger than the total mass $M_{\rm total}^{\rm GW170817}=2.74^{+0.04}_{-0.01}M_\odot$ inferred from GW170817, our results are consistent with a non-prompt collapse outcome of this BNS merger as suggested by the observed EM emission~\cite{Kasen:2017sxr,LIGOScientific:2017ync, Kashyap:2021wzs}.

The critical and threshold masses scale in the expected way with the stiffness of the EOS in the sense that soft models with low TOV-mass, compactness and stellar radii lead to smaller limiting masses than stiffer models, which provide more internal pressure to prevent black hole formation at high densities.
This suggests a tight and unique quasi-universal relation between the TOV-mass 
$M_{\rm TOV}$ of static neutron stars and the total BNS mass beyond which a prompt collapse after merger is inevitable 
\begin{equation} \label{eq:massratioest}
M_{\rm crit}/M_{\rm TOV}\approx 1.41\pm0.06\,,
\end{equation}
where the lower and upper bounds in this relation are inferred from the stiff and soft model, respectively, while the central value is approximated with the corresponding arithmetic average.
The 
uncertainty estimate in this relation 
is therefore a consequence of selecting V-QCD EOSs that coincide well with the boundaries of the $95\%$ confidence interval from generic EOS inference shown in Fig.\ref{fig:EOS}.
Because of this we expect that this  estimate 
reflects well the uncertainty for all possible consistent choices of the EOSs, while some specific choices may lead to mass ratios slightly outside the error band. 

Comparing the results for the intermediate EOS variants with and without quark matter 
shows only a small decrease of the limiting masses due to the presence of quark matter in the merger (see Table~\ref{tab:table1}).
This means the formation of quark matter during and after the merger has almost negligible impact on the value of the critical mass and the threshold masses, which we are nevertheless able to accurately resolve in our simulations as demonstrated in App.~\ref{app:MoreResults}.

\begin{figure*}[htb]
   \centering
    \includegraphics[height=0.3\textwidth]{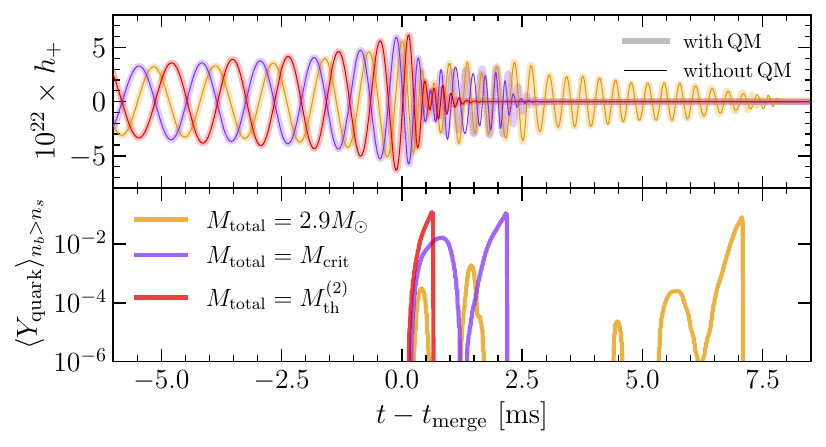}\quad
    \includegraphics[height=0.3\textwidth]{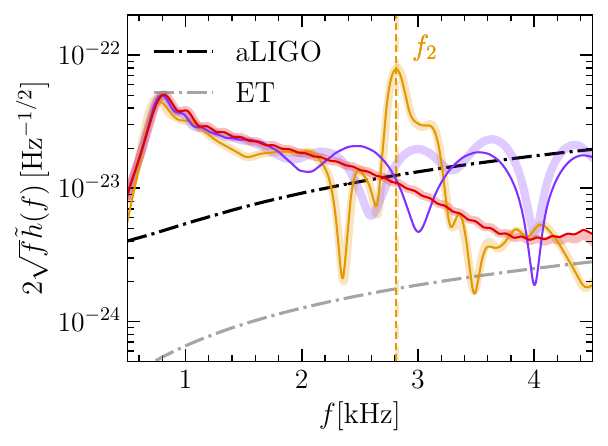}
    \caption{Left: On top we show the plus-polarization of the GW strain extrapolated to the estimated luminosity distance of $40\,{\rm Mpc}$ of the GW170817 event~\cite{LIGOScientific:2017vwq} and viewing angle $\theta=15^{\circ}$ determined from the jet of GW170817~\cite{Ghirlanda:2018uyx} (using $\phi=0^{\circ}$ without loss of generality). Shown are results for binaries with sub-critical ($2.9M_\odot$), critical ($M_{\rm crit}$) and supercritical mass ($M_{\rm th}^{(2)}$), using orange, violet, and red colors respectively (as in Fig.~\protect\ref{fig:3D}), 
    with cases with and without quark matter plotted with thick lines and thin lines, respectively. 
    On the bottom we show the corresponding quark matter fraction averaged over densities $n_b>n_s$ inside the merger remnant.
    Right: Corresponding PSD with the dominant $f_2$ mode marked by a dashed line whenever it is identifiable, together with sensitivity curves of the advanced LIGO (aLIGO) detector and the Einstein Telescope (ET).}
\label{fig:GWqm}
\end{figure*}

Let us next discuss the impact of quark matter on the GW signal.
In Fig.~\ref{fig:GWqm} we show in the left panel the GW strain (top), the quark volume fraction $Y_{\rm quark} \in [0,1]$ averaged over a region inside the merger remnant $\langle Y_{\rm quark} \rangle_{n_{b} \geq n_{s}} = \int_{n_{b} \geq n_{s}}d^{3}x\sqrt{\gamma}~Y_{\rm quark}/ \int_{n_{b} \geq n_{s}}d^{3}x\sqrt{\gamma}$ with number density $n_b\geq n_s$ (bottom) and in the right panel the corresponding power spectral density (PSD) together with sensitivity curves for the advanced LIGO (aLIGO) detector and the Einstein Telescope (ET).
Shown are curves for a sub-critical mass $M_{\rm total}=2.9M_\odot$ in orange, for the critical mass $M_{\rm crit}$ in violet and for the threshold mass $M_{\rm th}^{(2)}$ of promptness $p=2$ in red, with thick and thin lines corresponding to the model with and without quark matter phase, respectively.
As expected, larger masses lead to a clear truncation of the waveform due to the earlier black hole formation.
The post-merger signal for $M_{\rm th}^{(2)}$ is almost entirely dominated by the exponential ring-down of the black hole, while the critical ($M_{\rm total}=M_{\rm crit}$) and the subcritical ($M_{\rm total}=2.9M_\odot$) binaries survive for approximately $2\rm ms$ and $8\rm ms$ after the merger, respectively.

As shown in the bottom panel, there is a significant amount of quark matter formed during the various post merger phases.
This demonstrates that models like V-QCD that do not allow quark matter cores inside static neutron stars, because their strong phase transition destabilises the stars, can still lead to extended periods on the order of $10\rm ms$ during which a significant amount of quark matter is present inside a post-merger remnant. 
The spectral properties of the emitted GW signal during this period are potentially accessible with existing detectors.
Furthermore, what all cases have in common is the formation of a pure quark matter core right before the collapse.
In mergers with slightly sub-critical mass $M_{\rm total}=2.9M_\odot\lesssim M_{\rm crit}$ a curious phenomenon can be observed. Namely, an approximately $10\rm ms$ long stage is present, with periodic reoccurrence of quark matter inside the core of the merger remnant before it ultimately collapses into a black hole.
As shown in the Appendix, this stage appears generically for $M_{\rm total}\lesssim M_{\rm crit}$ in all our simulations with EOS models that include quark matter.
Nevertheless, the impact of quark matter on the GW spectrum, as shown in the power spectral density (PSD) in the right panel of the plot, depends most strongly on the total binary mass.
Here we see that only the critical mass mergers (violet curves) are impacted by the presence of quark matter while sub- and supercritical masses result in essentially no noticeable difference in the GW spectrum.
However, because of the short duration ($\approx2\rm ms$), accurately accessing this information presents an extreme challenge to current GW detectors, but may be accessible to those of the next generation such as the Einstein Telescope. 

\begin{figure}[htb]
    \centering
    \includegraphics[width=0.5\textwidth]{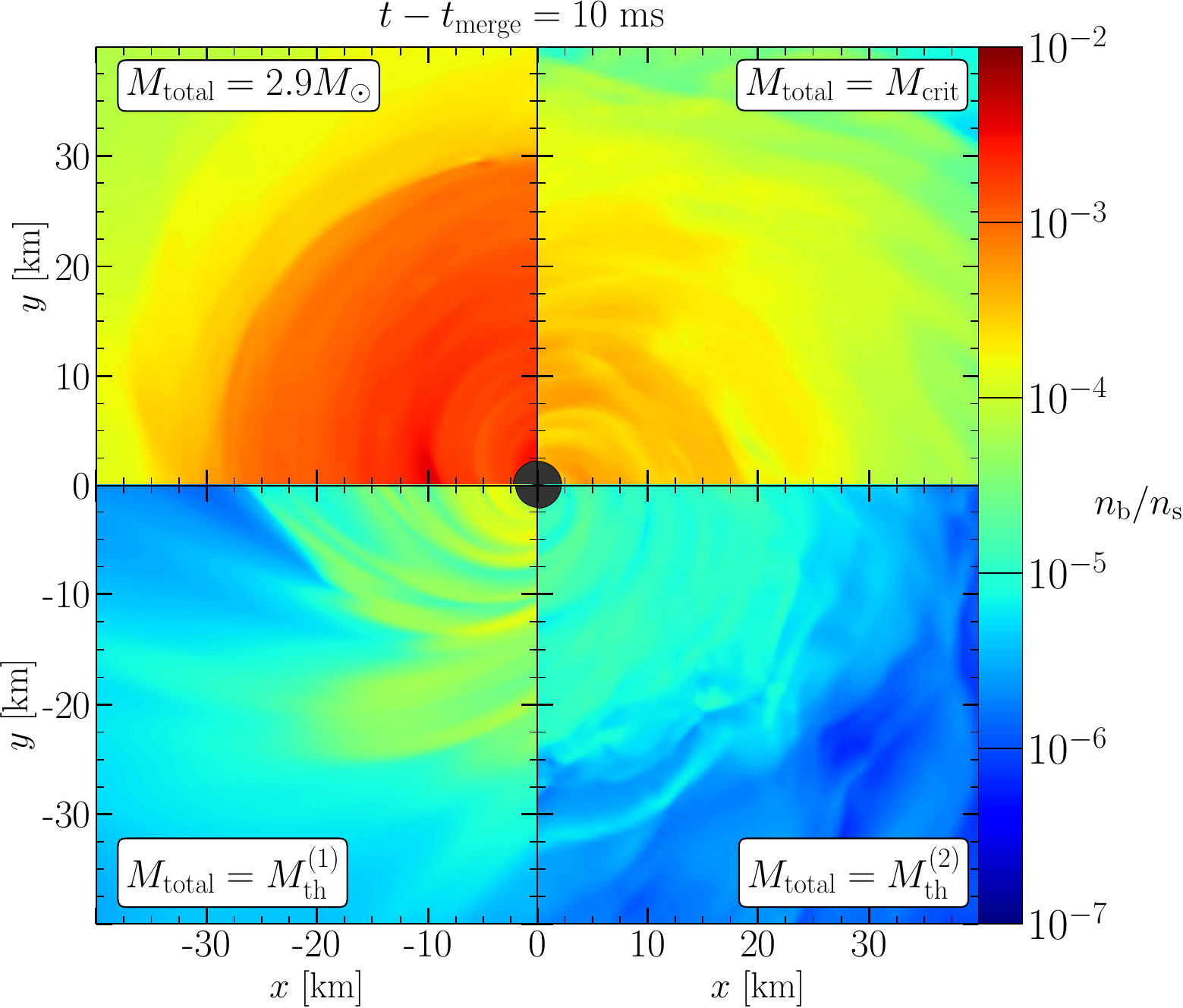}
    \caption{Post-collapse distribution and abundance of baryonic matter in the sub-critical (top left), critical (top right), and super-critical (bottom) regimes as obtained $10\rm ms$ after the merger.
    Notably, for $M_{\rm total} \geq M_{\rm crit}$, the highest baryon densities within the region of $\approx 30\rm km$ radius around the central black hole, drop by several orders of magnitude.
    }\label{fig:Postmerger_XY_tori_10ms}
\end{figure}

As argued above, the impact of quark matter in the vicinity of the prompt collapse regime is rather subtle and it is currently not clear if it will be possible in the near future to infer the formation of quark matter in BNS mergers just by analysing their GW signal.
For lack of clear GW signatures, it is essential to harvest the potential of other information channels in identifying quark matter formation, such as EM counterparts of merger events.
The prospects of observing an EM signal are crucially determined by the amount of mass left outside of the black hole.
Fig.~\ref{fig:Postmerger_XY_tori_10ms} shows how the residual amount of matter available for EM emission after the merger depends on the total mass of the binary.
Shown here is the number density in the orbital plane outside the black hole horizon, indicated by the black region in the centre, $10\rm ms$ after the merger for binaries with sub-critical mass $M_{\rm total}=2.9M_\odot$ (top left), critical mass $M_{\rm total}=M_{\rm crit}$ (top right) and for supercritical masses $M_{\rm th}^{(1)}$ (bottom left) and $M_{\rm th}^{(2)}$ (bottom right).
There is a clearly visible drop in the number density outside the horizon when crossing the critical mass bound.
In order to quantify this drop, we evaluate the total remnant baryonic mass outside of the apparent horizon in all our simulations by evaluating the integral
\begin{equation}
 M_{b}= \int_{r>r_{\rm AH}} W \rho \sqrt{\gamma}d^{3}x 
\end{equation}
at $10\rm ms$  after the collapse within a sphere of radius $\sim 74 \rm km$, where $W$ stands for the Lorentz factor, $\rho$ is the matter density, and $\sqrt{\gamma}$ the square root of the determinant of the spatial metric $\gamma_{ij}$.  

\begin{figure}[htb]
 \centering
 \includegraphics[width=0.48\textwidth]{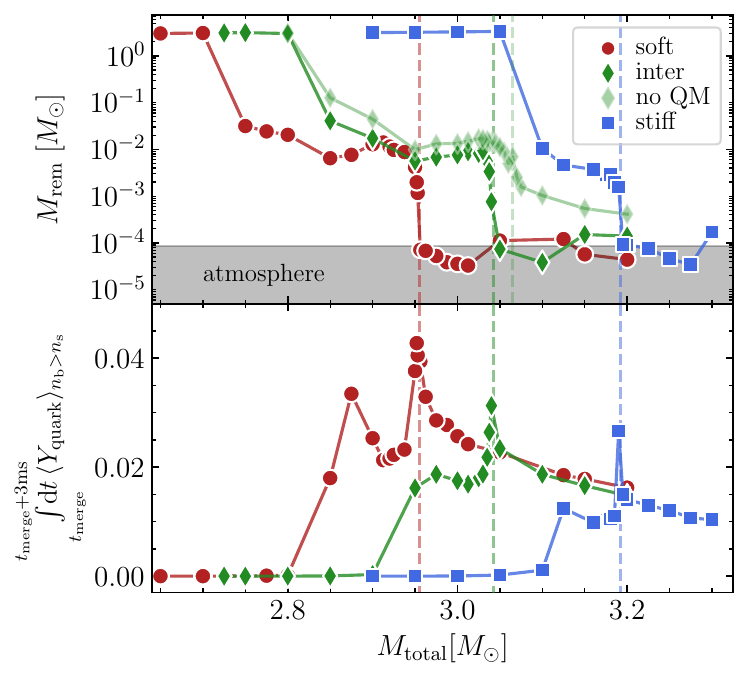}
 \caption{Top: Remnant baryon mass $M_{\rm rem}$ outside the horizon at $10\rm ms$ after black hole formation as a function of the total mass for different EOS models as indicated in the legend.
 Dashed lines mark the critical masses and the grey band indicates the region dominated by the density of the atmosphere in the simulation $\rho_{\rm atm}=10^{-14}M_\odot^{-2}$ as estimated by integrating $\int d^3x~\rho_{\rm atm}\approx 8.6 \times 10^{-5}M_\odot$ over the entire simulation domain of size $(2048M_\odot)^3$.
 Bottom: Time integral over $3\rm ms$ after merger of the volume averaged quark fraction inside regions with number density larger than nuclear saturation density.  
 }
\label{fig:remnant_mass_and_quark_time}
\end{figure}

We present the results of this analysis in the top panel of Fig.~\ref{fig:remnant_mass_and_quark_time}.
The critical mass of the binary separates those BNS systems that result in small tori masses  $5 \times 10^{-3} - 10^{-2}M_{\odot}$ from the supercritical binaries which leave only negligible amount of mass outside the apparent horizon, $M_{b}<10^{-3}M_{\odot}$.
Moreover, the resulting torus masses decrease by a factor of a few with the increasing stiffness of the EOS.
One can check that our 
findings are also 
consistent with the results for mass ejection in other studies on equal-mass~\cite{Hotokezaka2011} and unequal-mass (see, e.g.,~\cite{Cokluk:2023xio}) binary mergers.
Because the collapse can occur in different phases of the oscillating hypermassive neutron star (HMNS), small variations of the tori masses are present on both sides of the critical line.

Importantly, we find that the drop of the torus mass at criticality is much less pronounced and smaller in magnitude for the intermediate model where the quark matter phase is neglected (light green data points and curves in the top panel of Fig.~\ref{fig:remnant_mass_and_quark_time}). Since the nuclear to quark matter transition in our models is relatively strong, we expect that a similar reduction is found for EOS models that feature a weaker nuclear to quark matter transition. We stress however that a clear signal is visible even in the case where the phase transition is removed altogether.

For all the EOS employed in this study, dimensionless spins of the formed black holes of at least $\chi_{\rm BH} \geq 0.65$ are measured.
Assuming a large scale, ordered magnetic field, subcritical binaries should be capable of generating a short gamma-ray burst (sGRB) by launching a relativistic jet via the Blandford-Znajek mechanism (see \cite{Gottlieb:2023b} and references therein).
 
Furthermore, binaries closer to the critical mass (i.e., as $M_{\rm total}$ approaches $M_{\rm crit}$ from below) result in higher spins $\chi_{\rm BH} \sim 0.8$, which leads to a more efficient energy extraction via the Blandford-Znajek process~\cite{Lowell:2023kyu}. 
On the other hand, the tori masses for supercritical binaries are too small  to give rise to a detectable EM emission driven by the mass accretion onto the black hole~\cite{Ruiz:2017inq} and have also not been seen in the BNS event GW190425~\cite{LIGOScientific:2020aai} with a high total mass of $3.4 M_{\odot}$.
Formation of a detectable kilonova is also disfavoured, because in prompt-collapse scenarios and short-lived HMNS only negligible ejecta masses are present~\cite{Hotokezaka2013,Radice:2017lry}. 
We conclude that the only EM counterpart for equal-mass binaries with $M_{\rm total} \gtrsim M_{\rm crit}$ should be seen in the late inspiral or at the moment of merger, for example in the form of fast radio bursts~\cite{Paschalidis:2018tsa,Most:2022ayk}, quasi-periodic X-ray bursts in magnetars~\cite{Beloborodov:2020ylo} and precursor flares~\cite{Lyutikov:2018nti,Most:2020ami}.

Even though our simulations were carried out without magnetic fields, we expect our predictions regarding the disc masses in the prompt vs.~non-prompt regimes to remain unchanged for realistic magnetic field strengths $B \lesssim 10^{13-14}\rm G$.
The reason for this is that the decisive mechanisms preventing the prompt-collapse of a post-merger HMNS are hydrodynamical in nature, i.e., dominated by the rotational energy and the pressure support of the fluid and less by the magnetic field.

Finally, we show in the bottom panel of Fig.~\ref{fig:remnant_mass_and_quark_time} the \textit{persistence} of the quark matter, which is a diagnostic we introduce to simultaneously quantify the temporal extent and the abundance of the quark phase after the merger.
More specifically, we evaluate the integral of the volume-averaged quark fraction $\int \langle Y_{\rm quark} \rangle_{n_{b} \geq n_{s}} dt$ over a $3\rm ms$ long time interval after the merger.
This auxiliary quantity may be used to measure the overall contribution of the quark matter regions to the global dynamics of the system.
We find the largest amount of quark matter to be formed in critical mass mergers. 
In addition, for each EOS a secondary peak at $\sim 0.1 M_{\odot}$ smaller total mass is present. 
In consequence, we expect also more quark matter to be formed in binaries  that are slightly lighter than the critical mass because of the corresponding delay in collapse time.

\section{Conclusion}\label{sec:Conclusion}
In this work we performed an extensive numerical analysis of the prompt collapse behaviour of BNS mergers for a family of realistic EOS models that satisfy known observational bounds for behaviour of QCD at the highest densities realised inside neutron stars and their mergers.
In particular, we explored three EOS variants of different stiffness (from soft to stiff) constructed with the holographic V-QCD framework which provides a non-perturbative description of the phase transition between dense nuclear and quark matter within the same approach.
This framework also allows us to suppress the quark matter phase and therefore isolate in a consistent way the effect of the phase transition in our simulations.

As the main novel feature we introduce a set of criteria,  based on gauge independent curvature invariants of the Riemann tensor, to identify the critical mass $M_{\rm crit}$ that separates the prompt from the non-prompt collapse regime as well as a one-parameter family of threshold masses $M_{\rm th}^{(p)}$ with promptness $p$ which further characterises the prompt collapse regime of BNS mergers.
Using these criteria we uncover a rather tight relation $M_{\rm crit}/M_{\rm TOV}\approx 1.41\pm0.06$ between the TOV-mass and the critical binary mass beyond which any BNS system undergoes a prompt collapse after merger.

Furthermore, we demonstrate that the critical mass $M_{\rm crit}$ plays a crucial role in analysing the impact of quark matter in BNS mergers.
Importantly, we find that most quark matter is formed in critical mass mergers and not in sub- or supercritical ones.
This can be explained by the subtle metastable equilibrium realised at the critical mass which results in a few millisecond long prolongation of the remnant lifetime compared to non-critical binaries.
This leads to a clearly visible impact on the spectral properties of the post-merger GW spectrum in simulations near the critical mass, which we nonetheless expect to be challenging to isolate from future observations with existing detectors because of the short duration of the signal.

As the most promising indicator for the presence of quark matter we identify the residual mass outside the black hole formed after the merger.
We find that the amount of baryonic mass outside the horizon not only drops by several orders of magnitude right at the critical mass, but also that this amount depends on whether the collapse was induced by a quark matter or by a pure nuclear matter core.
The results of our simulations suggest that a black hole collapse induced by a softer quark matter core cleans the black hole exterior from residual matter more efficiently than in the case where the collapse happens against the resistance of a stiffer nuclear matter core.
In consequence, we find approximately one order of magnitude less baryonic mass outside the black hole produced in supercritical mass mergers with quark matter compared to those formed without quark matter.

There exists a number of possible extensions of our work.
A straightforward generalisation involves extending the BNS parameter space by taking into account different mass ratios and spins of the binary components similar to what has been done in~\cite{Tootle:2021umi}.
We have not attempted this yet because of the considerable computational costs required, but plan to do so in future work.
Another interesting avenue to explore are dissipative effects due to viscosity and their impact on the critical mass.
Bulk viscosity in particular could be studied using the recent developments in microphysics calculations of the relevant transport properties in holographic and perturbative QCD~\cite{CruzRojas:2024etx}.
Finally, it would be important to rigorously quantify the actual impact of quark matter on the expected EM emission in the prompt collapse regime, which could be done by adding magnetic fields together with a prescription for neutrino transport and a nuclear reaction network to our simulations.
Such a study would clearly require significantly higher resolution than it has been used in the present work in order to faithfully resolve the dynamics and amplification mechanism of the magnetic field. 
In combination with the necessary mass scan to identify the critical mass, however, such a study would result in computational costs that are larger than we can currently afford, but may be addressed in future.

\begin{acknowledgments}
We thank Tyler Gorda, Carlo Musolino and Luciano Rezzolla for useful discussions. 
C.~E. acknowledges support by the Deutsche Forschungsgemeinschaft (DFG, German Research Foundation) through the CRC-TR 211 ``Strong-interaction matter under extreme conditions''-- project number 315477589 -- TRR 211.
C.~E. and K.~T. acknowledge the support by the State of Hesse within the Research Cluster ELEMENTS (Project ID 500/10.006).
M.~J. has been supported by an appointment to the JRG Program at the APCTP through the Science and Technology Promotion Fund and Lottery Fund of the Korean Government, by the Korean Local Governments -- Gyeong\-sang\-buk-do Province and Pohang City -- and by the National Research Foundation of Korea (NRF) funded by the Korean government (MSIT) (grant number 2021R1A2C1010834).
Some calculations were performed on the local ITP Supercomputing Clusters Iboga and Calea. 
The merger simulations were performed on HPE Apollo HAWK at the High Performance Computing Center Stuttgart (HLRS) under the grant BNSMIC. 
For our visualisations we made use of the Kuibit library~\cite{Bozzola:2021hus}, as well as Mathematica~\cite{Mathematica} and Thermo Scientific\textsuperscript{\texttrademark} Amira 3D visualization software.
\end{acknowledgments}

\bibliography{main.bib}

\begin{thebibliography}{104}%
\makeatletter
\providecommand \@ifxundefined [1]{%
 \@ifx{#1\undefined}
}%
\providecommand \@ifnum [1]{%
 \ifnum #1\expandafter \@firstoftwo
 \else \expandafter \@secondoftwo
 \fi
}%
\providecommand \@ifx [1]{%
 \ifx #1\expandafter \@firstoftwo
 \else \expandafter \@secondoftwo
 \fi
}%
\providecommand \natexlab [1]{#1}%
\providecommand \enquote  [1]{``#1''}%
\providecommand \bibnamefont  [1]{#1}%
\providecommand \bibfnamefont [1]{#1}%
\providecommand \citenamefont [1]{#1}%
\providecommand \href@noop [0]{\@secondoftwo}%
\providecommand \href [0]{\begingroup \@sanitize@url \@href}%
\providecommand \@href[1]{\@@startlink{#1}\@@href}%
\providecommand \@@href[1]{\endgroup#1\@@endlink}%
\providecommand \@sanitize@url [0]{\catcode `\\12\catcode `\$12\catcode
  `\&12\catcode `\#12\catcode `\^12\catcode `\_12\catcode `\%12\relax}%
\providecommand \@@startlink[1]{}%
\providecommand \@@endlink[0]{}%
\providecommand \url  [0]{\begingroup\@sanitize@url \@url }%
\providecommand \@url [1]{\endgroup\@href {#1}{\urlprefix }}%
\providecommand \urlprefix  [0]{URL }%
\providecommand \Eprint [0]{\href }%
\providecommand \doibase [0]{https://doi.org/}%
\providecommand \selectlanguage [0]{\@gobble}%
\providecommand \bibinfo  [0]{\@secondoftwo}%
\providecommand \bibfield  [0]{\@secondoftwo}%
\providecommand \translation [1]{[#1]}%
\providecommand \BibitemOpen [0]{}%
\providecommand \bibitemStop [0]{}%
\providecommand \bibitemNoStop [0]{.\EOS\space}%
\providecommand \EOS [0]{\spacefactor3000\relax}%
\providecommand \BibitemShut  [1]{\csname bibitem#1\endcsname}%
\let\auto@bib@innerbib\@empty
\bibitem [{\citenamefont {Abbott}\ \emph
  {et~al.}(2017{\natexlab{a}})\citenamefont {Abbott} \emph
  {et~al.}}]{LIGOScientific:2017vwq}%
  \BibitemOpen
  \bibfield  {author} {\bibinfo {author} {\bibfnamefont {B.~P.}\ \bibnamefont
  {Abbott}} \emph {et~al.} (\bibinfo {collaboration} {LIGO Scientific,
  Virgo}),\ }\bibfield  {title} {\bibinfo {title} {{GW170817: Observation of
  Gravitational Waves from a Binary Neutron Star Inspiral}},\ }\href
  {https://doi.org/10.1103/PhysRevLett.119.161101} {\bibfield  {journal}
  {\bibinfo  {journal} {Phys. Rev. Lett.}\ }\textbf {\bibinfo {volume} {119}},\
  \bibinfo {pages} {161101} (\bibinfo {year} {2017}{\natexlab{a}})},\ \Eprint
  {https://arxiv.org/abs/1710.05832} {arXiv:1710.05832 [gr-qc]} \BibitemShut
  {NoStop}%
\bibitem [{\citenamefont {Abbott}\ \emph {et~al.}(2018)\citenamefont {Abbott}
  \emph {et~al.}}]{LIGOScientific:2018cki}%
  \BibitemOpen
  \bibfield  {author} {\bibinfo {author} {\bibfnamefont {B.~P.}\ \bibnamefont
  {Abbott}} \emph {et~al.} (\bibinfo {collaboration} {LIGO Scientific,
  Virgo}),\ }\bibfield  {title} {\bibinfo {title} {{GW170817: Measurements of
  neutron star radii and equation of state}},\ }\href
  {https://doi.org/10.1103/PhysRevLett.121.161101} {\bibfield  {journal}
  {\bibinfo  {journal} {Phys. Rev. Lett.}\ }\textbf {\bibinfo {volume} {121}},\
  \bibinfo {pages} {161101} (\bibinfo {year} {2018})},\ \Eprint
  {https://arxiv.org/abs/1805.11581} {arXiv:1805.11581 [gr-qc]} \BibitemShut
  {NoStop}%
\bibitem [{\citenamefont {Antoniadis}\ \emph {et~al.}(2013)\citenamefont
  {Antoniadis} \emph {et~al.}}]{Antoniadis:2013pzd}%
  \BibitemOpen
  \bibfield  {author} {\bibinfo {author} {\bibfnamefont {J.}~\bibnamefont
  {Antoniadis}} \emph {et~al.},\ }\bibfield  {title} {\bibinfo {title} {{A
  Massive Pulsar in a Compact Relativistic Binary}},\ }\href
  {https://doi.org/10.1126/science.1233232} {\bibfield  {journal} {\bibinfo
  {journal} {Science}\ }\textbf {\bibinfo {volume} {340}},\ \bibinfo {pages}
  {6131} (\bibinfo {year} {2013})},\ \Eprint {https://arxiv.org/abs/1304.6875}
  {arXiv:1304.6875 [astro-ph.HE]} \BibitemShut {NoStop}%
\bibitem [{\citenamefont {Cromartie}\ \emph {et~al.}(2019)\citenamefont
  {Cromartie} \emph {et~al.}}]{NANOGrav:2019jur}%
  \BibitemOpen
  \bibfield  {author} {\bibinfo {author} {\bibfnamefont {H.~T.}\ \bibnamefont
  {Cromartie}} \emph {et~al.} (\bibinfo {collaboration} {NANOGrav}),\
  }\bibfield  {title} {\bibinfo {title} {{Relativistic Shapiro delay
  measurements of an extremely massive millisecond pulsar}},\ }\href
  {https://doi.org/10.1038/s41550-019-0880-2} {\bibfield  {journal} {\bibinfo
  {journal} {Nature Astron.}\ }\textbf {\bibinfo {volume} {4}},\ \bibinfo
  {pages} {72} (\bibinfo {year} {2019})},\ \Eprint
  {https://arxiv.org/abs/1904.06759} {arXiv:1904.06759 [astro-ph.HE]}
  \BibitemShut {NoStop}%
\bibitem [{\citenamefont {Fonseca}\ \emph {et~al.}(2021)\citenamefont {Fonseca}
  \emph {et~al.}}]{Fonseca:2021wxt}%
  \BibitemOpen
  \bibfield  {author} {\bibinfo {author} {\bibfnamefont {E.}~\bibnamefont
  {Fonseca}} \emph {et~al.},\ }\bibfield  {title} {\bibinfo {title} {{Refined
  Mass and Geometric Measurements of the High-mass PSR J0740+6620}},\ }\href
  {https://doi.org/10.3847/2041-8213/ac03b8} {\bibfield  {journal} {\bibinfo
  {journal} {Astrophys. J. Lett.}\ }\textbf {\bibinfo {volume} {915}},\
  \bibinfo {pages} {L12} (\bibinfo {year} {2021})},\ \Eprint
  {https://arxiv.org/abs/2104.00880} {arXiv:2104.00880 [astro-ph.HE]}
  \BibitemShut {NoStop}%
\bibitem [{\citenamefont {Romani}\ \emph {et~al.}(2022)\citenamefont {Romani},
  \citenamefont {Kandel}, \citenamefont {Filippenko}, \citenamefont {Brink},\
  and\ \citenamefont {Zheng}}]{Romani:2022jhd}%
  \BibitemOpen
  \bibfield  {author} {\bibinfo {author} {\bibfnamefont {R.~W.}\ \bibnamefont
  {Romani}}, \bibinfo {author} {\bibfnamefont {D.}~\bibnamefont {Kandel}},
  \bibinfo {author} {\bibfnamefont {A.~V.}\ \bibnamefont {Filippenko}},
  \bibinfo {author} {\bibfnamefont {T.~G.}\ \bibnamefont {Brink}},\ and\
  \bibinfo {author} {\bibfnamefont {W.}~\bibnamefont {Zheng}},\ }\bibfield
  {title} {\bibinfo {title} {{PSR J0952\ensuremath{-}0607: The Fastest and
  Heaviest Known Galactic Neutron Star}},\ }\href
  {https://doi.org/10.3847/2041-8213/ac8007} {\bibfield  {journal} {\bibinfo
  {journal} {Astrophys. J. Lett.}\ }\textbf {\bibinfo {volume} {934}},\
  \bibinfo {pages} {L17} (\bibinfo {year} {2022})},\ \Eprint
  {https://arxiv.org/abs/2207.05124} {arXiv:2207.05124 [astro-ph.HE]}
  \BibitemShut {NoStop}%
\bibitem [{\citenamefont {Miller}\ \emph {et~al.}(2019)\citenamefont {Miller}
  \emph {et~al.}}]{Miller:2019cac}%
  \BibitemOpen
  \bibfield  {author} {\bibinfo {author} {\bibfnamefont {M.~C.}\ \bibnamefont
  {Miller}} \emph {et~al.},\ }\bibfield  {title} {\bibinfo {title} {{PSR
  J0030+0451 Mass and Radius from $NICER$ Data and Implications for the
  Properties of Neutron Star Matter}},\ }\href
  {https://doi.org/10.3847/2041-8213/ab50c5} {\bibfield  {journal} {\bibinfo
  {journal} {Astrophys. J. Lett.}\ }\textbf {\bibinfo {volume} {887}},\
  \bibinfo {pages} {L24} (\bibinfo {year} {2019})},\ \Eprint
  {https://arxiv.org/abs/1912.05705} {arXiv:1912.05705 [astro-ph.HE]}
  \BibitemShut {NoStop}%
\bibitem [{\citenamefont {Riley}\ \emph {et~al.}(2019)\citenamefont {Riley}
  \emph {et~al.}}]{Riley:2019yda}%
  \BibitemOpen
  \bibfield  {author} {\bibinfo {author} {\bibfnamefont {T.~E.}\ \bibnamefont
  {Riley}} \emph {et~al.},\ }\bibfield  {title} {\bibinfo {title} {{A $NICER$
  View of PSR J0030+0451: Millisecond Pulsar Parameter Estimation}},\ }\href
  {https://doi.org/10.3847/2041-8213/ab481c} {\bibfield  {journal} {\bibinfo
  {journal} {Astrophys. J. Lett.}\ }\textbf {\bibinfo {volume} {887}},\
  \bibinfo {pages} {L21} (\bibinfo {year} {2019})},\ \Eprint
  {https://arxiv.org/abs/1912.05702} {arXiv:1912.05702 [astro-ph.HE]}
  \BibitemShut {NoStop}%
\bibitem [{\citenamefont {Riley}\ \emph {et~al.}(2021)\citenamefont {Riley}
  \emph {et~al.}}]{Riley:2021pdl}%
  \BibitemOpen
  \bibfield  {author} {\bibinfo {author} {\bibfnamefont {T.~E.}\ \bibnamefont
  {Riley}} \emph {et~al.},\ }\bibfield  {title} {\bibinfo {title} {{A NICER
  View of the Massive Pulsar PSR J0740+6620 Informed by Radio Timing and
  XMM-Newton Spectroscopy}},\ }\href {https://doi.org/10.3847/2041-8213/ac0a81}
  {\bibfield  {journal} {\bibinfo  {journal} {Astrophys. J. Lett.}\ }\textbf
  {\bibinfo {volume} {918}},\ \bibinfo {pages} {L27} (\bibinfo {year}
  {2021})},\ \Eprint {https://arxiv.org/abs/2105.06980} {arXiv:2105.06980
  [astro-ph.HE]} \BibitemShut {NoStop}%
\bibitem [{\citenamefont {Miller}\ \emph {et~al.}(2021)\citenamefont {Miller}
  \emph {et~al.}}]{Miller:2021qha}%
  \BibitemOpen
  \bibfield  {author} {\bibinfo {author} {\bibfnamefont {M.~C.}\ \bibnamefont
  {Miller}} \emph {et~al.},\ }\bibfield  {title} {\bibinfo {title} {{The Radius
  of PSR J0740+6620 from NICER and XMM-Newton Data}},\ }\href
  {https://doi.org/10.3847/2041-8213/ac089b} {\bibfield  {journal} {\bibinfo
  {journal} {Astrophys. J. Lett.}\ }\textbf {\bibinfo {volume} {918}},\
  \bibinfo {pages} {L28} (\bibinfo {year} {2021})},\ \Eprint
  {https://arxiv.org/abs/2105.06979} {arXiv:2105.06979 [astro-ph.HE]}
  \BibitemShut {NoStop}%
\bibitem [{\citenamefont {Barr}\ \emph {et~al.}(2024)\citenamefont {Barr},
  \citenamefont {Dutta}, \citenamefont {Freire}, \citenamefont {Cadelano},
  \citenamefont {Gautam}, \citenamefont {Kramer}, \citenamefont {Pallanca},
  \citenamefont {Ransom}, \citenamefont {Ridolfi}, \citenamefont {Stappers},
  \citenamefont {Tauris}, \citenamefont {Krishnan}, \citenamefont {Wex},
  \citenamefont {Bailes}, \citenamefont {Behrend}, \citenamefont {Buchner},
  \citenamefont {Burgay}, \citenamefont {Chen}, \citenamefont {Champion},
  \citenamefont {Chen}, \citenamefont {Corongiu}, \citenamefont {Geyer},
  \citenamefont {Men}, \citenamefont {Padmanabh},\ and\ \citenamefont
  {Possenti}}]{Barr:2024wwl}%
  \BibitemOpen
  \bibfield  {author} {\bibinfo {author} {\bibfnamefont {E.~D.}\ \bibnamefont
  {Barr}}, \bibinfo {author} {\bibfnamefont {A.}~\bibnamefont {Dutta}},
  \bibinfo {author} {\bibfnamefont {P.~C.~C.}\ \bibnamefont {Freire}}, \bibinfo
  {author} {\bibfnamefont {M.}~\bibnamefont {Cadelano}}, \bibinfo {author}
  {\bibfnamefont {T.}~\bibnamefont {Gautam}}, \bibinfo {author} {\bibfnamefont
  {M.}~\bibnamefont {Kramer}}, \bibinfo {author} {\bibfnamefont
  {C.}~\bibnamefont {Pallanca}}, \bibinfo {author} {\bibfnamefont {S.~M.}\
  \bibnamefont {Ransom}}, \bibinfo {author} {\bibfnamefont {A.}~\bibnamefont
  {Ridolfi}}, \bibinfo {author} {\bibfnamefont {B.~W.}\ \bibnamefont
  {Stappers}}, \bibinfo {author} {\bibfnamefont {T.~M.}\ \bibnamefont
  {Tauris}}, \bibinfo {author} {\bibfnamefont {V.~V.}\ \bibnamefont
  {Krishnan}}, \bibinfo {author} {\bibfnamefont {N.}~\bibnamefont {Wex}},
  \bibinfo {author} {\bibfnamefont {M.}~\bibnamefont {Bailes}}, \bibinfo
  {author} {\bibfnamefont {J.}~\bibnamefont {Behrend}}, \bibinfo {author}
  {\bibfnamefont {S.}~\bibnamefont {Buchner}}, \bibinfo {author} {\bibfnamefont
  {M.}~\bibnamefont {Burgay}}, \bibinfo {author} {\bibfnamefont
  {W.}~\bibnamefont {Chen}}, \bibinfo {author} {\bibfnamefont {D.~J.}\
  \bibnamefont {Champion}}, \bibinfo {author} {\bibfnamefont {C.-H.~R.}\
  \bibnamefont {Chen}}, \bibinfo {author} {\bibfnamefont {A.}~\bibnamefont
  {Corongiu}}, \bibinfo {author} {\bibfnamefont {M.}~\bibnamefont {Geyer}},
  \bibinfo {author} {\bibfnamefont {Y.~P.}\ \bibnamefont {Men}}, \bibinfo
  {author} {\bibfnamefont {P.~V.}\ \bibnamefont {Padmanabh}},\ and\ \bibinfo
  {author} {\bibfnamefont {A.}~\bibnamefont {Possenti}},\ }\bibfield  {title}
  {\bibinfo {title} {A pulsar in a binary with a compact object in the mass gap
  between neutron stars and black holes},\ }\href
  {https://doi.org/10.1126/science.adg3005} {\bibfield  {journal} {\bibinfo
  {journal} {Science}\ }\textbf {\bibinfo {volume} {383}},\ \bibinfo {pages}
  {275} (\bibinfo {year} {2024})},\ \Eprint
  {https://arxiv.org/abs/https://www.science.org/doi/pdf/10.1126/science.adg3005}
  {https://www.science.org/doi/pdf/10.1126/science.adg3005} \BibitemShut
  {NoStop}%
\bibitem [{\citenamefont {Hebeler}\ \emph {et~al.}(2013)\citenamefont
  {Hebeler}, \citenamefont {Lattimer}, \citenamefont {Pethick},\ and\
  \citenamefont {Schwenk}}]{Hebeler:2013nza}%
  \BibitemOpen
  \bibfield  {author} {\bibinfo {author} {\bibfnamefont {K.}~\bibnamefont
  {Hebeler}}, \bibinfo {author} {\bibfnamefont {J.~M.}\ \bibnamefont
  {Lattimer}}, \bibinfo {author} {\bibfnamefont {C.~J.}\ \bibnamefont
  {Pethick}},\ and\ \bibinfo {author} {\bibfnamefont {A.}~\bibnamefont
  {Schwenk}},\ }\bibfield  {title} {\bibinfo {title} {{Equation of state and
  neutron star properties constrained by nuclear physics and observation}},\
  }\href {https://doi.org/10.1088/0004-637X/773/1/11} {\bibfield  {journal}
  {\bibinfo  {journal} {Astrophys. J.}\ }\textbf {\bibinfo {volume} {773}},\
  \bibinfo {pages} {11} (\bibinfo {year} {2013})},\ \Eprint
  {https://arxiv.org/abs/1303.4662} {arXiv:1303.4662 [astro-ph.SR]}
  \BibitemShut {NoStop}%
\bibitem [{\citenamefont {Gandolfi}\ \emph {et~al.}(2019)\citenamefont
  {Gandolfi}, \citenamefont {Lippuner}, \citenamefont {Steiner}, \citenamefont
  {Tews}, \citenamefont {Du},\ and\ \citenamefont
  {Al-Mamun}}]{Gandolfi:2019zpj}%
  \BibitemOpen
  \bibfield  {author} {\bibinfo {author} {\bibfnamefont {S.}~\bibnamefont
  {Gandolfi}}, \bibinfo {author} {\bibfnamefont {J.}~\bibnamefont {Lippuner}},
  \bibinfo {author} {\bibfnamefont {A.~W.}\ \bibnamefont {Steiner}}, \bibinfo
  {author} {\bibfnamefont {I.}~\bibnamefont {Tews}}, \bibinfo {author}
  {\bibfnamefont {X.}~\bibnamefont {Du}},\ and\ \bibinfo {author}
  {\bibfnamefont {M.}~\bibnamefont {Al-Mamun}},\ }\bibfield  {title} {\bibinfo
  {title} {{From the microscopic to the macroscopic world: from nucleons to
  neutron stars}},\ }\href {https://doi.org/10.1088/1361-6471/ab29b3}
  {\bibfield  {journal} {\bibinfo  {journal} {J. Phys. G}\ }\textbf {\bibinfo
  {volume} {46}},\ \bibinfo {pages} {103001} (\bibinfo {year} {2019})},\
  \Eprint {https://arxiv.org/abs/1903.06730} {arXiv:1903.06730 [nucl-th]}
  \BibitemShut {NoStop}%
\bibitem [{\citenamefont {Keller}\ \emph {et~al.}(2021)\citenamefont {Keller},
  \citenamefont {Wellenhofer}, \citenamefont {Hebeler},\ and\ \citenamefont
  {Schwenk}}]{Keller:2020qhx}%
  \BibitemOpen
  \bibfield  {author} {\bibinfo {author} {\bibfnamefont {J.}~\bibnamefont
  {Keller}}, \bibinfo {author} {\bibfnamefont {C.}~\bibnamefont {Wellenhofer}},
  \bibinfo {author} {\bibfnamefont {K.}~\bibnamefont {Hebeler}},\ and\ \bibinfo
  {author} {\bibfnamefont {A.}~\bibnamefont {Schwenk}},\ }\bibfield  {title}
  {\bibinfo {title} {{Neutron matter at finite temperature based on chiral
  effective field theory interactions}},\ }\href
  {https://doi.org/10.1103/PhysRevC.103.055806} {\bibfield  {journal} {\bibinfo
   {journal} {Phys. Rev. C}\ }\textbf {\bibinfo {volume} {103}},\ \bibinfo
  {pages} {055806} (\bibinfo {year} {2021})},\ \Eprint
  {https://arxiv.org/abs/2011.05855} {arXiv:2011.05855 [nucl-th]} \BibitemShut
  {NoStop}%
\bibitem [{\citenamefont {Drischler}\ \emph {et~al.}(2020)\citenamefont
  {Drischler}, \citenamefont {Melendez}, \citenamefont {Furnstahl},\ and\
  \citenamefont {Phillips}}]{Drischler:2020yad}%
  \BibitemOpen
  \bibfield  {author} {\bibinfo {author} {\bibfnamefont {C.}~\bibnamefont
  {Drischler}}, \bibinfo {author} {\bibfnamefont {J.~A.}\ \bibnamefont
  {Melendez}}, \bibinfo {author} {\bibfnamefont {R.~J.}\ \bibnamefont
  {Furnstahl}},\ and\ \bibinfo {author} {\bibfnamefont {D.~R.}\ \bibnamefont
  {Phillips}},\ }\bibfield  {title} {\bibinfo {title} {{Quantifying
  uncertainties and correlations in the nuclear-matter equation of state}},\
  }\href {https://doi.org/10.1103/PhysRevC.102.054315} {\bibfield  {journal}
  {\bibinfo  {journal} {Phys. Rev. C}\ }\textbf {\bibinfo {volume} {102}},\
  \bibinfo {pages} {054315} (\bibinfo {year} {2020})},\ \Eprint
  {https://arxiv.org/abs/2004.07805} {arXiv:2004.07805 [nucl-th]} \BibitemShut
  {NoStop}%
\bibitem [{\citenamefont {Freedman}\ and\ \citenamefont
  {McLerran}(1977)}]{Freedman:1976ub}%
  \BibitemOpen
  \bibfield  {author} {\bibinfo {author} {\bibfnamefont {B.~A.}\ \bibnamefont
  {Freedman}}\ and\ \bibinfo {author} {\bibfnamefont {L.~D.}\ \bibnamefont
  {McLerran}},\ }\bibfield  {title} {\bibinfo {title} {{Fermions and Gauge
  Vector Mesons at Finite Temperature and Density. 3. The Ground State Energy
  of a Relativistic Quark Gas}},\ }\href
  {https://doi.org/10.1103/PhysRevD.16.1169} {\bibfield  {journal} {\bibinfo
  {journal} {Phys. Rev. D}\ }\textbf {\bibinfo {volume} {16}},\ \bibinfo
  {pages} {1169} (\bibinfo {year} {1977})}\BibitemShut {NoStop}%
\bibitem [{\citenamefont {Vuorinen}(2003)}]{Vuorinen:2003fs}%
  \BibitemOpen
  \bibfield  {author} {\bibinfo {author} {\bibfnamefont {A.}~\bibnamefont
  {Vuorinen}},\ }\bibfield  {title} {\bibinfo {title} {{The Pressure of QCD at
  finite temperatures and chemical potentials}},\ }\href
  {https://doi.org/10.1103/PhysRevD.68.054017} {\bibfield  {journal} {\bibinfo
  {journal} {Phys. Rev. D}\ }\textbf {\bibinfo {volume} {68}},\ \bibinfo
  {pages} {054017} (\bibinfo {year} {2003})},\ \Eprint
  {https://arxiv.org/abs/hep-ph/0305183} {arXiv:hep-ph/0305183} \BibitemShut
  {NoStop}%
\bibitem [{\citenamefont {Gorda}\ \emph
  {et~al.}(2021{\natexlab{a}})\citenamefont {Gorda}, \citenamefont {Kurkela},
  \citenamefont {Paatelainen}, \citenamefont {S\"appi},\ and\ \citenamefont
  {Vuorinen}}]{Gorda:2021kme}%
  \BibitemOpen
  \bibfield  {author} {\bibinfo {author} {\bibfnamefont {T.}~\bibnamefont
  {Gorda}}, \bibinfo {author} {\bibfnamefont {A.}~\bibnamefont {Kurkela}},
  \bibinfo {author} {\bibfnamefont {R.}~\bibnamefont {Paatelainen}}, \bibinfo
  {author} {\bibfnamefont {S.}~\bibnamefont {S\"appi}},\ and\ \bibinfo {author}
  {\bibfnamefont {A.}~\bibnamefont {Vuorinen}},\ }\bibfield  {title} {\bibinfo
  {title} {{Cold quark matter at N3LO: Soft contributions}},\ }\href
  {https://doi.org/10.1103/PhysRevD.104.074015} {\bibfield  {journal} {\bibinfo
   {journal} {Phys. Rev. D}\ }\textbf {\bibinfo {volume} {104}},\ \bibinfo
  {pages} {074015} (\bibinfo {year} {2021}{\natexlab{a}})},\ \Eprint
  {https://arxiv.org/abs/2103.07427} {arXiv:2103.07427 [hep-ph]} \BibitemShut
  {NoStop}%
\bibitem [{\citenamefont {Gorda}\ \emph
  {et~al.}(2021{\natexlab{b}})\citenamefont {Gorda}, \citenamefont {Kurkela},
  \citenamefont {Paatelainen}, \citenamefont {S\"appi},\ and\ \citenamefont
  {Vuorinen}}]{Gorda:2021znl}%
  \BibitemOpen
  \bibfield  {author} {\bibinfo {author} {\bibfnamefont {T.}~\bibnamefont
  {Gorda}}, \bibinfo {author} {\bibfnamefont {A.}~\bibnamefont {Kurkela}},
  \bibinfo {author} {\bibfnamefont {R.}~\bibnamefont {Paatelainen}}, \bibinfo
  {author} {\bibfnamefont {S.}~\bibnamefont {S\"appi}},\ and\ \bibinfo {author}
  {\bibfnamefont {A.}~\bibnamefont {Vuorinen}},\ }\bibfield  {title} {\bibinfo
  {title} {{Soft Interactions in Cold Quark Matter}},\ }\href
  {https://doi.org/10.1103/PhysRevLett.127.162003} {\bibfield  {journal}
  {\bibinfo  {journal} {Phys. Rev. Lett.}\ }\textbf {\bibinfo {volume} {127}},\
  \bibinfo {pages} {162003} (\bibinfo {year} {2021}{\natexlab{b}})},\ \Eprint
  {https://arxiv.org/abs/2103.05658} {arXiv:2103.05658 [hep-ph]} \BibitemShut
  {NoStop}%
\bibitem [{\citenamefont {{Friedman}}\ \emph {et~al.}(1988)\citenamefont
  {{Friedman}}, \citenamefont {{Ipser}},\ and\ \citenamefont
  {{Sorkin}}}]{1988ApJ...325..722F}%
  \BibitemOpen
  \bibfield  {author} {\bibinfo {author} {\bibfnamefont {J.~L.}\ \bibnamefont
  {{Friedman}}}, \bibinfo {author} {\bibfnamefont {J.~R.}\ \bibnamefont
  {{Ipser}}},\ and\ \bibinfo {author} {\bibfnamefont {R.~D.}\ \bibnamefont
  {{Sorkin}}},\ }\bibfield  {title} {\bibinfo {title} {{Turning Point Method
  for Axisymmetric Stability of Rotating Relativistic Stars}},\ }\href
  {https://doi.org/10.1086/166043} {\bibfield  {journal} {\bibinfo  {journal}
  {\apj}\ }\textbf {\bibinfo {volume} {325}},\ \bibinfo {pages} {722} (\bibinfo
  {year} {1988})}\BibitemShut {NoStop}%
\bibitem [{\citenamefont {Breu}\ and\ \citenamefont
  {Rezzolla}(2016)}]{Breu:2016ufb}%
  \BibitemOpen
  \bibfield  {author} {\bibinfo {author} {\bibfnamefont {C.}~\bibnamefont
  {Breu}}\ and\ \bibinfo {author} {\bibfnamefont {L.}~\bibnamefont
  {Rezzolla}},\ }\bibfield  {title} {\bibinfo {title} {{Maximum mass, moment of
  inertia and compactness of relativistic stars}},\ }\href
  {https://doi.org/10.1093/mnras/stw575} {\bibfield  {journal} {\bibinfo
  {journal} {Mon. Not. Roy. Astron. Soc.}\ }\textbf {\bibinfo {volume} {459}},\
  \bibinfo {pages} {646} (\bibinfo {year} {2016})},\ \Eprint
  {https://arxiv.org/abs/1601.06083} {arXiv:1601.06083 [gr-qc]} \BibitemShut
  {NoStop}%
\bibitem [{\citenamefont {Demircik}\ \emph {et~al.}(2021)\citenamefont
  {Demircik}, \citenamefont {Ecker},\ and\ \citenamefont
  {J\"arvinen}}]{Demircik:2020jkc}%
  \BibitemOpen
  \bibfield  {author} {\bibinfo {author} {\bibfnamefont {T.}~\bibnamefont
  {Demircik}}, \bibinfo {author} {\bibfnamefont {C.}~\bibnamefont {Ecker}},\
  and\ \bibinfo {author} {\bibfnamefont {M.}~\bibnamefont {J\"arvinen}},\
  }\bibfield  {title} {\bibinfo {title} {{Rapidly Spinning Compact Stars with
  Deconfinement Phase Transition}},\ }\href
  {https://doi.org/10.3847/2041-8213/abd853} {\bibfield  {journal} {\bibinfo
  {journal} {Astrophys. J. Lett.}\ }\textbf {\bibinfo {volume} {907}},\
  \bibinfo {pages} {L37} (\bibinfo {year} {2021})},\ \Eprint
  {https://arxiv.org/abs/2009.10731} {arXiv:2009.10731 [astro-ph.HE]}
  \BibitemShut {NoStop}%
\bibitem [{\citenamefont {Musolino}\ \emph {et~al.}(2023)\citenamefont
  {Musolino}, \citenamefont {Ecker},\ and\ \citenamefont
  {Rezzolla}}]{Musolino:2023edi}%
  \BibitemOpen
  \bibfield  {author} {\bibinfo {author} {\bibfnamefont {C.}~\bibnamefont
  {Musolino}}, \bibinfo {author} {\bibfnamefont {C.}~\bibnamefont {Ecker}},\
  and\ \bibinfo {author} {\bibfnamefont {L.}~\bibnamefont {Rezzolla}},\
  }\bibfield  {title} {\bibinfo {title} {{On the maximum mass and oblateness of
  rotating neutron stars with generic equations of state}},\ }\href@noop {} {\
  (\bibinfo {year} {2023})},\ \Eprint {https://arxiv.org/abs/2307.03225}
  {arXiv:2307.03225 [gr-qc]} \BibitemShut {NoStop}%
\bibitem [{\citenamefont {Gottlieb}\ \emph {et~al.}(2023)\citenamefont
  {Gottlieb}, \citenamefont {Metzger}, \citenamefont {Quataert}, \citenamefont
  {Issa}, \citenamefont {Martineau}, \citenamefont {Foucart}, \citenamefont
  {Duez}, \citenamefont {Kidder}, \citenamefont {Pfeiffer},\ and\ \citenamefont
  {Scheel}}]{Gottlieb:2023b}%
  \BibitemOpen
  \bibfield  {author} {\bibinfo {author} {\bibfnamefont {O.}~\bibnamefont
  {Gottlieb}}, \bibinfo {author} {\bibfnamefont {B.~D.}\ \bibnamefont
  {Metzger}}, \bibinfo {author} {\bibfnamefont {E.}~\bibnamefont {Quataert}},
  \bibinfo {author} {\bibfnamefont {D.}~\bibnamefont {Issa}}, \bibinfo {author}
  {\bibfnamefont {T.}~\bibnamefont {Martineau}}, \bibinfo {author}
  {\bibfnamefont {F.}~\bibnamefont {Foucart}}, \bibinfo {author} {\bibfnamefont
  {M.~D.}\ \bibnamefont {Duez}}, \bibinfo {author} {\bibfnamefont {L.~E.}\
  \bibnamefont {Kidder}}, \bibinfo {author} {\bibfnamefont {H.~P.}\
  \bibnamefont {Pfeiffer}},\ and\ \bibinfo {author} {\bibfnamefont {M.~A.}\
  \bibnamefont {Scheel}},\ }\bibfield  {title} {\bibinfo {title} {{A Unified
  Picture of Short and Long Gamma-Ray Bursts from Compact Binary Mergers}},\
  }\href {https://doi.org/10.3847/2041-8213/ad096e} {\bibfield  {journal}
  {\bibinfo  {journal} {Astrophys. J. Lett.}\ }\textbf {\bibinfo {volume}
  {958}},\ \bibinfo {pages} {L33} (\bibinfo {year} {2023})},\ \Eprint
  {https://arxiv.org/abs/2309.00038} {arXiv:2309.00038 [astro-ph.HE]}
  \BibitemShut {NoStop}%
\bibitem [{\citenamefont {Novak}(2001)}]{Novak:2001ck}%
  \BibitemOpen
  \bibfield  {author} {\bibinfo {author} {\bibfnamefont {J.}~\bibnamefont
  {Novak}},\ }\bibfield  {title} {\bibinfo {title} {{Velocity induced collapses
  of stable neutron stars}},\ }\href
  {https://doi.org/10.1051/0004-6361:20011037} {\bibfield  {journal} {\bibinfo
  {journal} {Astron. Astrophys.}\ }\textbf {\bibinfo {volume} {376}},\ \bibinfo
  {pages} {606} (\bibinfo {year} {2001})},\ \Eprint
  {https://arxiv.org/abs/gr-qc/0107045} {arXiv:gr-qc/0107045} \BibitemShut
  {NoStop}%
\bibitem [{\citenamefont {Noble}\ and\ \citenamefont
  {Choptuik}(2008)}]{Noble:2007vf}%
  \BibitemOpen
  \bibfield  {author} {\bibinfo {author} {\bibfnamefont {S.~C.}\ \bibnamefont
  {Noble}}\ and\ \bibinfo {author} {\bibfnamefont {M.~W.}\ \bibnamefont
  {Choptuik}},\ }\bibfield  {title} {\bibinfo {title} {{Type II critical
  phenomena of neutron star collapse}},\ }\href
  {https://doi.org/10.1103/PhysRevD.78.064059} {\bibfield  {journal} {\bibinfo
  {journal} {Phys. Rev. D}\ }\textbf {\bibinfo {volume} {78}},\ \bibinfo
  {pages} {064059} (\bibinfo {year} {2008})},\ \Eprint
  {https://arxiv.org/abs/0709.3527} {arXiv:0709.3527 [gr-qc]} \BibitemShut
  {NoStop}%
\bibitem [{\citenamefont {Radice}\ \emph {et~al.}(2010)\citenamefont {Radice},
  \citenamefont {Rezzolla},\ and\ \citenamefont {Kellermann}}]{Radice:2010rw}%
  \BibitemOpen
  \bibfield  {author} {\bibinfo {author} {\bibfnamefont {D.}~\bibnamefont
  {Radice}}, \bibinfo {author} {\bibfnamefont {L.}~\bibnamefont {Rezzolla}},\
  and\ \bibinfo {author} {\bibfnamefont {T.}~\bibnamefont {Kellermann}},\
  }\bibfield  {title} {\bibinfo {title} {{Critical Phenomena in Neutron Stars
  I: Linearly Unstable Nonrotating Models}},\ }\href
  {https://doi.org/10.1088/0264-9381/27/23/235015} {\bibfield  {journal}
  {\bibinfo  {journal} {Class. Quant. Grav.}\ }\textbf {\bibinfo {volume}
  {27}},\ \bibinfo {pages} {235015} (\bibinfo {year} {2010})},\ \Eprint
  {https://arxiv.org/abs/1007.2809} {arXiv:1007.2809 [gr-qc]} \BibitemShut
  {NoStop}%
\bibitem [{\citenamefont {Jin}\ and\ \citenamefont {Suen}(2007)}]{Jin:2006gm}%
  \BibitemOpen
  \bibfield  {author} {\bibinfo {author} {\bibfnamefont {K.-J.}\ \bibnamefont
  {Jin}}\ and\ \bibinfo {author} {\bibfnamefont {W.-M.}\ \bibnamefont {Suen}},\
  }\bibfield  {title} {\bibinfo {title} {{Critical phenomena in head-on
  collisions of neutron stars}},\ }\href
  {https://doi.org/10.1103/PhysRevLett.98.131101} {\bibfield  {journal}
  {\bibinfo  {journal} {Phys. Rev. Lett.}\ }\textbf {\bibinfo {volume} {98}},\
  \bibinfo {pages} {131101} (\bibinfo {year} {2007})},\ \Eprint
  {https://arxiv.org/abs/gr-qc/0603094} {arXiv:gr-qc/0603094} \BibitemShut
  {NoStop}%
\bibitem [{\citenamefont {Kellermann}\ \emph {et~al.}(2010)\citenamefont
  {Kellermann}, \citenamefont {Rezzolla},\ and\ \citenamefont
  {Radice}}]{Kellermann:2010rt}%
  \BibitemOpen
  \bibfield  {author} {\bibinfo {author} {\bibfnamefont {T.}~\bibnamefont
  {Kellermann}}, \bibinfo {author} {\bibfnamefont {L.}~\bibnamefont
  {Rezzolla}},\ and\ \bibinfo {author} {\bibfnamefont {D.}~\bibnamefont
  {Radice}},\ }\bibfield  {title} {\bibinfo {title} {{Critical Phenomena in
  Neutron Stars II: Head-on Collisions}},\ }\href
  {https://doi.org/10.1088/0264-9381/27/23/235016} {\bibfield  {journal}
  {\bibinfo  {journal} {Class. Quant. Grav.}\ }\textbf {\bibinfo {volume}
  {27}},\ \bibinfo {pages} {235016} (\bibinfo {year} {2010})},\ \Eprint
  {https://arxiv.org/abs/1007.2797} {arXiv:1007.2797 [gr-qc]} \BibitemShut
  {NoStop}%
\bibitem [{\citenamefont {Bauswein}\ \emph {et~al.}(2013)\citenamefont
  {Bauswein}, \citenamefont {Baumgarte},\ and\ \citenamefont
  {Janka}}]{Bauswein:2013jpa}%
  \BibitemOpen
  \bibfield  {author} {\bibinfo {author} {\bibfnamefont {A.}~\bibnamefont
  {Bauswein}}, \bibinfo {author} {\bibfnamefont {T.~W.}\ \bibnamefont
  {Baumgarte}},\ and\ \bibinfo {author} {\bibfnamefont {H.~T.}\ \bibnamefont
  {Janka}},\ }\bibfield  {title} {\bibinfo {title} {{Prompt merger collapse and
  the maximum mass of neutron stars}},\ }\href
  {https://doi.org/10.1103/PhysRevLett.111.131101} {\bibfield  {journal}
  {\bibinfo  {journal} {Phys. Rev. Lett.}\ }\textbf {\bibinfo {volume} {111}},\
  \bibinfo {pages} {131101} (\bibinfo {year} {2013})},\ \Eprint
  {https://arxiv.org/abs/1307.5191} {arXiv:1307.5191 [astro-ph.SR]}
  \BibitemShut {NoStop}%
\bibitem [{\citenamefont {K\"oppel}\ \emph {et~al.}(2019)\citenamefont
  {K\"oppel}, \citenamefont {Bovard},\ and\ \citenamefont
  {Rezzolla}}]{Koppel:2019pys}%
  \BibitemOpen
  \bibfield  {author} {\bibinfo {author} {\bibfnamefont {S.}~\bibnamefont
  {K\"oppel}}, \bibinfo {author} {\bibfnamefont {L.}~\bibnamefont {Bovard}},\
  and\ \bibinfo {author} {\bibfnamefont {L.}~\bibnamefont {Rezzolla}},\
  }\bibfield  {title} {\bibinfo {title} {{A General-relativistic Determination
  of the Threshold Mass to Prompt Collapse in Binary Neutron Star Mergers}},\
  }\href {https://doi.org/10.3847/2041-8213/ab0210} {\bibfield  {journal}
  {\bibinfo  {journal} {Astrophys. J. Lett.}\ }\textbf {\bibinfo {volume}
  {872}},\ \bibinfo {pages} {L16} (\bibinfo {year} {2019})},\ \Eprint
  {https://arxiv.org/abs/1901.09977} {arXiv:1901.09977 [gr-qc]} \BibitemShut
  {NoStop}%
\bibitem [{\citenamefont {Agathos}\ \emph {et~al.}(2020)\citenamefont
  {Agathos}, \citenamefont {Zappa}, \citenamefont {Bernuzzi}, \citenamefont
  {Perego}, \citenamefont {Breschi},\ and\ \citenamefont
  {Radice}}]{Agathos:2019sah}%
  \BibitemOpen
  \bibfield  {author} {\bibinfo {author} {\bibfnamefont {M.}~\bibnamefont
  {Agathos}}, \bibinfo {author} {\bibfnamefont {F.}~\bibnamefont {Zappa}},
  \bibinfo {author} {\bibfnamefont {S.}~\bibnamefont {Bernuzzi}}, \bibinfo
  {author} {\bibfnamefont {A.}~\bibnamefont {Perego}}, \bibinfo {author}
  {\bibfnamefont {M.}~\bibnamefont {Breschi}},\ and\ \bibinfo {author}
  {\bibfnamefont {D.}~\bibnamefont {Radice}},\ }\bibfield  {title} {\bibinfo
  {title} {{Inferring Prompt Black-Hole Formation in Neutron Star Mergers from
  Gravitational-Wave Data}},\ }\href
  {https://doi.org/10.1103/PhysRevD.101.044006} {\bibfield  {journal} {\bibinfo
   {journal} {Phys. Rev. D}\ }\textbf {\bibinfo {volume} {101}},\ \bibinfo
  {pages} {044006} (\bibinfo {year} {2020})},\ \Eprint
  {https://arxiv.org/abs/1908.05442} {arXiv:1908.05442 [gr-qc]} \BibitemShut
  {NoStop}%
\bibitem [{\citenamefont {Bauswein}\ \emph {et~al.}(2021)\citenamefont
  {Bauswein}, \citenamefont {Blacker}, \citenamefont {Lioutas}, \citenamefont
  {Soultanis}, \citenamefont {Vijayan},\ and\ \citenamefont
  {Stergioulas}}]{Bauswein:2020xlt}%
  \BibitemOpen
  \bibfield  {author} {\bibinfo {author} {\bibfnamefont {A.}~\bibnamefont
  {Bauswein}}, \bibinfo {author} {\bibfnamefont {S.}~\bibnamefont {Blacker}},
  \bibinfo {author} {\bibfnamefont {G.}~\bibnamefont {Lioutas}}, \bibinfo
  {author} {\bibfnamefont {T.}~\bibnamefont {Soultanis}}, \bibinfo {author}
  {\bibfnamefont {V.}~\bibnamefont {Vijayan}},\ and\ \bibinfo {author}
  {\bibfnamefont {N.}~\bibnamefont {Stergioulas}},\ }\bibfield  {title}
  {\bibinfo {title} {{Systematics of prompt black-hole formation in neutron
  star mergers}},\ }\href {https://doi.org/10.1103/PhysRevD.103.123004}
  {\bibfield  {journal} {\bibinfo  {journal} {Phys. Rev. D}\ }\textbf {\bibinfo
  {volume} {103}},\ \bibinfo {pages} {123004} (\bibinfo {year} {2021})},\
  \Eprint {https://arxiv.org/abs/2010.04461} {arXiv:2010.04461 [astro-ph.HE]}
  \BibitemShut {NoStop}%
\bibitem [{\citenamefont {Tootle}\ \emph {et~al.}(2021)\citenamefont {Tootle},
  \citenamefont {Papenfort}, \citenamefont {Most},\ and\ \citenamefont
  {Rezzolla}}]{Tootle:2021umi}%
  \BibitemOpen
  \bibfield  {author} {\bibinfo {author} {\bibfnamefont {S.~D.}\ \bibnamefont
  {Tootle}}, \bibinfo {author} {\bibfnamefont {L.~J.}\ \bibnamefont
  {Papenfort}}, \bibinfo {author} {\bibfnamefont {E.~R.}\ \bibnamefont
  {Most}},\ and\ \bibinfo {author} {\bibfnamefont {L.}~\bibnamefont
  {Rezzolla}},\ }\bibfield  {title} {\bibinfo {title} {{Quasi-universal
  Behavior of the Threshold Mass in Unequal-mass, Spinning Binary Neutron Star
  Mergers}},\ }\href {https://doi.org/10.3847/2041-8213/ac350d} {\bibfield
  {journal} {\bibinfo  {journal} {Astrophys. J. Lett.}\ }\textbf {\bibinfo
  {volume} {922}},\ \bibinfo {pages} {L19} (\bibinfo {year} {2021})},\ \Eprint
  {https://arxiv.org/abs/2109.00940} {arXiv:2109.00940 [gr-qc]} \BibitemShut
  {NoStop}%
\bibitem [{\citenamefont {Kashyap}\ \emph {et~al.}(2022)\citenamefont {Kashyap}
  \emph {et~al.}}]{Kashyap:2021wzs}%
  \BibitemOpen
  \bibfield  {author} {\bibinfo {author} {\bibfnamefont {R.}~\bibnamefont
  {Kashyap}} \emph {et~al.},\ }\bibfield  {title} {\bibinfo {title} {{Numerical
  relativity simulations of prompt collapse mergers: Threshold mass and
  phenomenological constraints on neutron star properties after GW170817}},\
  }\href {https://doi.org/10.1103/PhysRevD.105.103022} {\bibfield  {journal}
  {\bibinfo  {journal} {Phys. Rev. D}\ }\textbf {\bibinfo {volume} {105}},\
  \bibinfo {pages} {103022} (\bibinfo {year} {2022})},\ \Eprint
  {https://arxiv.org/abs/2111.05183} {arXiv:2111.05183 [astro-ph.HE]}
  \BibitemShut {NoStop}%
\bibitem [{\citenamefont {Bauswein}\ \emph {et~al.}(2017)\citenamefont
  {Bauswein}, \citenamefont {Just}, \citenamefont {Janka},\ and\ \citenamefont
  {Stergioulas}}]{Bauswein:2017vtn}%
  \BibitemOpen
  \bibfield  {author} {\bibinfo {author} {\bibfnamefont {A.}~\bibnamefont
  {Bauswein}}, \bibinfo {author} {\bibfnamefont {O.}~\bibnamefont {Just}},
  \bibinfo {author} {\bibfnamefont {H.-T.}\ \bibnamefont {Janka}},\ and\
  \bibinfo {author} {\bibfnamefont {N.}~\bibnamefont {Stergioulas}},\
  }\bibfield  {title} {\bibinfo {title} {{Neutron-star radius constraints from
  GW170817 and future detections}},\ }\href
  {https://doi.org/10.3847/2041-8213/aa9994} {\bibfield  {journal} {\bibinfo
  {journal} {Astrophys. J. Lett.}\ }\textbf {\bibinfo {volume} {850}},\
  \bibinfo {pages} {L34} (\bibinfo {year} {2017})},\ \Eprint
  {https://arxiv.org/abs/1710.06843} {arXiv:1710.06843 [astro-ph.HE]}
  \BibitemShut {NoStop}%
\bibitem [{\citenamefont {Altiparmak}\ \emph {et~al.}(2022)\citenamefont
  {Altiparmak}, \citenamefont {Ecker},\ and\ \citenamefont
  {Rezzolla}}]{Altiparmak:2022bke}%
  \BibitemOpen
  \bibfield  {author} {\bibinfo {author} {\bibfnamefont {S.}~\bibnamefont
  {Altiparmak}}, \bibinfo {author} {\bibfnamefont {C.}~\bibnamefont {Ecker}},\
  and\ \bibinfo {author} {\bibfnamefont {L.}~\bibnamefont {Rezzolla}},\
  }\bibfield  {title} {\bibinfo {title} {{On the Sound Speed in Neutron
  Stars}},\ }\href {https://doi.org/10.3847/2041-8213/ac9b2a} {\bibfield
  {journal} {\bibinfo  {journal} {Astrophys. J. Lett.}\ }\textbf {\bibinfo
  {volume} {939}},\ \bibinfo {pages} {L34} (\bibinfo {year} {2022})},\ \Eprint
  {https://arxiv.org/abs/2203.14974} {arXiv:2203.14974 [astro-ph.HE]}
  \BibitemShut {NoStop}%
\bibitem [{\citenamefont {Demircik}\ \emph {et~al.}(2022)\citenamefont
  {Demircik}, \citenamefont {Ecker},\ and\ \citenamefont
  {J\"arvinen}}]{Demircik:2021zll}%
  \BibitemOpen
  \bibfield  {author} {\bibinfo {author} {\bibfnamefont {T.}~\bibnamefont
  {Demircik}}, \bibinfo {author} {\bibfnamefont {C.}~\bibnamefont {Ecker}},\
  and\ \bibinfo {author} {\bibfnamefont {M.}~\bibnamefont {J\"arvinen}},\
  }\bibfield  {title} {\bibinfo {title} {{Dense and Hot QCD at Strong
  Coupling}},\ }\href {https://doi.org/10.1103/PhysRevX.12.041012} {\bibfield
  {journal} {\bibinfo  {journal} {Phys. Rev. X}\ }\textbf {\bibinfo {volume}
  {12}},\ \bibinfo {pages} {041012} (\bibinfo {year} {2022})},\ \Eprint
  {https://arxiv.org/abs/2112.12157} {arXiv:2112.12157 [hep-ph]} \BibitemShut
  {NoStop}%
\bibitem [{\citenamefont {Ecker}\ and\ \citenamefont
  {Rezzolla}(2022{\natexlab{a}})}]{Ecker:2022dlg}%
  \BibitemOpen
  \bibfield  {author} {\bibinfo {author} {\bibfnamefont {C.}~\bibnamefont
  {Ecker}}\ and\ \bibinfo {author} {\bibfnamefont {L.}~\bibnamefont
  {Rezzolla}},\ }\bibfield  {title} {\bibinfo {title} {{Impact of large-mass
  constraints on the properties of neutron stars}},\ }\href
  {https://doi.org/10.1093/mnras/stac3755} {\bibfield  {journal} {\bibinfo
  {journal} {Mon. Not. Roy. Astron. Soc.}\ }\textbf {\bibinfo {volume} {519}},\
  \bibinfo {pages} {2615} (\bibinfo {year} {2022}{\natexlab{a}})},\ \Eprint
  {https://arxiv.org/abs/2209.08101} {arXiv:2209.08101 [astro-ph.HE]}
  \BibitemShut {NoStop}%
\bibitem [{\citenamefont {Tews}\ \emph {et~al.}(2013)\citenamefont {Tews},
  \citenamefont {Kr\"uger}, \citenamefont {Hebeler},\ and\ \citenamefont
  {Schwenk}}]{Tews:2012fj}%
  \BibitemOpen
  \bibfield  {author} {\bibinfo {author} {\bibfnamefont {I.}~\bibnamefont
  {Tews}}, \bibinfo {author} {\bibfnamefont {T.}~\bibnamefont {Kr\"uger}},
  \bibinfo {author} {\bibfnamefont {K.}~\bibnamefont {Hebeler}},\ and\ \bibinfo
  {author} {\bibfnamefont {A.}~\bibnamefont {Schwenk}},\ }\bibfield  {title}
  {\bibinfo {title} {{Neutron matter at next-to-next-to-next-to-leading order
  in chiral effective field theory}},\ }\href
  {https://doi.org/10.1103/PhysRevLett.110.032504} {\bibfield  {journal}
  {\bibinfo  {journal} {Phys. Rev. Lett.}\ }\textbf {\bibinfo {volume} {110}},\
  \bibinfo {pages} {032504} (\bibinfo {year} {2013})},\ \Eprint
  {https://arxiv.org/abs/1206.0025} {arXiv:1206.0025 [nucl-th]} \BibitemShut
  {NoStop}%
\bibitem [{\citenamefont {Oertel}\ \emph {et~al.}(2017)\citenamefont {Oertel},
  \citenamefont {Hempel}, \citenamefont {Kl\"ahn},\ and\ \citenamefont
  {Typel}}]{Oertel:2016bki}%
  \BibitemOpen
  \bibfield  {author} {\bibinfo {author} {\bibfnamefont {M.}~\bibnamefont
  {Oertel}}, \bibinfo {author} {\bibfnamefont {M.}~\bibnamefont {Hempel}},
  \bibinfo {author} {\bibfnamefont {T.}~\bibnamefont {Kl\"ahn}},\ and\ \bibinfo
  {author} {\bibfnamefont {S.}~\bibnamefont {Typel}},\ }\bibfield  {title}
  {\bibinfo {title} {{Equations of state for supernovae and compact stars}},\
  }\href {https://doi.org/10.1103/RevModPhys.89.015007} {\bibfield  {journal}
  {\bibinfo  {journal} {Rev. Mod. Phys.}\ }\textbf {\bibinfo {volume} {89}},\
  \bibinfo {pages} {015007} (\bibinfo {year} {2017})},\ \Eprint
  {https://arxiv.org/abs/1610.03361} {arXiv:1610.03361 [astro-ph.HE]}
  \BibitemShut {NoStop}%
\bibitem [{\citenamefont {Tews}\ \emph {et~al.}(2018)\citenamefont {Tews},
  \citenamefont {Margueron},\ and\ \citenamefont {Reddy}}]{Tews:2018iwm}%
  \BibitemOpen
  \bibfield  {author} {\bibinfo {author} {\bibfnamefont {I.}~\bibnamefont
  {Tews}}, \bibinfo {author} {\bibfnamefont {J.}~\bibnamefont {Margueron}},\
  and\ \bibinfo {author} {\bibfnamefont {S.}~\bibnamefont {Reddy}},\ }\bibfield
   {title} {\bibinfo {title} {{Critical examination of constraints on the
  equation of state of dense matter obtained from GW170817}},\ }\href
  {https://doi.org/10.1103/PhysRevC.98.045804} {\bibfield  {journal} {\bibinfo
  {journal} {Phys. Rev. C}\ }\textbf {\bibinfo {volume} {98}},\ \bibinfo
  {pages} {045804} (\bibinfo {year} {2018})},\ \Eprint
  {https://arxiv.org/abs/1804.02783} {arXiv:1804.02783 [nucl-th]} \BibitemShut
  {NoStop}%
\bibitem [{\citenamefont {Ecker}\ and\ \citenamefont
  {Rezzolla}(2022{\natexlab{b}})}]{Ecker:2022xxj}%
  \BibitemOpen
  \bibfield  {author} {\bibinfo {author} {\bibfnamefont {C.}~\bibnamefont
  {Ecker}}\ and\ \bibinfo {author} {\bibfnamefont {L.}~\bibnamefont
  {Rezzolla}},\ }\bibfield  {title} {\bibinfo {title} {{A General,
  Scale-independent Description of the Sound Speed in Neutron Stars}},\ }\href
  {https://doi.org/10.3847/2041-8213/ac8674} {\bibfield  {journal} {\bibinfo
  {journal} {Astrophys. J. Lett.}\ }\textbf {\bibinfo {volume} {939}},\
  \bibinfo {pages} {L35} (\bibinfo {year} {2022}{\natexlab{b}})},\ \Eprint
  {https://arxiv.org/abs/2207.04417} {arXiv:2207.04417 [gr-qc]} \BibitemShut
  {NoStop}%
\bibitem [{\citenamefont {Jiang}\ \emph {et~al.}(2023)\citenamefont {Jiang},
  \citenamefont {Ecker},\ and\ \citenamefont {Rezzolla}}]{Jiang:2022tps}%
  \BibitemOpen
  \bibfield  {author} {\bibinfo {author} {\bibfnamefont {J.-L.}\ \bibnamefont
  {Jiang}}, \bibinfo {author} {\bibfnamefont {C.}~\bibnamefont {Ecker}},\ and\
  \bibinfo {author} {\bibfnamefont {L.}~\bibnamefont {Rezzolla}},\ }\bibfield
  {title} {\bibinfo {title} {{Bayesian Analysis of Neutron-star Properties with
  Parameterized Equations of State: The Role of the Likelihood Functions}},\
  }\href {https://doi.org/10.3847/1538-4357/acc4be} {\bibfield  {journal}
  {\bibinfo  {journal} {Astrophys. J.}\ }\textbf {\bibinfo {volume} {949}},\
  \bibinfo {pages} {11} (\bibinfo {year} {2023})},\ \Eprint
  {https://arxiv.org/abs/2211.00018} {arXiv:2211.00018 [gr-qc]} \BibitemShut
  {NoStop}%
\bibitem [{\citenamefont {Bitaghsir~Fadafan}\ \emph {et~al.}(2021)\citenamefont
  {Bitaghsir~Fadafan}, \citenamefont {Cruz~Rojas},\ and\ \citenamefont
  {Evans}}]{BitaghsirFadafan:2020otb}%
  \BibitemOpen
  \bibfield  {author} {\bibinfo {author} {\bibfnamefont {K.}~\bibnamefont
  {Bitaghsir~Fadafan}}, \bibinfo {author} {\bibfnamefont {J.}~\bibnamefont
  {Cruz~Rojas}},\ and\ \bibinfo {author} {\bibfnamefont {N.}~\bibnamefont
  {Evans}},\ }\bibfield  {title} {\bibinfo {title} {{Holographic quark matter
  with colour superconductivity and a stiff equation of state for compact
  stars}},\ }\href {https://doi.org/10.1103/PhysRevD.103.026012} {\bibfield
  {journal} {\bibinfo  {journal} {Phys. Rev. D}\ }\textbf {\bibinfo {volume}
  {103}},\ \bibinfo {pages} {026012} (\bibinfo {year} {2021})},\ \Eprint
  {https://arxiv.org/abs/2009.14079} {arXiv:2009.14079 [hep-ph]} \BibitemShut
  {NoStop}%
\bibitem [{\citenamefont {Ghoroku}\ \emph {et~al.}(2021)\citenamefont
  {Ghoroku}, \citenamefont {Kashiwa}, \citenamefont {Nakano}, \citenamefont
  {Tachibana},\ and\ \citenamefont {Toyoda}}]{Ghoroku:2021fos}%
  \BibitemOpen
  \bibfield  {author} {\bibinfo {author} {\bibfnamefont {K.}~\bibnamefont
  {Ghoroku}}, \bibinfo {author} {\bibfnamefont {K.}~\bibnamefont {Kashiwa}},
  \bibinfo {author} {\bibfnamefont {Y.}~\bibnamefont {Nakano}}, \bibinfo
  {author} {\bibfnamefont {M.}~\bibnamefont {Tachibana}},\ and\ \bibinfo
  {author} {\bibfnamefont {F.}~\bibnamefont {Toyoda}},\ }\bibfield  {title}
  {\bibinfo {title} {{Stiff equation of state for a holographic nuclear matter
  as instanton gas}},\ }\href {https://doi.org/10.1103/PhysRevD.104.126002}
  {\bibfield  {journal} {\bibinfo  {journal} {Phys. Rev. D}\ }\textbf {\bibinfo
  {volume} {104}},\ \bibinfo {pages} {126002} (\bibinfo {year} {2021})},\
  \Eprint {https://arxiv.org/abs/2107.14450} {arXiv:2107.14450 [hep-th]}
  \BibitemShut {NoStop}%
\bibitem [{\citenamefont {J\"arvinen}(2022)}]{Jarvinen:2021jbd}%
  \BibitemOpen
  \bibfield  {author} {\bibinfo {author} {\bibfnamefont {M.}~\bibnamefont
  {J\"arvinen}},\ }\bibfield  {title} {\bibinfo {title} {{Holographic modeling
  of nuclear matter and neutron stars}},\ }\href
  {https://doi.org/10.1140/epjc/s10052-022-10227-x} {\bibfield  {journal}
  {\bibinfo  {journal} {Eur. Phys. J. C}\ }\textbf {\bibinfo {volume} {82}},\
  \bibinfo {pages} {282} (\bibinfo {year} {2022})},\ \Eprint
  {https://arxiv.org/abs/2110.08281} {arXiv:2110.08281 [hep-ph]} \BibitemShut
  {NoStop}%
\bibitem [{\citenamefont {Kovensky}\ \emph {et~al.}(2022)\citenamefont
  {Kovensky}, \citenamefont {Poole},\ and\ \citenamefont
  {Schmitt}}]{Kovensky:2021kzl}%
  \BibitemOpen
  \bibfield  {author} {\bibinfo {author} {\bibfnamefont {N.}~\bibnamefont
  {Kovensky}}, \bibinfo {author} {\bibfnamefont {A.}~\bibnamefont {Poole}},\
  and\ \bibinfo {author} {\bibfnamefont {A.}~\bibnamefont {Schmitt}},\
  }\bibfield  {title} {\bibinfo {title} {{Building a realistic neutron star
  from holography}},\ }\href {https://doi.org/10.1103/PhysRevD.105.034022}
  {\bibfield  {journal} {\bibinfo  {journal} {Phys. Rev. D}\ }\textbf {\bibinfo
  {volume} {105}},\ \bibinfo {pages} {034022} (\bibinfo {year} {2022})},\
  \Eprint {https://arxiv.org/abs/2111.03374} {arXiv:2111.03374 [hep-ph]}
  \BibitemShut {NoStop}%
\bibitem [{\citenamefont {Hoyos}\ \emph {et~al.}(2022)\citenamefont {Hoyos},
  \citenamefont {Jokela},\ and\ \citenamefont {Vuorinen}}]{Hoyos:2021uff}%
  \BibitemOpen
  \bibfield  {author} {\bibinfo {author} {\bibfnamefont {C.}~\bibnamefont
  {Hoyos}}, \bibinfo {author} {\bibfnamefont {N.}~\bibnamefont {Jokela}},\ and\
  \bibinfo {author} {\bibfnamefont {A.}~\bibnamefont {Vuorinen}},\ }\bibfield
  {title} {\bibinfo {title} {{Holographic approach to compact stars and their
  binary mergers}},\ }\href {https://doi.org/10.1016/j.ppnp.2022.103972}
  {\bibfield  {journal} {\bibinfo  {journal} {Prog. Part. Nucl. Phys.}\
  }\textbf {\bibinfo {volume} {126}},\ \bibinfo {pages} {103972} (\bibinfo
  {year} {2022})},\ \Eprint {https://arxiv.org/abs/2112.08422}
  {arXiv:2112.08422 [hep-th]} \BibitemShut {NoStop}%
\bibitem [{\citenamefont {Bartolini}\ \emph {et~al.}(2022)\citenamefont
  {Bartolini}, \citenamefont {Gudnason}, \citenamefont {Leutgeb},\ and\
  \citenamefont {Rebhan}}]{Bartolini:2022rkl}%
  \BibitemOpen
  \bibfield  {author} {\bibinfo {author} {\bibfnamefont {L.}~\bibnamefont
  {Bartolini}}, \bibinfo {author} {\bibfnamefont {S.~B.}\ \bibnamefont
  {Gudnason}}, \bibinfo {author} {\bibfnamefont {J.}~\bibnamefont {Leutgeb}},\
  and\ \bibinfo {author} {\bibfnamefont {A.}~\bibnamefont {Rebhan}},\
  }\bibfield  {title} {\bibinfo {title} {{Neutron stars and phase diagram in a
  hard-wall AdS/QCD model}},\ }\href
  {https://doi.org/10.1103/PhysRevD.105.126014} {\bibfield  {journal} {\bibinfo
   {journal} {Phys. Rev. D}\ }\textbf {\bibinfo {volume} {105}},\ \bibinfo
  {pages} {126014} (\bibinfo {year} {2022})},\ \Eprint
  {https://arxiv.org/abs/2202.12845} {arXiv:2202.12845 [hep-th]} \BibitemShut
  {NoStop}%
\bibitem [{\citenamefont {Fraga}\ \emph {et~al.}(2014)\citenamefont {Fraga},
  \citenamefont {Kurkela},\ and\ \citenamefont {Vuorinen}}]{Fraga:2013qra}%
  \BibitemOpen
  \bibfield  {author} {\bibinfo {author} {\bibfnamefont {E.~S.}\ \bibnamefont
  {Fraga}}, \bibinfo {author} {\bibfnamefont {A.}~\bibnamefont {Kurkela}},\
  and\ \bibinfo {author} {\bibfnamefont {A.}~\bibnamefont {Vuorinen}},\
  }\bibfield  {title} {\bibinfo {title} {{Interacting quark matter equation of
  state for compact stars}},\ }\href
  {https://doi.org/10.1088/2041-8205/781/2/L25} {\bibfield  {journal} {\bibinfo
   {journal} {Astrophys. J. Lett.}\ }\textbf {\bibinfo {volume} {781}},\
  \bibinfo {pages} {L25} (\bibinfo {year} {2014})},\ \Eprint
  {https://arxiv.org/abs/1311.5154} {arXiv:1311.5154 [nucl-th]} \BibitemShut
  {NoStop}%
\bibitem [{\citenamefont {Radice}\ \emph {et~al.}(2017)\citenamefont {Radice},
  \citenamefont {Bernuzzi}, \citenamefont {Del~Pozzo}, \citenamefont
  {Roberts},\ and\ \citenamefont {Ott}}]{Radice:2016rys}%
  \BibitemOpen
  \bibfield  {author} {\bibinfo {author} {\bibfnamefont {D.}~\bibnamefont
  {Radice}}, \bibinfo {author} {\bibfnamefont {S.}~\bibnamefont {Bernuzzi}},
  \bibinfo {author} {\bibfnamefont {W.}~\bibnamefont {Del~Pozzo}}, \bibinfo
  {author} {\bibfnamefont {L.~F.}\ \bibnamefont {Roberts}},\ and\ \bibinfo
  {author} {\bibfnamefont {C.~D.}\ \bibnamefont {Ott}},\ }\bibfield  {title}
  {\bibinfo {title} {{Probing Extreme-Density Matter with Gravitational Wave
  Observations of Binary Neutron Star Merger Remnants}},\ }\href
  {https://doi.org/10.3847/2041-8213/aa775f} {\bibfield  {journal} {\bibinfo
  {journal} {Astrophys. J. Lett.}\ }\textbf {\bibinfo {volume} {842}},\
  \bibinfo {pages} {L10} (\bibinfo {year} {2017})},\ \Eprint
  {https://arxiv.org/abs/1612.06429} {arXiv:1612.06429 [astro-ph.HE]}
  \BibitemShut {NoStop}%
\bibitem [{\citenamefont {Most}\ \emph
  {et~al.}(2019{\natexlab{a}})\citenamefont {Most}, \citenamefont {Papenfort},
  \citenamefont {Dexheimer}, \citenamefont {Hanauske}, \citenamefont {Schramm},
  \citenamefont {St\"ocker},\ and\ \citenamefont {Rezzolla}}]{Most:2018eaw}%
  \BibitemOpen
  \bibfield  {author} {\bibinfo {author} {\bibfnamefont {E.~R.}\ \bibnamefont
  {Most}}, \bibinfo {author} {\bibfnamefont {L.~J.}\ \bibnamefont {Papenfort}},
  \bibinfo {author} {\bibfnamefont {V.}~\bibnamefont {Dexheimer}}, \bibinfo
  {author} {\bibfnamefont {M.}~\bibnamefont {Hanauske}}, \bibinfo {author}
  {\bibfnamefont {S.}~\bibnamefont {Schramm}}, \bibinfo {author} {\bibfnamefont
  {H.}~\bibnamefont {St\"ocker}},\ and\ \bibinfo {author} {\bibfnamefont
  {L.}~\bibnamefont {Rezzolla}},\ }\bibfield  {title} {\bibinfo {title}
  {{Signatures of quark-hadron phase transitions in general-relativistic
  neutron-star mergers}},\ }\href
  {https://doi.org/10.1103/PhysRevLett.122.061101} {\bibfield  {journal}
  {\bibinfo  {journal} {Phys. Rev. Lett.}\ }\textbf {\bibinfo {volume} {122}},\
  \bibinfo {pages} {061101} (\bibinfo {year} {2019}{\natexlab{a}})},\ \Eprint
  {https://arxiv.org/abs/1807.03684} {arXiv:1807.03684 [astro-ph.HE]}
  \BibitemShut {NoStop}%
\bibitem [{\citenamefont {Bauswein}\ \emph {et~al.}(2019)\citenamefont
  {Bauswein}, \citenamefont {Bastian}, \citenamefont {Blaschke}, \citenamefont
  {Chatziioannou}, \citenamefont {Clark}, \citenamefont {Fischer},\ and\
  \citenamefont {Oertel}}]{Bauswein:2018bma}%
  \BibitemOpen
  \bibfield  {author} {\bibinfo {author} {\bibfnamefont {A.}~\bibnamefont
  {Bauswein}}, \bibinfo {author} {\bibfnamefont {N.-U.~F.}\ \bibnamefont
  {Bastian}}, \bibinfo {author} {\bibfnamefont {D.~B.}\ \bibnamefont
  {Blaschke}}, \bibinfo {author} {\bibfnamefont {K.}~\bibnamefont
  {Chatziioannou}}, \bibinfo {author} {\bibfnamefont {J.~A.}\ \bibnamefont
  {Clark}}, \bibinfo {author} {\bibfnamefont {T.}~\bibnamefont {Fischer}},\
  and\ \bibinfo {author} {\bibfnamefont {M.}~\bibnamefont {Oertel}},\
  }\bibfield  {title} {\bibinfo {title} {{Identifying a first-order phase
  transition in neutron star mergers through gravitational waves}},\ }\href
  {https://doi.org/10.1103/PhysRevLett.122.061102} {\bibfield  {journal}
  {\bibinfo  {journal} {Phys. Rev. Lett.}\ }\textbf {\bibinfo {volume} {122}},\
  \bibinfo {pages} {061102} (\bibinfo {year} {2019})},\ \Eprint
  {https://arxiv.org/abs/1809.01116} {arXiv:1809.01116 [astro-ph.HE]}
  \BibitemShut {NoStop}%
\bibitem [{\citenamefont {Weih}\ \emph {et~al.}(2020)\citenamefont {Weih},
  \citenamefont {Hanauske},\ and\ \citenamefont {Rezzolla}}]{Weih:2019xvw}%
  \BibitemOpen
  \bibfield  {author} {\bibinfo {author} {\bibfnamefont {L.~R.}\ \bibnamefont
  {Weih}}, \bibinfo {author} {\bibfnamefont {M.}~\bibnamefont {Hanauske}},\
  and\ \bibinfo {author} {\bibfnamefont {L.}~\bibnamefont {Rezzolla}},\
  }\bibfield  {title} {\bibinfo {title} {{Postmerger Gravitational-Wave
  Signatures of Phase Transitions in Binary Mergers}},\ }\href
  {https://doi.org/10.1103/PhysRevLett.124.171103} {\bibfield  {journal}
  {\bibinfo  {journal} {Phys. Rev. Lett.}\ }\textbf {\bibinfo {volume} {124}},\
  \bibinfo {pages} {171103} (\bibinfo {year} {2020})},\ \Eprint
  {https://arxiv.org/abs/1912.09340} {arXiv:1912.09340 [gr-qc]} \BibitemShut
  {NoStop}%
\bibitem [{\citenamefont {Bauswein}\ and\ \citenamefont
  {Blacker}(2020)}]{Bauswein:2020ggy}%
  \BibitemOpen
  \bibfield  {author} {\bibinfo {author} {\bibfnamefont {A.}~\bibnamefont
  {Bauswein}}\ and\ \bibinfo {author} {\bibfnamefont {S.}~\bibnamefont
  {Blacker}},\ }\bibfield  {title} {\bibinfo {title} {{Impact of quark
  deconfinement in neutron star mergers and hybrid star mergers}},\ }\href
  {https://doi.org/10.1140/epjst/e2020-000138-7} {\bibfield  {journal}
  {\bibinfo  {journal} {Eur. Phys. J. ST}\ }\textbf {\bibinfo {volume} {229}},\
  \bibinfo {pages} {3595} (\bibinfo {year} {2020})},\ \Eprint
  {https://arxiv.org/abs/2006.16183} {arXiv:2006.16183 [astro-ph.HE]}
  \BibitemShut {NoStop}%
\bibitem [{\citenamefont {Tootle}\ \emph {et~al.}(2022)\citenamefont {Tootle},
  \citenamefont {Ecker}, \citenamefont {Topolski}, \citenamefont {Demircik},
  \citenamefont {J\"arvinen},\ and\ \citenamefont {Rezzolla}}]{Tootle:2022pvd}%
  \BibitemOpen
  \bibfield  {author} {\bibinfo {author} {\bibfnamefont {S.}~\bibnamefont
  {Tootle}}, \bibinfo {author} {\bibfnamefont {C.}~\bibnamefont {Ecker}},
  \bibinfo {author} {\bibfnamefont {K.}~\bibnamefont {Topolski}}, \bibinfo
  {author} {\bibfnamefont {T.}~\bibnamefont {Demircik}}, \bibinfo {author}
  {\bibfnamefont {M.}~\bibnamefont {J\"arvinen}},\ and\ \bibinfo {author}
  {\bibfnamefont {L.}~\bibnamefont {Rezzolla}},\ }\bibfield  {title} {\bibinfo
  {title} {{Quark formation and phenomenology in binary neutron-star mergers
  using V-QCD}},\ }\href {https://doi.org/10.21468/SciPostPhys.13.5.109}
  {\bibfield  {journal} {\bibinfo  {journal} {SciPost Phys.}\ }\textbf
  {\bibinfo {volume} {13}},\ \bibinfo {pages} {109} (\bibinfo {year} {2022})},\
  \Eprint {https://arxiv.org/abs/2205.05691} {arXiv:2205.05691 [astro-ph.HE]}
  \BibitemShut {NoStop}%
\bibitem [{\citenamefont {Huang}\ \emph {et~al.}(2022)\citenamefont {Huang},
  \citenamefont {Baiotti}, \citenamefont {Kojo}, \citenamefont {Takami},
  \citenamefont {Sotani}, \citenamefont {Togashi}, \citenamefont {Hatsuda},
  \citenamefont {Nagataki},\ and\ \citenamefont {Fan}}]{Huang:2022mqp}%
  \BibitemOpen
  \bibfield  {author} {\bibinfo {author} {\bibfnamefont {Y.-J.}\ \bibnamefont
  {Huang}}, \bibinfo {author} {\bibfnamefont {L.}~\bibnamefont {Baiotti}},
  \bibinfo {author} {\bibfnamefont {T.}~\bibnamefont {Kojo}}, \bibinfo {author}
  {\bibfnamefont {K.}~\bibnamefont {Takami}}, \bibinfo {author} {\bibfnamefont
  {H.}~\bibnamefont {Sotani}}, \bibinfo {author} {\bibfnamefont
  {H.}~\bibnamefont {Togashi}}, \bibinfo {author} {\bibfnamefont
  {T.}~\bibnamefont {Hatsuda}}, \bibinfo {author} {\bibfnamefont
  {S.}~\bibnamefont {Nagataki}},\ and\ \bibinfo {author} {\bibfnamefont
  {Y.-Z.}\ \bibnamefont {Fan}},\ }\bibfield  {title} {\bibinfo {title} {{Merger
  and Postmerger of Binary Neutron Stars with a Quark-Hadron Crossover Equation
  of State}},\ }\href {https://doi.org/10.1103/PhysRevLett.129.181101}
  {\bibfield  {journal} {\bibinfo  {journal} {Phys. Rev. Lett.}\ }\textbf
  {\bibinfo {volume} {129}},\ \bibinfo {pages} {181101} (\bibinfo {year}
  {2022})},\ \Eprint {https://arxiv.org/abs/2203.04528} {arXiv:2203.04528
  [astro-ph.HE]} \BibitemShut {NoStop}%
\bibitem [{\citenamefont {Blacker}\ \emph {et~al.}(2023)\citenamefont
  {Blacker}, \citenamefont {Bauswein},\ and\ \citenamefont
  {Typel}}]{Blacker:2023afl}%
  \BibitemOpen
  \bibfield  {author} {\bibinfo {author} {\bibfnamefont {S.}~\bibnamefont
  {Blacker}}, \bibinfo {author} {\bibfnamefont {A.}~\bibnamefont {Bauswein}},\
  and\ \bibinfo {author} {\bibfnamefont {S.}~\bibnamefont {Typel}},\ }\bibfield
   {title} {\bibinfo {title} {{Exploring thermal effects of the hadron-quark
  matter transition in neutron star mergers}},\ }\href
  {https://doi.org/10.1103/PhysRevD.108.063032} {\bibfield  {journal} {\bibinfo
   {journal} {Phys. Rev. D}\ }\textbf {\bibinfo {volume} {108}},\ \bibinfo
  {pages} {063032} (\bibinfo {year} {2023})},\ \Eprint
  {https://arxiv.org/abs/2304.01971} {arXiv:2304.01971 [astro-ph.HE]}
  \BibitemShut {NoStop}%
\bibitem [{\citenamefont {Haque}\ \emph {et~al.}(2024)\citenamefont {Haque},
  \citenamefont {Mallick},\ and\ \citenamefont {Thakur}}]{Haque:2022dsc}%
  \BibitemOpen
  \bibfield  {author} {\bibinfo {author} {\bibfnamefont {S.}~\bibnamefont
  {Haque}}, \bibinfo {author} {\bibfnamefont {R.}~\bibnamefont {Mallick}},\
  and\ \bibinfo {author} {\bibfnamefont {S.~K.}\ \bibnamefont {Thakur}},\
  }\bibfield  {title} {\bibinfo {title} {{Effects of Onset of Phase Transition
  on Binary Neutron Star Mergers}},\ }\href
  {https://doi.org/10.1093/mnras/stad3839} {\bibfield  {journal} {\bibinfo
  {journal} {Mon. Not. Roy. Astron. Soc.}\ }\textbf {\bibinfo {volume} {527}},\
  \bibinfo {pages} {11575} (\bibinfo {year} {2024})},\ \Eprint
  {https://arxiv.org/abs/2207.14485} {arXiv:2207.14485 [astro-ph.HE]}
  \BibitemShut {NoStop}%
\bibitem [{\citenamefont {Prakash}\ \emph {et~al.}(2021)\citenamefont
  {Prakash}, \citenamefont {Radice}, \citenamefont {Logoteta}, \citenamefont
  {Perego}, \citenamefont {Nedora}, \citenamefont {Bombaci}, \citenamefont
  {Kashyap}, \citenamefont {Bernuzzi},\ and\ \citenamefont
  {Endrizzi}}]{Prakash:2021wpz}%
  \BibitemOpen
  \bibfield  {author} {\bibinfo {author} {\bibfnamefont {A.}~\bibnamefont
  {Prakash}}, \bibinfo {author} {\bibfnamefont {D.}~\bibnamefont {Radice}},
  \bibinfo {author} {\bibfnamefont {D.}~\bibnamefont {Logoteta}}, \bibinfo
  {author} {\bibfnamefont {A.}~\bibnamefont {Perego}}, \bibinfo {author}
  {\bibfnamefont {V.}~\bibnamefont {Nedora}}, \bibinfo {author} {\bibfnamefont
  {I.}~\bibnamefont {Bombaci}}, \bibinfo {author} {\bibfnamefont
  {R.}~\bibnamefont {Kashyap}}, \bibinfo {author} {\bibfnamefont
  {S.}~\bibnamefont {Bernuzzi}},\ and\ \bibinfo {author} {\bibfnamefont
  {A.}~\bibnamefont {Endrizzi}},\ }\bibfield  {title} {\bibinfo {title}
  {{Signatures of deconfined quark phases in binary neutron star mergers}},\
  }\href {https://doi.org/10.1103/PhysRevD.104.083029} {\bibfield  {journal}
  {\bibinfo  {journal} {Phys. Rev. D}\ }\textbf {\bibinfo {volume} {104}},\
  \bibinfo {pages} {083029} (\bibinfo {year} {2021})},\ \Eprint
  {https://arxiv.org/abs/2106.07885} {arXiv:2106.07885 [astro-ph.HE]}
  \BibitemShut {NoStop}%
\bibitem [{\citenamefont {J\"arvinen}\ and\ \citenamefont
  {Kiritsis}(2012)}]{Jarvinen:2011qe}%
  \BibitemOpen
  \bibfield  {author} {\bibinfo {author} {\bibfnamefont {M.}~\bibnamefont
  {J\"arvinen}}\ and\ \bibinfo {author} {\bibfnamefont {E.}~\bibnamefont
  {Kiritsis}},\ }\bibfield  {title} {\bibinfo {title} {{Holographic Models for
  QCD in the Veneziano Limit}},\ }\href
  {https://doi.org/10.1007/JHEP03(2012)002} {\bibfield  {journal} {\bibinfo
  {journal} {JHEP}\ }\textbf {\bibinfo {volume} {03}},\ \bibinfo {pages}
  {002}},\ \Eprint {https://arxiv.org/abs/1112.1261} {arXiv:1112.1261 [hep-ph]}
  \BibitemShut {NoStop}%
\bibitem [{\citenamefont {Jokela}\ \emph {et~al.}(2019)\citenamefont {Jokela},
  \citenamefont {J\"arvinen},\ and\ \citenamefont {Remes}}]{Jokela:2018ers}%
  \BibitemOpen
  \bibfield  {author} {\bibinfo {author} {\bibfnamefont {N.}~\bibnamefont
  {Jokela}}, \bibinfo {author} {\bibfnamefont {M.}~\bibnamefont {J\"arvinen}},\
  and\ \bibinfo {author} {\bibfnamefont {J.}~\bibnamefont {Remes}},\ }\bibfield
   {title} {\bibinfo {title} {{Holographic QCD in the Veneziano limit and
  neutron stars}},\ }\href {https://doi.org/10.1007/JHEP03(2019)041} {\bibfield
   {journal} {\bibinfo  {journal} {JHEP}\ }\textbf {\bibinfo {volume} {03}},\
  \bibinfo {pages} {041}},\ \Eprint {https://arxiv.org/abs/1809.07770}
  {arXiv:1809.07770 [hep-ph]} \BibitemShut {NoStop}%
\bibitem [{\citenamefont {Jokela}\ \emph {et~al.}(2021)\citenamefont {Jokela},
  \citenamefont {J\"arvinen}, \citenamefont {Nijs},\ and\ \citenamefont
  {Remes}}]{Jokela:2020piw}%
  \BibitemOpen
  \bibfield  {author} {\bibinfo {author} {\bibfnamefont {N.}~\bibnamefont
  {Jokela}}, \bibinfo {author} {\bibfnamefont {M.}~\bibnamefont {J\"arvinen}},
  \bibinfo {author} {\bibfnamefont {G.}~\bibnamefont {Nijs}},\ and\ \bibinfo
  {author} {\bibfnamefont {J.}~\bibnamefont {Remes}},\ }\bibfield  {title}
  {\bibinfo {title} {{Unified weak and strong coupling framework for nuclear
  matter and neutron stars}},\ }\href
  {https://doi.org/10.1103/PhysRevD.103.086004} {\bibfield  {journal} {\bibinfo
   {journal} {Phys. Rev. D}\ }\textbf {\bibinfo {volume} {103}},\ \bibinfo
  {pages} {086004} (\bibinfo {year} {2021})},\ \Eprint
  {https://arxiv.org/abs/2006.01141} {arXiv:2006.01141 [hep-ph]} \BibitemShut
  {NoStop}%
\bibitem [{\citenamefont {Alho}\ \emph {et~al.}(2013)\citenamefont {Alho},
  \citenamefont {J\"arvinen}, \citenamefont {Kajantie}, \citenamefont
  {Kiritsis},\ and\ \citenamefont {Tuominen}}]{Alho:2012mh}%
  \BibitemOpen
  \bibfield  {author} {\bibinfo {author} {\bibfnamefont {T.}~\bibnamefont
  {Alho}}, \bibinfo {author} {\bibfnamefont {M.}~\bibnamefont {J\"arvinen}},
  \bibinfo {author} {\bibfnamefont {K.}~\bibnamefont {Kajantie}}, \bibinfo
  {author} {\bibfnamefont {E.}~\bibnamefont {Kiritsis}},\ and\ \bibinfo
  {author} {\bibfnamefont {K.}~\bibnamefont {Tuominen}},\ }\bibfield  {title}
  {\bibinfo {title} {{On finite-temperature holographic QCD in the Veneziano
  limit}},\ }\href {https://doi.org/10.1007/JHEP01(2013)093} {\bibfield
  {journal} {\bibinfo  {journal} {JHEP}\ }\textbf {\bibinfo {volume} {01}},\
  \bibinfo {pages} {093}},\ \Eprint {https://arxiv.org/abs/1210.4516}
  {arXiv:1210.4516 [hep-ph]} \BibitemShut {NoStop}%
\bibitem [{\citenamefont {Alho}\ \emph {et~al.}(2015)\citenamefont {Alho},
  \citenamefont {J\"arvinen}, \citenamefont {Kajantie}, \citenamefont
  {Kiritsis},\ and\ \citenamefont {Tuominen}}]{Alho:2015zua}%
  \BibitemOpen
  \bibfield  {author} {\bibinfo {author} {\bibfnamefont {T.}~\bibnamefont
  {Alho}}, \bibinfo {author} {\bibfnamefont {M.}~\bibnamefont {J\"arvinen}},
  \bibinfo {author} {\bibfnamefont {K.}~\bibnamefont {Kajantie}}, \bibinfo
  {author} {\bibfnamefont {E.}~\bibnamefont {Kiritsis}},\ and\ \bibinfo
  {author} {\bibfnamefont {K.}~\bibnamefont {Tuominen}},\ }\bibfield  {title}
  {\bibinfo {title} {{Quantum and stringy corrections to the equation of state
  of holographic QCD matter and the nature of the chiral transition}},\ }\href
  {https://doi.org/10.1103/PhysRevD.91.055017} {\bibfield  {journal} {\bibinfo
  {journal} {Phys. Rev. D}\ }\textbf {\bibinfo {volume} {91}},\ \bibinfo
  {pages} {055017} (\bibinfo {year} {2015})},\ \Eprint
  {https://arxiv.org/abs/1501.06379} {arXiv:1501.06379 [hep-ph]} \BibitemShut
  {NoStop}%
\bibitem [{\citenamefont {Ishii}\ \emph {et~al.}(2019)\citenamefont {Ishii},
  \citenamefont {J\"arvinen},\ and\ \citenamefont {Nijs}}]{Ishii:2019gta}%
  \BibitemOpen
  \bibfield  {author} {\bibinfo {author} {\bibfnamefont {T.}~\bibnamefont
  {Ishii}}, \bibinfo {author} {\bibfnamefont {M.}~\bibnamefont {J\"arvinen}},\
  and\ \bibinfo {author} {\bibfnamefont {G.}~\bibnamefont {Nijs}},\ }\bibfield
  {title} {\bibinfo {title} {{Cool baryon and quark matter in holographic
  QCD}},\ }\href {https://doi.org/10.1007/JHEP07(2019)003} {\bibfield
  {journal} {\bibinfo  {journal} {JHEP}\ }\textbf {\bibinfo {volume} {07}},\
  \bibinfo {pages} {003}},\ \Eprint {https://arxiv.org/abs/1903.06169}
  {arXiv:1903.06169 [hep-ph]} \BibitemShut {NoStop}%
\bibitem [{\citenamefont {Rozali}\ \emph {et~al.}(2008)\citenamefont {Rozali},
  \citenamefont {Shieh}, \citenamefont {Van~Raamsdonk},\ and\ \citenamefont
  {Wu}}]{Rozali:2007rx}%
  \BibitemOpen
  \bibfield  {author} {\bibinfo {author} {\bibfnamefont {M.}~\bibnamefont
  {Rozali}}, \bibinfo {author} {\bibfnamefont {H.-H.}\ \bibnamefont {Shieh}},
  \bibinfo {author} {\bibfnamefont {M.}~\bibnamefont {Van~Raamsdonk}},\ and\
  \bibinfo {author} {\bibfnamefont {J.}~\bibnamefont {Wu}},\ }\bibfield
  {title} {\bibinfo {title} {{Cold Nuclear Matter In Holographic QCD}},\ }\href
  {https://doi.org/10.1088/1126-6708/2008/01/053} {\bibfield  {journal}
  {\bibinfo  {journal} {JHEP}\ }\textbf {\bibinfo {volume} {01}},\ \bibinfo
  {pages} {053}},\ \Eprint {https://arxiv.org/abs/0708.1322} {arXiv:0708.1322
  [hep-th]} \BibitemShut {NoStop}%
\bibitem [{\citenamefont {Ecker}\ \emph {et~al.}(2020)\citenamefont {Ecker},
  \citenamefont {J\"arvinen}, \citenamefont {Nijs},\ and\ \citenamefont
  {van~der Schee}}]{Ecker:2019xrw}%
  \BibitemOpen
  \bibfield  {author} {\bibinfo {author} {\bibfnamefont {C.}~\bibnamefont
  {Ecker}}, \bibinfo {author} {\bibfnamefont {M.}~\bibnamefont {J\"arvinen}},
  \bibinfo {author} {\bibfnamefont {G.}~\bibnamefont {Nijs}},\ and\ \bibinfo
  {author} {\bibfnamefont {W.}~\bibnamefont {van~der Schee}},\ }\bibfield
  {title} {\bibinfo {title} {{Gravitational waves from holographic neutron star
  mergers}},\ }\href {https://doi.org/10.1103/PhysRevD.101.103006} {\bibfield
  {journal} {\bibinfo  {journal} {Phys. Rev. D}\ }\textbf {\bibinfo {volume}
  {101}},\ \bibinfo {pages} {103006} (\bibinfo {year} {2020})},\ \Eprint
  {https://arxiv.org/abs/1908.03213} {arXiv:1908.03213 [astro-ph.HE]}
  \BibitemShut {NoStop}%
\bibitem [{\citenamefont {Jokela}\ \emph {et~al.}(2022)\citenamefont {Jokela},
  \citenamefont {J\"arvinen},\ and\ \citenamefont {Remes}}]{Jokela:2021vwy}%
  \BibitemOpen
  \bibfield  {author} {\bibinfo {author} {\bibfnamefont {N.}~\bibnamefont
  {Jokela}}, \bibinfo {author} {\bibfnamefont {M.}~\bibnamefont {J\"arvinen}},\
  and\ \bibinfo {author} {\bibfnamefont {J.}~\bibnamefont {Remes}},\ }\bibfield
   {title} {\bibinfo {title} {{Holographic QCD in the NICER era}},\ }\href
  {https://doi.org/10.1103/PhysRevD.105.086005} {\bibfield  {journal} {\bibinfo
   {journal} {Phys. Rev. D}\ }\textbf {\bibinfo {volume} {105}},\ \bibinfo
  {pages} {086005} (\bibinfo {year} {2022})},\ \Eprint
  {https://arxiv.org/abs/2111.12101} {arXiv:2111.12101 [hep-ph]} \BibitemShut
  {NoStop}%
\bibitem [{\citenamefont {Cherubini}\ \emph {et~al.}(2002)\citenamefont
  {Cherubini}, \citenamefont {Bini}, \citenamefont {Capozziello},\ and\
  \citenamefont {Ruffini}}]{Cherubini:2002gen}%
  \BibitemOpen
  \bibfield  {author} {\bibinfo {author} {\bibfnamefont {C.}~\bibnamefont
  {Cherubini}}, \bibinfo {author} {\bibfnamefont {D.}~\bibnamefont {Bini}},
  \bibinfo {author} {\bibfnamefont {S.}~\bibnamefont {Capozziello}},\ and\
  \bibinfo {author} {\bibfnamefont {R.}~\bibnamefont {Ruffini}},\ }\bibfield
  {title} {\bibinfo {title} {{Second order scalar invariants of the Riemann
  tensor: Applications to black hole space-times}},\ }\href
  {https://doi.org/10.1142/S0218271802002037} {\bibfield  {journal} {\bibinfo
  {journal} {Int. J. Mod. Phys. D}\ }\textbf {\bibinfo {volume} {11}},\
  \bibinfo {pages} {827} (\bibinfo {year} {2002})},\ \Eprint
  {https://arxiv.org/abs/gr-qc/0302095} {arXiv:gr-qc/0302095} \BibitemShut
  {NoStop}%
\bibitem [{Note1()}]{Note1}%
  \BibitemOpen
  \bibinfo {note} {Note, that it is possible to define $K_2$ using the right
  dual instead of the left dual, but this difference is irrelevant for our
  purposes.}\BibitemShut {Stop}%
\bibitem [{Note2()}]{Note2}%
  \BibitemOpen
  \bibinfo {note} {We expect $M_{\protect \rm crit}$ to also depend in a
  non-trivial way on the mass ratio, but leave an investigation of this
  dependence for future work.}\BibitemShut {Stop}%
\bibitem [{Note3()}]{Note3}%
  \BibitemOpen
  \bibinfo {note} {Strictly speaking, the criterion is not entirely covariant
  because of the coordinate time derivatives, but can probably be phrased in an
  entirely covariant way by replacing the time coordinate $t$ by the parameter
  of any non-spacelike geodesic that passes through the point $p$ where
  $K_1(p)={\protect \rm max}(K_1(x))$}\BibitemShut {NoStop}%
\bibitem [{\citenamefont {Loffler}\ \emph {et~al.}(2012)\citenamefont {Loffler}
  \emph {et~al.}}]{Loffler:2011ay}%
  \BibitemOpen
  \bibfield  {author} {\bibinfo {author} {\bibfnamefont {F.}~\bibnamefont
  {Loffler}} \emph {et~al.},\ }\bibfield  {title} {\bibinfo {title} {{The
  Einstein Toolkit: A Community Computational Infrastructure for Relativistic
  Astrophysics}},\ }\href {https://doi.org/10.1088/0264-9381/29/11/115001}
  {\bibfield  {journal} {\bibinfo  {journal} {Class. Quant. Grav.}\ }\textbf
  {\bibinfo {volume} {29}},\ \bibinfo {pages} {115001} (\bibinfo {year}
  {2012})},\ \Eprint {https://arxiv.org/abs/1111.3344} {arXiv:1111.3344
  [gr-qc]} \BibitemShut {NoStop}%
\bibitem [{\citenamefont {Inc.}()}]{Mathematica}%
  \BibitemOpen
  \bibfield  {author} {\bibinfo {author} {\bibfnamefont {W.~R.}\ \bibnamefont
  {Inc.}},\ }\href {https://www.wolfram.com/mathematica} {\bibinfo {title}
  {Mathematica, {V}ersion 14.0}},\ \bibinfo {note} {champaign, IL,
  2024}\BibitemShut {NoStop}%
\bibitem [{\citenamefont {Husa}\ \emph {et~al.}(2006)\citenamefont {Husa},
  \citenamefont {Hinder},\ and\ \citenamefont {Lechner}}]{Husa:2004ip}%
  \BibitemOpen
  \bibfield  {author} {\bibinfo {author} {\bibfnamefont {S.}~\bibnamefont
  {Husa}}, \bibinfo {author} {\bibfnamefont {I.}~\bibnamefont {Hinder}},\ and\
  \bibinfo {author} {\bibfnamefont {C.}~\bibnamefont {Lechner}},\ }\bibfield
  {title} {\bibinfo {title} {{Kranc: A Mathematica application to generate
  numerical codes for tensorial evolution equations}},\ }\href
  {https://doi.org/10.1016/j.cpc.2006.02.002} {\bibfield  {journal} {\bibinfo
  {journal} {Comput. Phys. Commun.}\ }\textbf {\bibinfo {volume} {174}},\
  \bibinfo {pages} {983} (\bibinfo {year} {2006})},\ \Eprint
  {https://arxiv.org/abs/gr-qc/0404023} {arXiv:gr-qc/0404023} \BibitemShut
  {NoStop}%
\bibitem [{\citenamefont {Papenfort}\ \emph {et~al.}(2021)\citenamefont
  {Papenfort}, \citenamefont {Tootle}, \citenamefont {Grandcl\'ement},
  \citenamefont {Most},\ and\ \citenamefont {Rezzolla}}]{Papenfort:2021hod}%
  \BibitemOpen
  \bibfield  {author} {\bibinfo {author} {\bibfnamefont {L.~J.}\ \bibnamefont
  {Papenfort}}, \bibinfo {author} {\bibfnamefont {S.~D.}\ \bibnamefont
  {Tootle}}, \bibinfo {author} {\bibfnamefont {P.}~\bibnamefont
  {Grandcl\'ement}}, \bibinfo {author} {\bibfnamefont {E.~R.}\ \bibnamefont
  {Most}},\ and\ \bibinfo {author} {\bibfnamefont {L.}~\bibnamefont
  {Rezzolla}},\ }\bibfield  {title} {\bibinfo {title} {{New public code for
  initial data of unequal-mass, spinning compact-object binaries}},\ }\href
  {https://doi.org/10.1103/PhysRevD.104.024057} {\bibfield  {journal} {\bibinfo
   {journal} {Phys. Rev. D}\ }\textbf {\bibinfo {volume} {104}},\ \bibinfo
  {pages} {024057} (\bibinfo {year} {2021})},\ \Eprint
  {https://arxiv.org/abs/2103.09911} {arXiv:2103.09911 [gr-qc]} \BibitemShut
  {NoStop}%
\bibitem [{\citenamefont {Grandclement}(2010)}]{Grandclement:2009ju}%
  \BibitemOpen
  \bibfield  {author} {\bibinfo {author} {\bibfnamefont {P.}~\bibnamefont
  {Grandclement}},\ }\bibfield  {title} {\bibinfo {title} {{Kadath: A Spectral
  solver for theoretical physics}},\ }\href
  {https://doi.org/10.1016/j.jcp.2010.01.005} {\bibfield  {journal} {\bibinfo
  {journal} {J. Comput. Phys.}\ }\textbf {\bibinfo {volume} {229}},\ \bibinfo
  {pages} {3334} (\bibinfo {year} {2010})},\ \Eprint
  {https://arxiv.org/abs/0909.1228} {arXiv:0909.1228 [gr-qc]} \BibitemShut
  {NoStop}%
\bibitem [{\citenamefont {Schnetter}\ \emph {et~al.}(2004)\citenamefont
  {Schnetter}, \citenamefont {Hawley},\ and\ \citenamefont
  {Hawke}}]{Schnetter:2003rb}%
  \BibitemOpen
  \bibfield  {author} {\bibinfo {author} {\bibfnamefont {E.}~\bibnamefont
  {Schnetter}}, \bibinfo {author} {\bibfnamefont {S.~H.}\ \bibnamefont
  {Hawley}},\ and\ \bibinfo {author} {\bibfnamefont {I.}~\bibnamefont
  {Hawke}},\ }\bibfield  {title} {\bibinfo {title} {{Evolutions in 3-D
  numerical relativity using fixed mesh refinement}},\ }\href
  {https://doi.org/10.1088/0264-9381/21/6/014} {\bibfield  {journal} {\bibinfo
  {journal} {Class. Quant. Grav.}\ }\textbf {\bibinfo {volume} {21}},\ \bibinfo
  {pages} {1465} (\bibinfo {year} {2004})},\ \Eprint
  {https://arxiv.org/abs/gr-qc/0310042} {arXiv:gr-qc/0310042} \BibitemShut
  {NoStop}%
\bibitem [{\citenamefont {Bernuzzi}\ and\ \citenamefont
  {Hilditch}(2010)}]{Bernuzzi:2009ex}%
  \BibitemOpen
  \bibfield  {author} {\bibinfo {author} {\bibfnamefont {S.}~\bibnamefont
  {Bernuzzi}}\ and\ \bibinfo {author} {\bibfnamefont {D.}~\bibnamefont
  {Hilditch}},\ }\bibfield  {title} {\bibinfo {title} {{Constraint violation in
  free evolution schemes: Comparing BSSNOK with a conformal decomposition of
  Z4}},\ }\href {https://doi.org/10.1103/PhysRevD.81.084003} {\bibfield
  {journal} {\bibinfo  {journal} {Phys. Rev. D}\ }\textbf {\bibinfo {volume}
  {81}},\ \bibinfo {pages} {084003} (\bibinfo {year} {2010})},\ \Eprint
  {https://arxiv.org/abs/0912.2920} {arXiv:0912.2920 [gr-qc]} \BibitemShut
  {NoStop}%
\bibitem [{\citenamefont {Alic}\ \emph {et~al.}(2012)\citenamefont {Alic},
  \citenamefont {Bona-Casas}, \citenamefont {Bona}, \citenamefont {Rezzolla},\
  and\ \citenamefont {Palenzuela}}]{Alic:2011gg}%
  \BibitemOpen
  \bibfield  {author} {\bibinfo {author} {\bibfnamefont {D.}~\bibnamefont
  {Alic}}, \bibinfo {author} {\bibfnamefont {C.}~\bibnamefont {Bona-Casas}},
  \bibinfo {author} {\bibfnamefont {C.}~\bibnamefont {Bona}}, \bibinfo {author}
  {\bibfnamefont {L.}~\bibnamefont {Rezzolla}},\ and\ \bibinfo {author}
  {\bibfnamefont {C.}~\bibnamefont {Palenzuela}},\ }\bibfield  {title}
  {\bibinfo {title} {{Conformal and covariant formulation of the Z4 system with
  constraint-violation damping}},\ }\href
  {https://doi.org/10.1103/PhysRevD.85.064040} {\bibfield  {journal} {\bibinfo
  {journal} {Phys. Rev. D}\ }\textbf {\bibinfo {volume} {85}},\ \bibinfo
  {pages} {064040} (\bibinfo {year} {2012})},\ \Eprint
  {https://arxiv.org/abs/1106.2254} {arXiv:1106.2254 [gr-qc]} \BibitemShut
  {NoStop}%
\bibitem [{\citenamefont {Most}\ \emph
  {et~al.}(2019{\natexlab{b}})\citenamefont {Most}, \citenamefont {Papenfort},\
  and\ \citenamefont {Rezzolla}}]{Most:2019kfe}%
  \BibitemOpen
  \bibfield  {author} {\bibinfo {author} {\bibfnamefont {E.~R.}\ \bibnamefont
  {Most}}, \bibinfo {author} {\bibfnamefont {L.~J.}\ \bibnamefont
  {Papenfort}},\ and\ \bibinfo {author} {\bibfnamefont {L.}~\bibnamefont
  {Rezzolla}},\ }\bibfield  {title} {\bibinfo {title} {{Beyond second-order
  convergence in simulations of magnetized binary neutron stars with realistic
  microphysics}},\ }\href {https://doi.org/10.1093/mnras/stz2809} {\bibfield
  {journal} {\bibinfo  {journal} {Mon. Not. Roy. Astron. Soc.}\ }\textbf
  {\bibinfo {volume} {490}},\ \bibinfo {pages} {3588} (\bibinfo {year}
  {2019}{\natexlab{b}})},\ \Eprint {https://arxiv.org/abs/1907.10328}
  {arXiv:1907.10328 [astro-ph.HE]} \BibitemShut {NoStop}%
\bibitem [{\citenamefont {Etienne}\ \emph {et~al.}(2015)\citenamefont
  {Etienne}, \citenamefont {Paschalidis}, \citenamefont {Haas}, \citenamefont
  {M\"osta},\ and\ \citenamefont {Shapiro}}]{Etienne:2015cea}%
  \BibitemOpen
  \bibfield  {author} {\bibinfo {author} {\bibfnamefont {Z.~B.}\ \bibnamefont
  {Etienne}}, \bibinfo {author} {\bibfnamefont {V.}~\bibnamefont
  {Paschalidis}}, \bibinfo {author} {\bibfnamefont {R.}~\bibnamefont {Haas}},
  \bibinfo {author} {\bibfnamefont {P.}~\bibnamefont {M\"osta}},\ and\ \bibinfo
  {author} {\bibfnamefont {S.~L.}\ \bibnamefont {Shapiro}},\ }\bibfield
  {title} {\bibinfo {title} {{IllinoisGRMHD: An Open-Source, User-Friendly
  GRMHD Code for Dynamical Spacetimes}},\ }\href
  {https://doi.org/10.1088/0264-9381/32/17/175009} {\bibfield  {journal}
  {\bibinfo  {journal} {Class. Quant. Grav.}\ }\textbf {\bibinfo {volume}
  {32}},\ \bibinfo {pages} {175009} (\bibinfo {year} {2015})},\ \Eprint
  {https://arxiv.org/abs/1501.07276} {arXiv:1501.07276 [astro-ph.HE]}
  \BibitemShut {NoStop}%
\bibitem [{\citenamefont {Kasen}\ \emph {et~al.}(2017)\citenamefont {Kasen},
  \citenamefont {Metzger}, \citenamefont {Barnes}, \citenamefont {Quataert},\
  and\ \citenamefont {Ramirez-Ruiz}}]{Kasen:2017sxr}%
  \BibitemOpen
  \bibfield  {author} {\bibinfo {author} {\bibfnamefont {D.}~\bibnamefont
  {Kasen}}, \bibinfo {author} {\bibfnamefont {B.}~\bibnamefont {Metzger}},
  \bibinfo {author} {\bibfnamefont {J.}~\bibnamefont {Barnes}}, \bibinfo
  {author} {\bibfnamefont {E.}~\bibnamefont {Quataert}},\ and\ \bibinfo
  {author} {\bibfnamefont {E.}~\bibnamefont {Ramirez-Ruiz}},\ }\bibfield
  {title} {\bibinfo {title} {{Origin of the heavy elements in binary
  neutron-star mergers from a gravitational wave event}},\ }\href
  {https://doi.org/10.1038/nature24453} {\bibfield  {journal} {\bibinfo
  {journal} {Nature}\ }\textbf {\bibinfo {volume} {551}},\ \bibinfo {pages}
  {80} (\bibinfo {year} {2017})},\ \Eprint {https://arxiv.org/abs/1710.05463}
  {arXiv:1710.05463 [astro-ph.HE]} \BibitemShut {NoStop}%
\bibitem [{\citenamefont {Abbott}\ \emph
  {et~al.}(2017{\natexlab{b}})\citenamefont {Abbott} \emph
  {et~al.}}]{LIGOScientific:2017ync}%
  \BibitemOpen
  \bibfield  {author} {\bibinfo {author} {\bibfnamefont {B.~P.}\ \bibnamefont
  {Abbott}} \emph {et~al.} (\bibinfo {collaboration} {LIGO Scientific, Virgo,
  Fermi GBM, INTEGRAL, IceCube, AstroSat Cadmium Zinc Telluride Imager Team,
  IPN, Insight-Hxmt, ANTARES, Swift, AGILE Team, 1M2H Team, Dark Energy Camera
  GW-EM, DES, DLT40, GRAWITA, Fermi-LAT, ATCA, ASKAP, Las Cumbres Observatory
  Group, OzGrav, DWF (Deeper Wider Faster Program), AST3, CAASTRO, VINROUGE,
  MASTER, J-GEM, GROWTH, JAGWAR, CaltechNRAO, TTU-NRAO, NuSTAR, Pan-STARRS,
  MAXI Team, TZAC Consortium, KU, Nordic Optical Telescope, ePESSTO, GROND,
  Texas Tech University, SALT Group, TOROS, BOOTES, MWA, CALET, IKI-GW
  Follow-up, H.E.S.S., LOFAR, LWA, HAWC, Pierre Auger, ALMA, Euro VLBI Team, Pi
  of Sky, Chandra Team at McGill University, DFN, ATLAS Telescopes, High Time
  Resolution Universe Survey, RIMAS, RATIR, SKA South Africa/MeerKAT}),\
  }\bibfield  {title} {\bibinfo {title} {{Multi-messenger Observations of a
  Binary Neutron Star Merger}},\ }\href
  {https://doi.org/10.3847/2041-8213/aa91c9} {\bibfield  {journal} {\bibinfo
  {journal} {Astrophys. J. Lett.}\ }\textbf {\bibinfo {volume} {848}},\
  \bibinfo {pages} {L12} (\bibinfo {year} {2017}{\natexlab{b}})},\ \Eprint
  {https://arxiv.org/abs/1710.05833} {arXiv:1710.05833 [astro-ph.HE]}
  \BibitemShut {NoStop}%
\bibitem [{\citenamefont {Ghirlanda}\ \emph {et~al.}(2019)\citenamefont
  {Ghirlanda} \emph {et~al.}}]{Ghirlanda:2018uyx}%
  \BibitemOpen
  \bibfield  {author} {\bibinfo {author} {\bibfnamefont {G.}~\bibnamefont
  {Ghirlanda}} \emph {et~al.},\ }\bibfield  {title} {\bibinfo {title} {{Compact
  radio emission indicates a structured jet was produced by a binary neutron
  star merger}},\ }\href {https://doi.org/10.1126/science.aau8815} {\bibfield
  {journal} {\bibinfo  {journal} {Science}\ }\textbf {\bibinfo {volume}
  {363}},\ \bibinfo {pages} {968} (\bibinfo {year} {2019})},\ \Eprint
  {https://arxiv.org/abs/1808.00469} {arXiv:1808.00469 [astro-ph.HE]}
  \BibitemShut {NoStop}%
\bibitem [{\citenamefont {Hotokezaka}\ \emph {et~al.}(2011)\citenamefont
  {Hotokezaka}, \citenamefont {Kyutoku}, \citenamefont {Okawa}, \citenamefont
  {Shibata},\ and\ \citenamefont {Kiuchi}}]{Hotokezaka2011}%
  \BibitemOpen
  \bibfield  {author} {\bibinfo {author} {\bibfnamefont {K.}~\bibnamefont
  {Hotokezaka}}, \bibinfo {author} {\bibfnamefont {K.}~\bibnamefont {Kyutoku}},
  \bibinfo {author} {\bibfnamefont {H.}~\bibnamefont {Okawa}}, \bibinfo
  {author} {\bibfnamefont {M.}~\bibnamefont {Shibata}},\ and\ \bibinfo {author}
  {\bibfnamefont {K.}~\bibnamefont {Kiuchi}},\ }\bibfield  {title} {\bibinfo
  {title} {Binary neutron star mergers: Dependence on the nuclear equation of
  state},\ }\href {https://doi.org/10.1103/PhysRevD.83.124008} {\bibfield
  {journal} {\bibinfo  {journal} {Phys. Rev. D}\ }\textbf {\bibinfo {volume}
  {83}},\ \bibinfo {pages} {124008} (\bibinfo {year} {2011})}\BibitemShut
  {NoStop}%
\bibitem [{\citenamefont {Çokluk}\ \emph {et~al.}(2023)\citenamefont
  {Çokluk}, \citenamefont {Yakut},\ and\ \citenamefont
  {Giacomazzo}}]{Cokluk:2023xio}%
  \BibitemOpen
  \bibfield  {author} {\bibinfo {author} {\bibfnamefont {K.~A.}\ \bibnamefont
  {Çokluk}}, \bibinfo {author} {\bibfnamefont {K.}~\bibnamefont {Yakut}},\
  and\ \bibinfo {author} {\bibfnamefont {B.}~\bibnamefont {Giacomazzo}},\
  }\bibfield  {title} {\bibinfo {title} {{General relativistic simulations of
  high-mass binary neutron star mergers: rapid formation of low-mass stellar
  black holes}},\ }\href {https://doi.org/10.1093/mnras/stad3752} {\bibfield
  {journal} {\bibinfo  {journal} {Monthly Notices of the Royal Astronomical
  Society}\ }\textbf {\bibinfo {volume} {527}},\ \bibinfo {pages} {8043}
  (\bibinfo {year} {2023})},\ \Eprint
  {https://arxiv.org/abs/https://academic.oup.com/mnras/article-pdf/527/3/8043/54532476/stad3752.pdf}
  {https://academic.oup.com/mnras/article-pdf/527/3/8043/54532476/stad3752.pdf}
  \BibitemShut {NoStop}%
\bibitem [{\citenamefont {Lowell}\ \emph {et~al.}(2024)\citenamefont {Lowell},
  \citenamefont {Jacquemin-Ide}, \citenamefont {Tchekhovskoy},\ and\
  \citenamefont {Duncan}}]{Lowell:2023kyu}%
  \BibitemOpen
  \bibfield  {author} {\bibinfo {author} {\bibfnamefont {B.}~\bibnamefont
  {Lowell}}, \bibinfo {author} {\bibfnamefont {J.}~\bibnamefont
  {Jacquemin-Ide}}, \bibinfo {author} {\bibfnamefont {A.}~\bibnamefont
  {Tchekhovskoy}},\ and\ \bibinfo {author} {\bibfnamefont {A.}~\bibnamefont
  {Duncan}},\ }\bibfield  {title} {\bibinfo {title} {{Rapid Black Hole
  Spin-down by Thick Magnetically Arrested Disks}},\ }\href
  {https://doi.org/10.3847/1538-4357/ad09af} {\bibfield  {journal} {\bibinfo
  {journal} {Astrophys. J.}\ }\textbf {\bibinfo {volume} {960}},\ \bibinfo
  {pages} {82} (\bibinfo {year} {2024})},\ \Eprint
  {https://arxiv.org/abs/2302.01351} {arXiv:2302.01351 [astro-ph.HE]}
  \BibitemShut {NoStop}%
\bibitem [{\citenamefont {Ruiz}\ and\ \citenamefont
  {Shapiro}(2017)}]{Ruiz:2017inq}%
  \BibitemOpen
  \bibfield  {author} {\bibinfo {author} {\bibfnamefont {M.}~\bibnamefont
  {Ruiz}}\ and\ \bibinfo {author} {\bibfnamefont {S.~L.}\ \bibnamefont
  {Shapiro}},\ }\bibfield  {title} {\bibinfo {title} {{General relativistic
  magnetohydrodynamics simulations of prompt-collapse neutron star mergers: The
  absence of jets}},\ }\href {https://doi.org/10.1103/PhysRevD.96.084063}
  {\bibfield  {journal} {\bibinfo  {journal} {Phys. Rev. D}\ }\textbf {\bibinfo
  {volume} {96}},\ \bibinfo {pages} {084063} (\bibinfo {year} {2017})},\
  \Eprint {https://arxiv.org/abs/1709.00414} {arXiv:1709.00414 [astro-ph.HE]}
  \BibitemShut {NoStop}%
\bibitem [{\citenamefont {Abbott}\ \emph {et~al.}(2020)\citenamefont {Abbott}
  \emph {et~al.}}]{LIGOScientific:2020aai}%
  \BibitemOpen
  \bibfield  {author} {\bibinfo {author} {\bibfnamefont {B.~P.}\ \bibnamefont
  {Abbott}} \emph {et~al.} (\bibinfo {collaboration} {LIGO Scientific,
  Virgo}),\ }\bibfield  {title} {\bibinfo {title} {{GW190425: Observation of a
  Compact Binary Coalescence with Total Mass $\sim 3.4 M_{\odot}$}},\ }\href
  {https://doi.org/10.3847/2041-8213/ab75f5} {\bibfield  {journal} {\bibinfo
  {journal} {Astrophys. J. Lett.}\ }\textbf {\bibinfo {volume} {892}},\
  \bibinfo {pages} {L3} (\bibinfo {year} {2020})},\ \Eprint
  {https://arxiv.org/abs/2001.01761} {arXiv:2001.01761 [astro-ph.HE]}
  \BibitemShut {NoStop}%
\bibitem [{\citenamefont {Hotokezaka}\ \emph {et~al.}(2013)\citenamefont
  {Hotokezaka}, \citenamefont {Kiuchi}, \citenamefont {Kyutoku}, \citenamefont
  {Okawa}, \citenamefont {Sekiguchi}, \citenamefont {Shibata},\ and\
  \citenamefont {Taniguchi}}]{Hotokezaka2013}%
  \BibitemOpen
  \bibfield  {author} {\bibinfo {author} {\bibfnamefont {K.}~\bibnamefont
  {Hotokezaka}}, \bibinfo {author} {\bibfnamefont {K.}~\bibnamefont {Kiuchi}},
  \bibinfo {author} {\bibfnamefont {K.}~\bibnamefont {Kyutoku}}, \bibinfo
  {author} {\bibfnamefont {H.}~\bibnamefont {Okawa}}, \bibinfo {author}
  {\bibfnamefont {Y.-i.}\ \bibnamefont {Sekiguchi}}, \bibinfo {author}
  {\bibfnamefont {M.}~\bibnamefont {Shibata}},\ and\ \bibinfo {author}
  {\bibfnamefont {K.}~\bibnamefont {Taniguchi}},\ }\bibfield  {title} {\bibinfo
  {title} {Mass ejection from the merger of binary neutron stars},\ }\href
  {https://doi.org/10.1103/PhysRevD.87.024001} {\bibfield  {journal} {\bibinfo
  {journal} {Phys. Rev. D}\ }\textbf {\bibinfo {volume} {87}},\ \bibinfo
  {pages} {024001} (\bibinfo {year} {2013})}\BibitemShut {NoStop}%
\bibitem [{\citenamefont {Radice}\ \emph {et~al.}(2018)\citenamefont {Radice},
  \citenamefont {Perego}, \citenamefont {Zappa},\ and\ \citenamefont
  {Bernuzzi}}]{Radice:2017lry}%
  \BibitemOpen
  \bibfield  {author} {\bibinfo {author} {\bibfnamefont {D.}~\bibnamefont
  {Radice}}, \bibinfo {author} {\bibfnamefont {A.}~\bibnamefont {Perego}},
  \bibinfo {author} {\bibfnamefont {F.}~\bibnamefont {Zappa}},\ and\ \bibinfo
  {author} {\bibfnamefont {S.}~\bibnamefont {Bernuzzi}},\ }\bibfield  {title}
  {\bibinfo {title} {{GW170817: Joint Constraint on the Neutron Star Equation
  of State from Multimessenger Observations}},\ }\href
  {https://doi.org/10.3847/2041-8213/aaa402} {\bibfield  {journal} {\bibinfo
  {journal} {Astrophys. J. Lett.}\ }\textbf {\bibinfo {volume} {852}},\
  \bibinfo {pages} {L29} (\bibinfo {year} {2018})},\ \Eprint
  {https://arxiv.org/abs/1711.03647} {arXiv:1711.03647 [astro-ph.HE]}
  \BibitemShut {NoStop}%
\bibitem [{\citenamefont {Paschalidis}\ and\ \citenamefont
  {Ruiz}(2019)}]{Paschalidis:2018tsa}%
  \BibitemOpen
  \bibfield  {author} {\bibinfo {author} {\bibfnamefont {V.}~\bibnamefont
  {Paschalidis}}\ and\ \bibinfo {author} {\bibfnamefont {M.}~\bibnamefont
  {Ruiz}},\ }\bibfield  {title} {\bibinfo {title} {{Are fast radio bursts the
  most likely electromagnetic counterpart of neutron star mergers resulting in
  prompt collapse?}},\ }\href {https://doi.org/10.1103/PhysRevD.100.043001}
  {\bibfield  {journal} {\bibinfo  {journal} {Phys. Rev. D}\ }\textbf {\bibinfo
  {volume} {100}},\ \bibinfo {pages} {043001} (\bibinfo {year} {2019})},\
  \Eprint {https://arxiv.org/abs/1808.04822} {arXiv:1808.04822 [astro-ph.HE]}
  \BibitemShut {NoStop}%
\bibitem [{\citenamefont {Most}\ and\ \citenamefont
  {Philippov}(2023)}]{Most:2022ayk}%
  \BibitemOpen
  \bibfield  {author} {\bibinfo {author} {\bibfnamefont {E.~R.}\ \bibnamefont
  {Most}}\ and\ \bibinfo {author} {\bibfnamefont {A.~A.}\ \bibnamefont
  {Philippov}},\ }\bibfield  {title} {\bibinfo {title} {{Reconnection-Powered
  Fast Radio Transients from Coalescing Neutron Star Binaries}},\ }\href
  {https://doi.org/10.1103/PhysRevLett.130.245201} {\bibfield  {journal}
  {\bibinfo  {journal} {Phys. Rev. Lett.}\ }\textbf {\bibinfo {volume} {130}},\
  \bibinfo {pages} {245201} (\bibinfo {year} {2023})},\ \Eprint
  {https://arxiv.org/abs/2207.14435} {arXiv:2207.14435 [astro-ph.HE]}
  \BibitemShut {NoStop}%
\bibitem [{\citenamefont {Beloborodov}(2021)}]{Beloborodov:2020ylo}%
  \BibitemOpen
  \bibfield  {author} {\bibinfo {author} {\bibfnamefont {A.~M.}\ \bibnamefont
  {Beloborodov}},\ }\bibfield  {title} {\bibinfo {title} {{Emission of Magnetar
  Bursts and Precursors of Neutron Star Mergers}},\ }\href
  {https://doi.org/10.3847/1538-4357/ac17e7} {\bibfield  {journal} {\bibinfo
  {journal} {Astrophys. J.}\ }\textbf {\bibinfo {volume} {921}},\ \bibinfo
  {pages} {92} (\bibinfo {year} {2021})},\ \Eprint
  {https://arxiv.org/abs/2011.07310} {arXiv:2011.07310 [astro-ph.HE]}
  \BibitemShut {NoStop}%
\bibitem [{\citenamefont {Lyutikov}(2019)}]{Lyutikov:2018nti}%
  \BibitemOpen
  \bibfield  {author} {\bibinfo {author} {\bibfnamefont {M.}~\bibnamefont
  {Lyutikov}},\ }\bibfield  {title} {\bibinfo {title} {{Electrodynamics of
  binary neutron star mergers}},\ }\href
  {https://doi.org/10.1093/mnras/sty3303} {\bibfield  {journal} {\bibinfo
  {journal} {Mon. Not. Roy. Astron. Soc.}\ }\textbf {\bibinfo {volume} {483}},\
  \bibinfo {pages} {2766} (\bibinfo {year} {2019})},\ \Eprint
  {https://arxiv.org/abs/1809.10478} {arXiv:1809.10478 [astro-ph.HE]}
  \BibitemShut {NoStop}%
\bibitem [{\citenamefont {Most}\ and\ \citenamefont
  {Philippov}(2020)}]{Most:2020ami}%
  \BibitemOpen
  \bibfield  {author} {\bibinfo {author} {\bibfnamefont {E.~R.}\ \bibnamefont
  {Most}}\ and\ \bibinfo {author} {\bibfnamefont {A.~A.}\ \bibnamefont
  {Philippov}},\ }\bibfield  {title} {\bibinfo {title} {{Electromagnetic
  precursors to gravitational wave events: Numerical simulations of flaring in
  pre-merger binary neutron star magnetospheres}},\ }\href
  {https://doi.org/10.3847/2041-8213/ab8196} {\bibfield  {journal} {\bibinfo
  {journal} {Astrophys. J. Lett.}\ }\textbf {\bibinfo {volume} {893}},\
  \bibinfo {pages} {L6} (\bibinfo {year} {2020})},\ \Eprint
  {https://arxiv.org/abs/2001.06037} {arXiv:2001.06037 [astro-ph.HE]}
  \BibitemShut {NoStop}%
\bibitem [{\citenamefont {Cruz~Rojas}\ \emph {et~al.}(2024)\citenamefont
  {Cruz~Rojas}, \citenamefont {Gorda}, \citenamefont {Hoyos}, \citenamefont
  {Jokela}, \citenamefont {J\"arvinen}, \citenamefont {Kurkela}, \citenamefont
  {Paatelainen}, \citenamefont {S\"appi},\ and\ \citenamefont
  {Vuorinen}}]{CruzRojas:2024etx}%
  \BibitemOpen
  \bibfield  {author} {\bibinfo {author} {\bibfnamefont {J.}~\bibnamefont
  {Cruz~Rojas}}, \bibinfo {author} {\bibfnamefont {T.}~\bibnamefont {Gorda}},
  \bibinfo {author} {\bibfnamefont {C.}~\bibnamefont {Hoyos}}, \bibinfo
  {author} {\bibfnamefont {N.}~\bibnamefont {Jokela}}, \bibinfo {author}
  {\bibfnamefont {M.}~\bibnamefont {J\"arvinen}}, \bibinfo {author}
  {\bibfnamefont {A.}~\bibnamefont {Kurkela}}, \bibinfo {author} {\bibfnamefont
  {R.}~\bibnamefont {Paatelainen}}, \bibinfo {author} {\bibfnamefont
  {S.}~\bibnamefont {S\"appi}},\ and\ \bibinfo {author} {\bibfnamefont
  {A.}~\bibnamefont {Vuorinen}},\ }\bibfield  {title} {\bibinfo {title}
  {{Estimate for the bulk viscosity of strongly coupled quark matter}},\
  }\Eprint {https://arxiv.org/abs/2402.00621} {arXiv:2402.00621 [hep-ph]}
  (\bibinfo {year} {2024})\BibitemShut {NoStop}%
\bibitem [{\citenamefont {Bozzola}(2021)}]{Bozzola:2021hus}%
  \BibitemOpen
  \bibfield  {author} {\bibinfo {author} {\bibfnamefont {G.}~\bibnamefont
  {Bozzola}},\ }\bibfield  {title} {\bibinfo {title} {{kuibit: Analyzing
  Einstein Toolkit simulations with Python}},\ }\href
  {https://doi.org/10.21105/joss.03099} {\bibfield  {journal} {\bibinfo
  {journal} {J. Open Source Softw.}\ }\textbf {\bibinfo {volume} {6}},\
  \bibinfo {pages} {3099} (\bibinfo {year} {2021})},\ \Eprint
  {https://arxiv.org/abs/2104.06376} {arXiv:2104.06376 [gr-qc]} \BibitemShut
  {NoStop}%
\bibitem [{\citenamefont {Hagedorn}\ and\ \citenamefont
  {Rafelski}(1980)}]{Hagedorn:1980kb}%
  \BibitemOpen
  \bibfield  {author} {\bibinfo {author} {\bibfnamefont {R.}~\bibnamefont
  {Hagedorn}}\ and\ \bibinfo {author} {\bibfnamefont {J.}~\bibnamefont
  {Rafelski}},\ }\bibfield  {title} {\bibinfo {title} {{Hot Hadronic Matter and
  Nuclear Collisions}},\ }\href {https://doi.org/10.1016/0370-2693(80)90566-3}
  {\bibfield  {journal} {\bibinfo  {journal} {Phys. Lett. B}\ }\textbf
  {\bibinfo {volume} {97}},\ \bibinfo {pages} {136} (\bibinfo {year}
  {1980})}\BibitemShut {NoStop}%
\bibitem [{\citenamefont {Rischke}\ \emph {et~al.}(1991)\citenamefont
  {Rischke}, \citenamefont {Gorenstein}, \citenamefont {Stoecker},\ and\
  \citenamefont {Greiner}}]{Rischke:1991ke}%
  \BibitemOpen
  \bibfield  {author} {\bibinfo {author} {\bibfnamefont {D.~H.}\ \bibnamefont
  {Rischke}}, \bibinfo {author} {\bibfnamefont {M.~I.}\ \bibnamefont
  {Gorenstein}}, \bibinfo {author} {\bibfnamefont {H.}~\bibnamefont
  {Stoecker}},\ and\ \bibinfo {author} {\bibfnamefont {W.}~\bibnamefont
  {Greiner}},\ }\bibfield  {title} {\bibinfo {title} {{Excluded volume effect
  for the nuclear matter equation of state}},\ }\href
  {https://doi.org/10.1007/BF01548574} {\bibfield  {journal} {\bibinfo
  {journal} {Z. Phys. C}\ }\textbf {\bibinfo {volume} {51}},\ \bibinfo {pages}
  {485} (\bibinfo {year} {1991})}\BibitemShut {NoStop}%
\bibitem [{\citenamefont {Vovchenko}\ \emph {et~al.}(2017)\citenamefont
  {Vovchenko}, \citenamefont {Gorenstein},\ and\ \citenamefont
  {Stoecker}}]{Vovchenko:2016rkn}%
  \BibitemOpen
  \bibfield  {author} {\bibinfo {author} {\bibfnamefont {V.}~\bibnamefont
  {Vovchenko}}, \bibinfo {author} {\bibfnamefont {M.~I.}\ \bibnamefont
  {Gorenstein}},\ and\ \bibinfo {author} {\bibfnamefont {H.}~\bibnamefont
  {Stoecker}},\ }\bibfield  {title} {\bibinfo {title} {{van der Waals
  Interactions in Hadron Resonance Gas: From Nuclear Matter to Lattice QCD}},\
  }\href {https://doi.org/10.1103/PhysRevLett.118.182301} {\bibfield  {journal}
  {\bibinfo  {journal} {Phys. Rev. Lett.}\ }\textbf {\bibinfo {volume} {118}},\
  \bibinfo {pages} {182301} (\bibinfo {year} {2017})},\ \Eprint
  {https://arxiv.org/abs/1609.03975} {arXiv:1609.03975 [hep-ph]} \BibitemShut
  {NoStop}%
\end{thebibliography}%

\appendix
\section{Construction of the EOS variant without quark matter} \label{app:NMext}

In order to isolate the effects of quark matter formation during the merger on the collapse behavior, we construct a hybrid V-QCD EOS where the quark matter phase has been removed by hand. To accomplish this we start from the intermediate EOS variant of~\cite{Demircik:2021zll}. Obtaining a consistent model is, however, more involved than simply neglecting the quark matter phase, because the nuclear matter component of this model is well-defined only up to about ten saturation densities. This is because of the van der Waals setup that was used to define the nuclear matter EOS at finite temperature: the excluded volume correction in the setup immediately sets the maximum density to the inverse of the value of the excluded volume per nucleon. Notice that the densities at which this issue appears are so high that they are not expected to play a major role in our simulations. That is, when densities exceed ten saturation densities, this happens in a limited part of the system, and collapse to a black hole is imminent, thus ending the HMNS stage which is the most interesting stage for us. Nevertheless, we want to construct a fully consistent extension of the EOS (rather than an ad-hoc extension) to densities above the limiting value set by the van der Waals setup.

In order to remove this limitation, 
we modify the excluded volume correction in the van der Waals setup such that it is smoothly turned off at extremely high densities. Because the van der Waals model is only used to extrapolate the V-QCD nuclear matter EOS from zero temperature to nonzero temperature, so that the EOS at zero temperature (both at $\beta$-equilibrium and out-of-equilibrium) remains unchanged. Therefore the modification solely affects the temperature dependence of the nuclear matter EOS above five saturation densities.

Let us then discuss the precise definition in detail. In~\cite{Demircik:2021zll} we defined the excluded volume corrected pressure though the set of equations (see also~\cite{Hagedorn:1980kb,Rischke:1991ke,Vovchenko:2016rkn})
\begin{eqnarray}
&&p_\mathrm{ex}(T,\{\mu_k\})=p_\mathrm{id}(T,\{\tilde{\mu}_k\})\,,\label{pexc} \\&&\tilde{\mu}_i=\mu_i-v_ip_\mathrm{ex}(T,\{\mu_k\})\,,
\label{eq:mucond}
\end{eqnarray}
where $p_\mathrm{ex}$ is the corrected pressure, $p_\mathrm{id}$ is the ideal gas pressure, and the indices $i,k$ run over the particles included in the van der Waals model, i.e., the nucleons, the elecrons as well as their antiparticles: $i,k \in \{n,p,e,\bar{n},\bar{p},\bar{e}\}$. We chose $v_p=v_{\bar{p}}=v_n=v_{\bar{n}}=v_0$, $v_e=v_{\bar{e}}=0$, and set $v_0=0.56\,\text{fm}^3$. However, without the need of changing the framework in any other way, we may consider a generalized definition where we replace Eq.~\eqref{eq:mucond} by
\begin{equation}
\tilde{\mu}_i=\mu_i-F_i(p_\mathrm{ex}(T,\{\mu_k\}))\, , 
\end{equation}
where $F_i(p)$ is an arbitrary function. A reasonable choice for the nucleons employs the earlier definition at small densities, which means that $F_i$ is linear in $p$, and turns off the correction smoothly at large densities, which is achieved by making $F_i$ constant.  We choose
\begin{equation} \label{eq:Fidef}
 F_i(p) = v_0\, p\, \frac{1+\alpha \left(\frac{p}{p_c}\right)^{w} +\frac{p_c}{p}\left(\frac{p}{p_c}\right)^{2w} }{1+\left(\frac{p}{p_c}\right)^{2w}} \equiv F(p) \ ,
\end{equation}
for the nucleons ($i \in \{n,p,\bar{n},\bar{p}\}$), where
\begin{equation}
 \alpha = -0.1 \ , \qquad w = 15 \ ,\qquad  p_c = 911~\textrm{MeV/fm}^3 \, .
\end{equation}
For the electrons and positrons the function $F_i$ is set to zero. 

The expression in Eq.~\eqref{eq:Fidef} may appear overly complicated. Nevertheless, we wished to choose a function that represents a rather rapid step between the linear behavior at small densities and a constant one at high densities, which can be expressed in terms of elementary functions, and for which the excluded volume $F'(p)$ decreases monotonically with increasing $p$. Using a simpler Ansatz than that of Eq.~\eqref{eq:Fidef} typically fails to satisfy some of these properties.

No other changes are necessary in the construction of the EOS. The modified correction implies that, for example, the corrected number densities are given by
\begin{eqnarray}
n_\mathrm{ex}^{(i)}(T,\{\mu_k\})&=&\frac{\partial p_\mathrm{ex}(T,\{\mu_k\})}{\partial\mu_i}\\\nonumber
&=&\frac{n_\mathrm{id}^{(i)}(T,\{\tilde\mu_k\})}{1+F'(p_\mathrm{ex}(T,\{\mu_k\}))\,\sum_l n_\mathrm{id}^{(l)}(T,\{\tilde\mu_k\})}\,,\;\;\label{nex}
\end{eqnarray}
where $n_\mathrm{id}^{(i)}=\partial p_\mathrm{id}(T,\{\tilde\mu_k\})/\partial \mu_i$ and $i$ denotes all the fermion species, while the index $l$ runs only over nucleons and antinucleons.

\begin{figure}[tbh]
 \centering
 \includegraphics[width=0.5\textwidth]{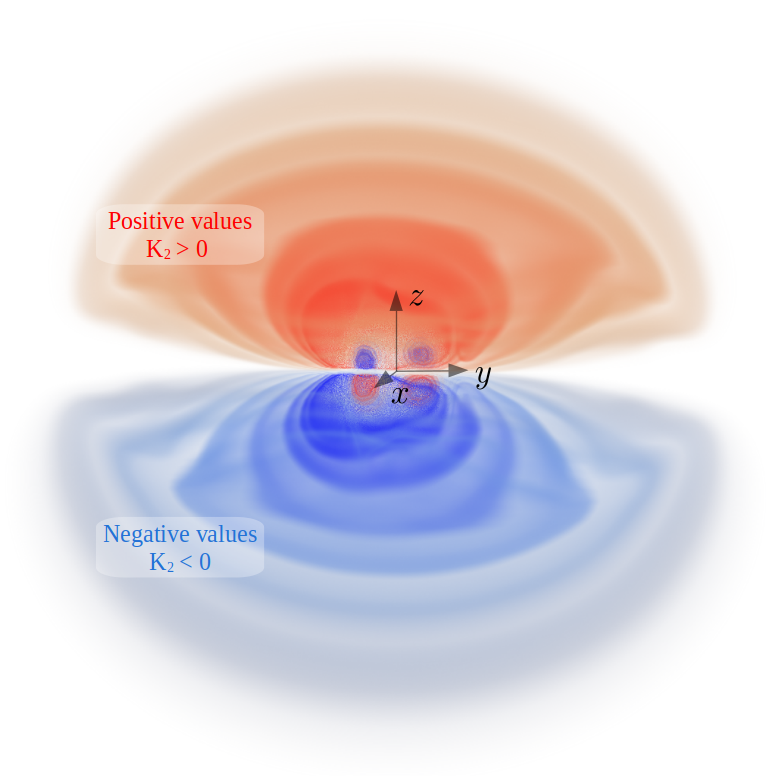}
 \caption{Iso-contours of the $K_{2}$ scalar illustrating the negative parity with respect to the equatorial plane. Positive and negative values are colored with shades of red and blue respectively, with the intensity dictated by the absolute value.}
\label{fig:curvK2_3d}
\end{figure}
\section{Additional Simulation Results}\label{app:MoreResults}
This appendix is a collection of the simulation results which we have not shown in the main text.
We start with the 3D rendering in Fig.~\ref{fig:curvK2_3d} which illustrates the global structure of the Chern-Pontryagin scalar $K_2$, in particular its negative parity with respect to the orbital plane. 
Here, the positive values of $K_2$ are shown in red and the negative values in blue colours, while values close to or identical to zero, present in the vicinity of the orbital plane and far away from the centre, are left white.
While $K_2$ displays a non-trivial structure outside the orbital plane, the fact that it vanishes in the centre clearly renders this quantity impractical to characterise the collapse properties of a BNS merger.   

\begin{figure*}[htb]
    \centering
    \includegraphics[width=0.32\textwidth]{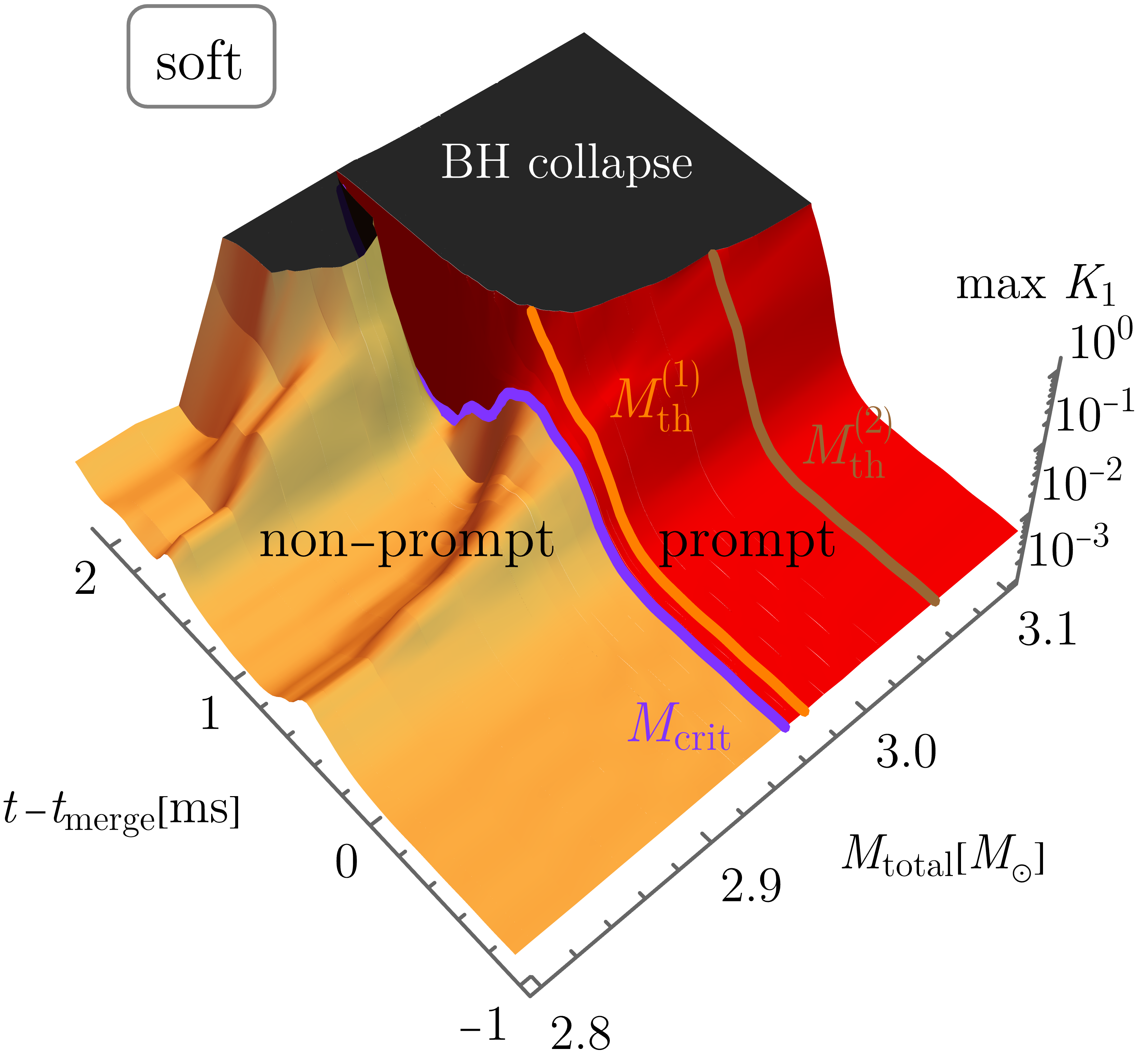}
    \includegraphics[width=0.32\textwidth]{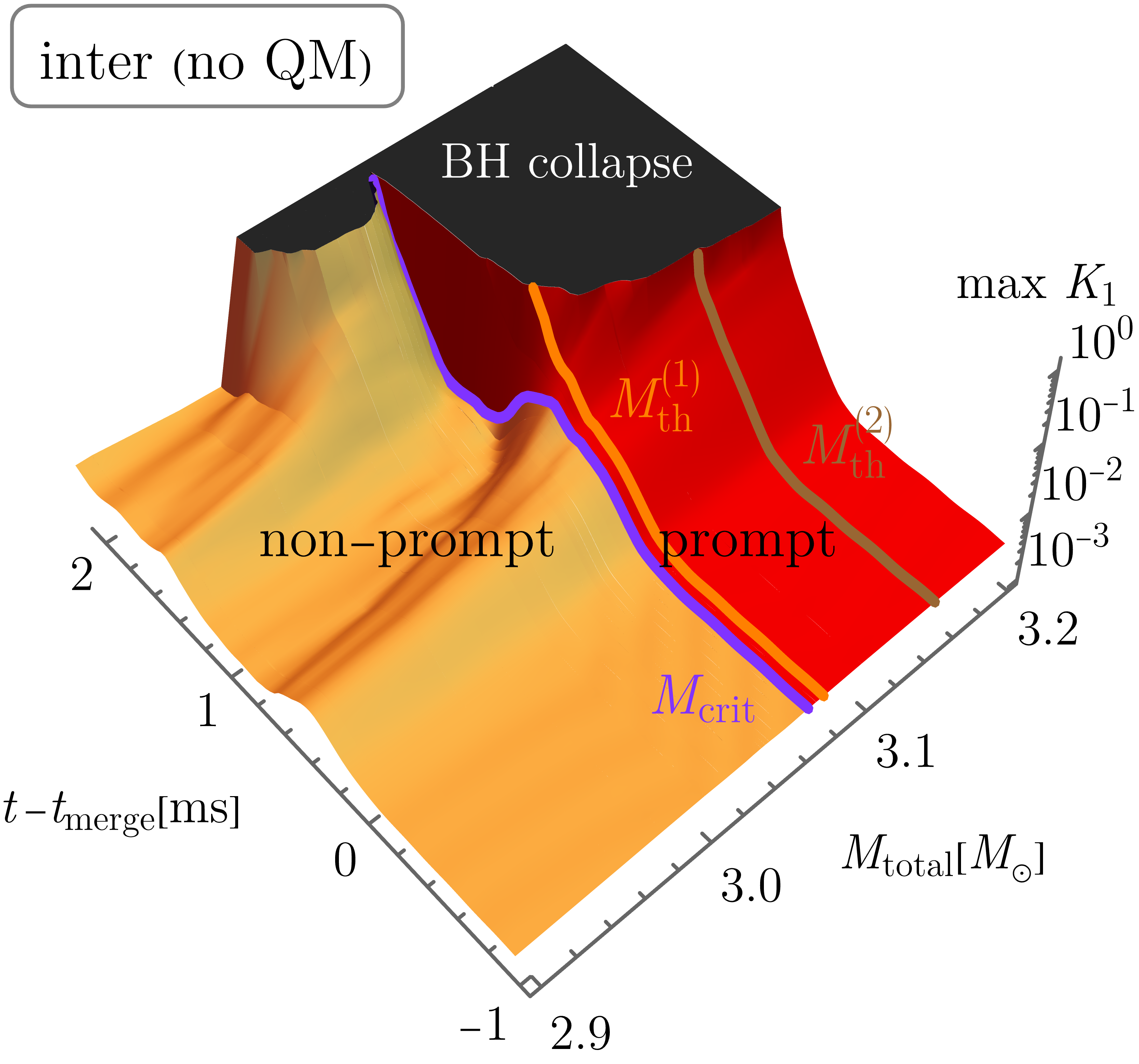}
    \includegraphics[width=0.32\textwidth]{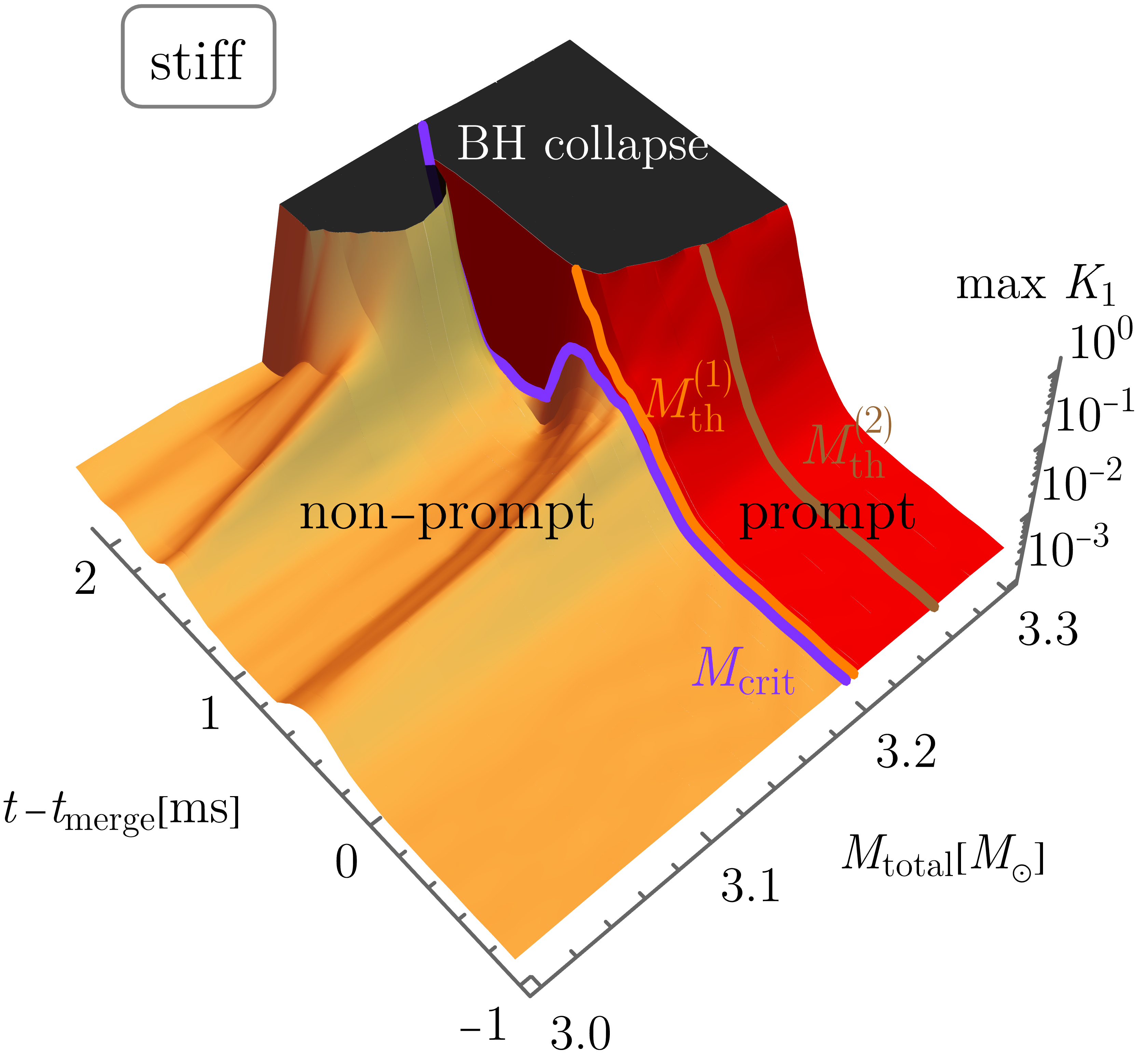}
    \caption{Maximum of $K_1$ as a function of $t-t_{\rm merge}$ and total binary mass $M_{\rm total}$ of the soft (left), intermediate without quark matter (middle) and stiff (right) V-QCD model.
    Prompt and non-prompt collapse regions are coloured in red and orange, respectively, and are separated by the critical mass $M_{\rm crit}$ marked in violet, while the black regions indicate where a black hole horizon is formed.}
    \label{fig:3Dothers}
\end{figure*}
We next compare in Fig.~\ref{fig:3Dothers} the maximal values of the Kretschmann scalar $K_1$ as a function of the total mass $M_{\rm total}$ and the post-merger time $t-t_{\rm merge}$ for the soft, intermediate without quark matter and stiff V-QCD model EOSs.  
This figure demonstrates the universality, i.e., EOS independence of the overall structure of $K_1$.
It also clearly shows the trend of $M_{\rm crit}$ with respect to the EOS-stiffness: soft models result in smaller values of $M_{\rm crit}$ than stiff models.
Furthermore, a careful comparison of the plot in the middle for the intermediate EOS without quark matter to Fig.~\ref{fig:3D} (main text) for the corresponding EOS with quark matter shows a small but clearly resolved positive shift of $\Delta M_{\rm crit}\approx 0.028 M_\odot$ in the model without quark matter.
While this small shift is probably irrelevant for any observational purposes, it nevertheless demonstrates the generic trend that quark matter tends to destabilise the post-merger remnant, as indicated by the lower value of $M_{\rm crit}$.

\begin{figure*}[htb]
\center
  \includegraphics[width=0.9\textwidth]{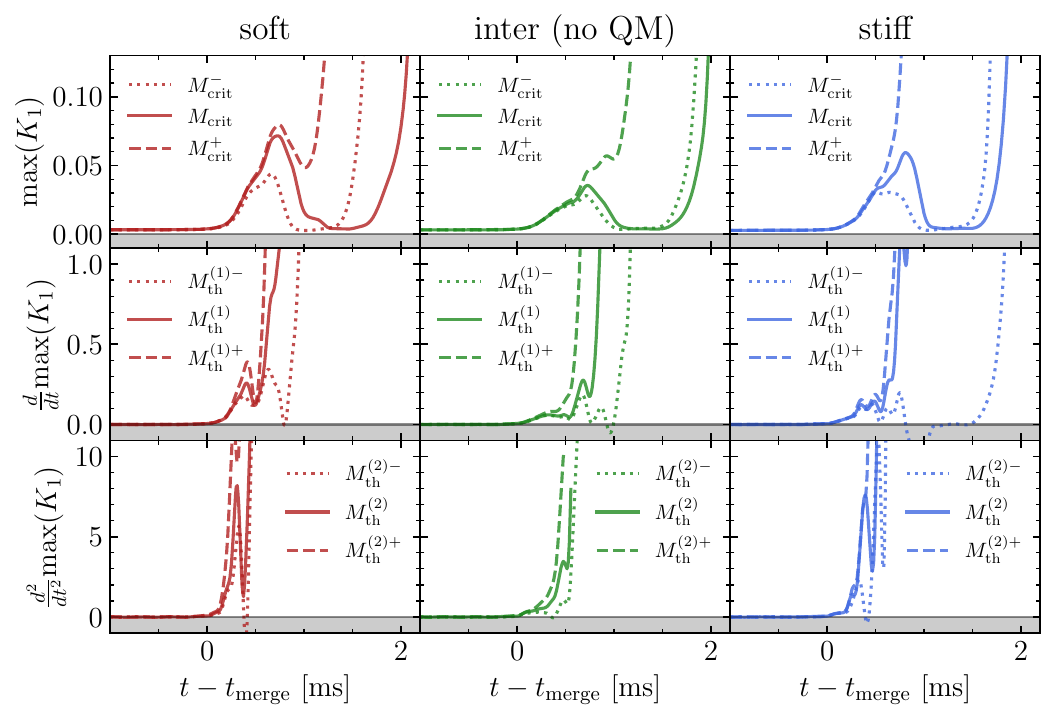}
 \caption{Spatial maxima of $K_1$ (top) and their first (middle) and second derivatives (bottom) as function of the post merger time $t-t_{\rm merge}$ of the soft (left), intermediate without quark matter (center) and stiff model (right). The central values of the various limiting masses, as determined by Eq.~\eqref{eq:crit_mass_def} and Eq.~\eqref{eq:Mth}, are plotted as solid lines, while dotted and dashed lines indicate their nearest neighbours that bracket the central values in our mass scan and the region of negative values is indicated by grey areas.}
 \label{fig:K1dtappendix}
\end{figure*}
For completeness, we show in Fig.~\ref{fig:K1dtappendix} the details of our limiting mass inference for the other EOS models considered in this work.
These figures follow the same logic as Fig.~\ref{fig:K1dt} in the main text. 
Lastly, the largest time delay in black hole formation at the critical mass takes place for the soft EOS model, as can be seen from the plot in the top left corner.

\begin{figure*}[htb]
 \centering
 \includegraphics[height=0.3\textwidth]{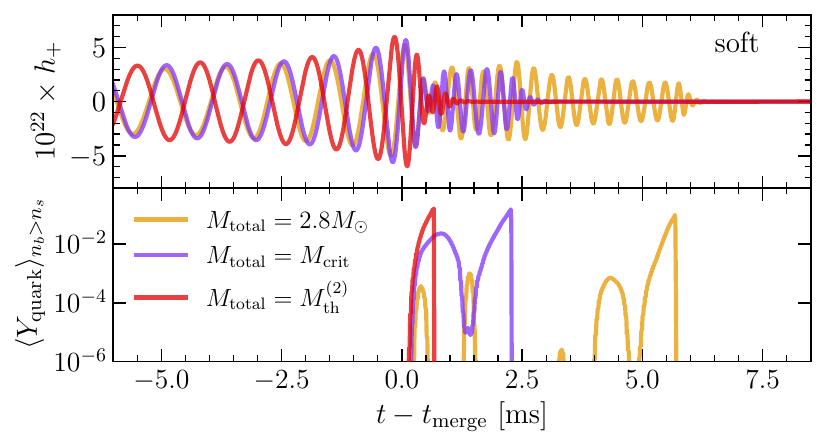}\quad
 \includegraphics[height=0.3\textwidth]{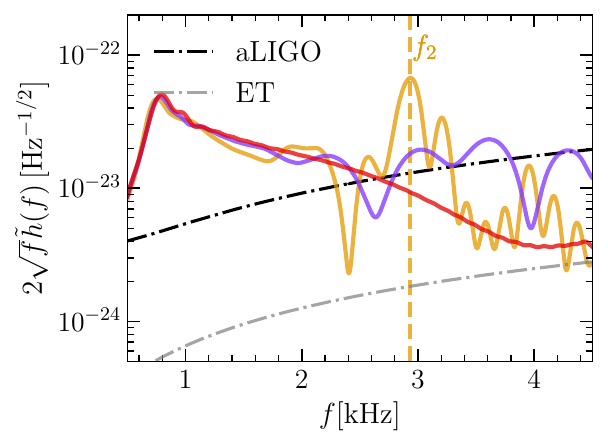}
 \includegraphics[height=0.3\textwidth]{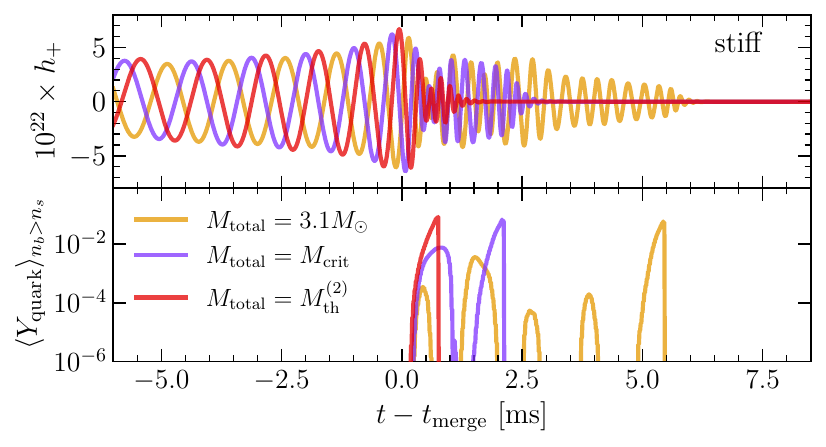}\quad
  \includegraphics[height=0.3\textwidth]{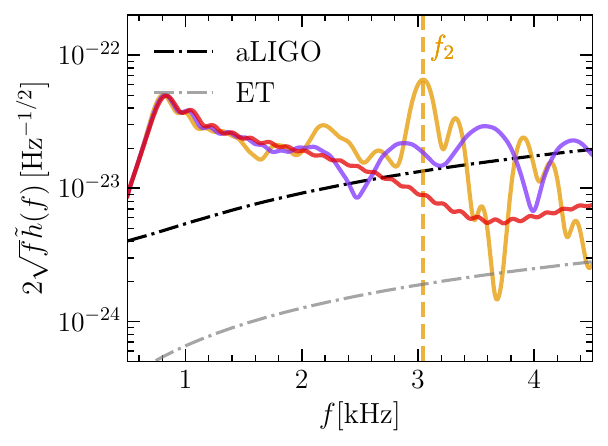}
 \caption{Left: In the top of each panel we show the plus-polarization of the GW strain extrapolated to the estimated luminosity distance of $40\,{\rm Mpc}$ of the GW170817 event~\cite{LIGOScientific:2017vwq} and viewing angle $\theta=15^{\circ}$ determined from the jet of GW170817~\cite{Ghirlanda:2018uyx} (using $\phi=0^{\circ}$ without loss of generality). Binaries with sub-critical, critical ($M_{\rm crit}$) and supercritical mass ($M_{\rm th}^{(2)}$) are shown in different colors. 
 In the bottom panels we show the corresponding quark matter fraction averaged over densities $n_b>n_s$ inside the merger remnant.
 The plots in the top row correspond to the soft and those in the bottom row to the stiff EOS model.
 Right: Corresponding PSD where we marked the dominant $f_2$ mode where it is identifiable, together with sensitivity curves of the advanced LIGO (aLIGO) detector and the Einstein Telescope (ET).
}
\label{fig:GWqm2}
\end{figure*}

Finally, we end this appendix with a discussion of the GW features and the quark matter production for the soft and stiff EOS models shown in Fig.~\ref{fig:GWqm2}.  
The results shown here are qualitatively similar to those in the corresponding Fig.~\ref{fig:GWqm} in the main text, with one notable exception, namely that the stiff model leads to the clearest periodic recurrence of quark matter in sub critical mass mergers.
As mentioned in our conclusion, it would be interesting to study this curious recurrence pattern further by including bulk viscosity for quark matter in further simulations, in particular for the stiff model where this feature appears in a very clear form.  

\section{Check of the Numeric Implementation}\label{app:RN}
As a non-trivial check of the implementation of the curvature scalars we compare the closed-form expressions for the Reissner-Nordström (RN) black hole to the numerical results obtained with our numeric code.
The line element of the RN geometry can be written as follows
\begin{eqnarray}
ds^2&=&g_{ab}dx^adx^b\nonumber\\
    &=&-f(r)dt^2+f^{-1}(r)dr^2+r^2 d\Omega_2\,,\\
    f(r)&=&1-\frac{2M}{r}+\frac{Q^2}{r^2}\,,
\end{eqnarray}
where the conserved charges $M$ and $Q$ parameterise the mass and the electric charge of the black hole.
The geometry has an inner and an outer horizon located at $r_{BH}=M\pm \sqrt{M^2-Q^2}$.
The case $|Q|=M>0$ corresponds to the critical solution where both horizons coincide, while for $Q=0$ the geometry reduces to the Schwarzschild solution with horizon radius $r_{BH}=2M$ and the second horizon coincides with the singularity at $r=0$.
For our comparison below we only consider the case $0<Q<M$ and the part of the geometry beyond the outer horizon $r\geq M+\sqrt{M^2-Q^2}$.

The RN geometry is a solution to the Einstein equations with the following energy momentum tensor
\begin{eqnarray}
T_{ab}&=&F_{ac}F^c_b-\frac{1}{4}g_{ab}F^2\,,\\
F_{ab}&=&\partial_a A_b-\partial_b A_a\,,\\
   A_a&=&\left(\frac{Q}{r},0,0,0\right)\,.
\end{eqnarray}
The energy momentum tensor is traceless $g_{ab}T^{ab}=0$, but the contraction $T_{ab}T^{ab}=\frac{Q^4}{16\pi^2 r^8}$ is non-trivial and rapidly decreases with $r$. 
The three relevant curvature invariants evaluate for the RN metric to
\begin{eqnarray}
	K_1&=&\frac{8 \left(6 M^2 r^2-12 M Q^2 r+7 Q^4\right)}{r^8}\,,\\
	K_2&=&0\,,\\
	K_3&=&-\frac{8 \left(6 M^2 r^2-12 M Q^2 r+5 Q^4\right)}{r^8}\,,
\end{eqnarray}
where $K_2=0$ follows from the spherical symmetry of the RN metric in combination with negative parity of $K_2$. 

\begin{figure}[htb]
   \centering
    \includegraphics[width=0.45\textwidth]{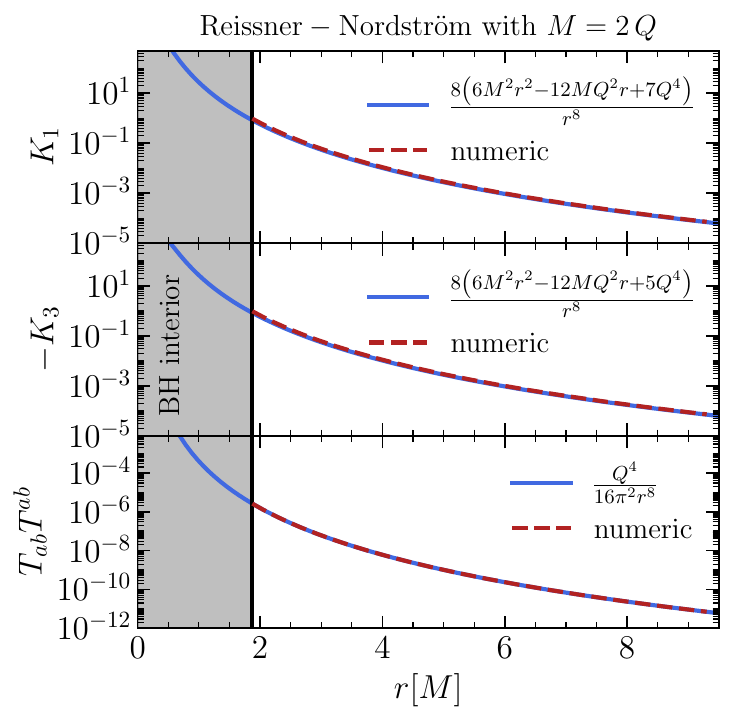}
    \caption{Comparison between the exact (blue) the numerically obtained (dashed red) solutions for $K_1$ (top), $K_3$ (middle) and $T_{ab}T^{ab}$ (bottom) for $M=1$ and $Q=1/2$.
    The gray-shaded area marks the region inside the outer black hole horizon (black line) located at $r_{\rm BH}\approx 1.866~M$.} 
\label{fig:check}
\end{figure}

In Fig.~\ref{fig:check} we compare the analytic expressions of $K_1$ (left), $K_2$ (middle) and $T_{ab}T^{ab}$ (right) for $M=1$ and $Q=1/2$ to the numerically evaluated solutions computed with the code which we also use in our binary merger simulations.
This comparison shows perfect agreement between the numeric and closed-form solution and therefore provides a non-trivial consistency check of our implementation.

\section{Free-Fall Criterion}\label{app:freeFall}
The approach of~\cite{Koppel:2019pys} uses the lapse function $\alpha$ to determine the threshold mass $M_{\rm th}$.
The criterion that defines $M_{\rm th}$ requires to determine the global minimum of the lapse function $\alpha(t)$ together with the Newtonian free-fall time $\tau_{\rm TOV}=\frac{\pi}{2}\sqrt{\frac{R_{\rm TOV}^3}{2\,M_{\rm TOV}}}$, where $M_{\rm TOV}$ and $R_{\rm TOV}$ are the maximum mass and the corresponding radius of the individual star.
In this approach, the threshold mass $M_{\rm th}$ is defined by the limit in which the binary collapses on the Newtonian free-fall timescale:
\begin{equation}
    M/M_{\rm TOV}\to M_{\rm th}/M_{\rm TOV}\,\quad {\rm for}\,\quad t_{\rm coll}\to\tau_{\rm TOV}\,, 
\end{equation}
where the collapse time $t_{\rm coll}=t_{\rm BH}-\bar{t}_{\rm merge}$ is defined as the difference between the time of black hole formation and the merger. 
The merger and black hole formation times are in general ambiguous.
However, the following definitions in terms of the global minimum of the lapse function were shown~\cite{Koppel:2019pys} to lead to robust predictions for approximately symmetric binaries
\begin{eqnarray}
\bar{t}_{\rm merge}&:&\quad {\rm min}(\alpha(t))=0.35\,,\label{eq:alphamerg} \\
t_{\rm BH}&:&\quad {\rm min}(\alpha(t))=0.2\,,\label{eq:alphaBH}\,
\end{eqnarray}
where the time $\bar{t}_{\rm merge}$ defined here via the lapse function criterion should not be confused with the one determined by the global maximum of the GW strain amplitude $t_{\rm merge}$. 
In practice, a series of different binaries with $M_{\rm total}\approx M_{\rm th}$ is simulated and used to extrapolate $M_{\rm total}$ to the point where $t_{\rm coll}/\tau_{\rm TOV}=1$, assuming vanishing slope at this point.

\begin{figure}[htb]
   \centering
    \includegraphics[width=0.45\textwidth]{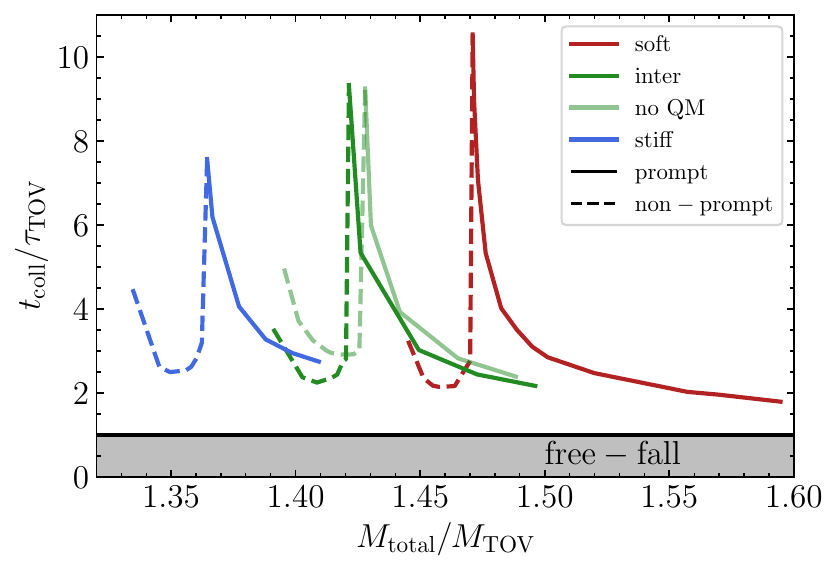}
    \caption{ Evaluation of the free-fall. Solid and dashed lines are the ratios of collapse and free fall time in the prompt and non-prompt regime, respectively. 
    The solid black line indicates the free-fall limit of equal collapse and free-fall time.}
\label{fig:freefall}
\end{figure}

In Fig.~\ref{fig:freefall} we evaluate the free fall criterion for our family of V-QCD EOSs and collect the relevant quantities in Tab.~\ref{tab:freefall}.
\begin{table}[htb]
\begin{tabular}{ccccccc}
\centering
EOS & $\tfrac{M_{\rm TOV}}{M_\odot}$ & $\tfrac{R_{\rm TOV}}{\rm km}$ & $\tfrac{\tau_{\rm TOV}}{\rm \mu s}$ & $\tfrac{\bar{t}_{\rm merge}-t_{\rm merge}}{\rm \mu s}$ & $\tfrac{t_{BH}-t_{\rm merge}}{\rm \mu s}$ & $\tfrac{t_{\rm coll}}{\tau_{\rm TOV}}$\\
 \hline
 soft          & $2.01$ & $12.41$ & $88.4$ & $324\,(252)$ & $795\,(471)$ & $5.33\,(2.47)$\\
 interm.  & $2.14$ & $12.50$ & $84.7$ & $296\,(220)$ & $768\,(436)$ & $5.34\,(2.44)$\\
 no QM         & $2.15$ & $12.50$ & $81.4$ & $272\,(219)$ & $760\,(450)$ & $5.99\,(2.83)$\\
 stiff         & $2.34$ & $12.64$ & $80.8$ & $245\,(217)$ & $857\,(482)$ & $7.57\,(3.28)$\\
 \hline
\end{tabular}
\caption{
Quantities necessary to evaluate the free-fall criterion.
The values listed are for $M_{\rm th}^{(1)}$ $(M_{\rm th}^{(2)})$ as evaluated from the corresponding min-max criteria for $K_1$.
}\label{tab:freefall}
\end{table}
In our set of simulations we find no evidence that the slope of the collapse time vanishes as the free-fall time scale is approached.
This is because free-fall is never realised in a collapsing fluid with finite internal pressure.
We therefore conclude that the threshold mass criterion defined in terms of specific values of the gauge function $\alpha(t)$ such as Eqs.\eqref{eq:alphamerg}-\eqref{eq:alphaBH}, together with the Newtonian notion of the free-fall, does not lead to a well-defined threshold mass, even though it may give rough approximations to the threshold masses defined in this article. 
\end{document}